## Roadmap

# Towards Quantum Entanglement Positron Emission Tomography (QE-PET)

Paweł Moskal[1,2,15,16], Shiva Abbaszadeh[3], Harmanjeet Singh Bilkhu[4], Peter Caradonna[5], Pragya Das[4], Praveen Gurunath Bharathi[3], Ana Marija Kožuljević[6], Zdenka Kuncic[7], Deepak Kumar[1,2], Mihael Makek[8], Marek Nowakowski[9,10], Siddharth Parashari[11], Kaustav Prasad[4], Gregory R. Romanchek[3,12], Sushil Sharma[1,2], Kenji Shimazoe[5], Ewa Stępień[1,2], Sodai Takyu[13], Miwako Takahashi[13], Mizuki Uenomachi[14], Ritesh Verma[4], and Taiga Yamaya[13]

[1]Institute of Physics, Jagiellonian University, Poland
[2]Center for Theranostics, Jagiellonian University, Poland
[3]University of California Santa Cruz, United States of America
[4]Indian Institute of Technology Bombay, Mumbai, India
[5]The University of Tokyo, Tokyo, Japan
[6]Institute for Medical Research and Occupational Health, Division of Radiation Protection, Zagreb, Croatia
[7]University of Sydney, Australia
[8]University of Zagreb Faculty of Science, Department of Physics, Zagreb, Croatia
[9]ICTP South American Institute for Fundamental Research, São Paulo, Brazil
[10]Universidade Federal de São Paulo, São Paulo, Brazil
[11]Instituto de Fisica Corpuscular (IFIC), CSIC-UV, Catedrático José Beltrán, 2, E-46980 Paterna, Spain
[12]University of Illinois Urbana–Champaign, United States of America
[13]National Institutes for Quantum Science and Technology, Chiba, Japan
[14]Institute of Science Tokyo, Tokyo, Japan
[15]Guest editor of the Roadmap.
[16]Author to whom any correspondence should be addressed. E-mail: p.moskal@uj.edu.pl

## Abstract

Annihilation photons are quantum entangled in their polarization, a property that is not accessible in state-of-the-art clinical Positron Emission Tomography (PET). This roadmap describes the current status of research in the emerging field of quantum entanglement applications involving these photons for medical diagnosis. It outlines the underlying physics phenomena and the development of detector systems that can serve as a foundation for future Quantum Entanglement PET (QE-PET) scanners. These scanners will be capable of utilizing the entanglement between annihilation photons by measuring their Compton scattering on electrons.

This roadmap comprises up-to-date experimental results on the study of quantum entanglement and the decoherence of annihilation photons, alongside the current theoretical understanding of these phenomena. The methods and detector technologies described herein are being developed (i) in order to enhance standard PET imaging by suppressing random coincidences in the reconstruction of annihilation site density distributions, (ii) in order to establish the degree of quantum entanglement as a diagnostic biomarker for tissue pathology and oxygenation, and (iii) in order to elaborate a method for pH imaging.

Whether entanglement-based imaging and pH mapping can be successfully translated into clinical practice remains an open question and a subject of exciting ongoing research. This roadmap serves as an invitation to the scientific community to join this burgeoning field.

## Contents

# 1. Introduction

*Pawel Moskal*
Marian Smoluchowski Institute of Physics
Center for Theranostics, Jagiellonian University, 31-007 Kraków, Poland
p.moskal@uj.edu.pl

**Status**

Positron Emission Tomography (PET) has revolutionized modern medicine by enabling the non-invasive quantification of physiological processes, such as metabolic rates (Standardized Uptake Value, SUV) and receptor expression (Alavi *et al* 2021). Recent advancements, particularly the development of total-body systems like the uEXPLORER, have extended these capabilities into the temporal domain, allowing for dynamic PET "movies" that track tracer kinetics throughout the entire body simultaneously (Badawi *et al* 2019). At its physical core, PET relies on electron-positron annihilation, which produces two back-to-back photons (Fig. 1A). By recording the position, energy deposition, and arrival time of these photons the annihilation event can be localized. However, image reconstruction is challenged by Compton scattering within the patient and accidental coincidences (Fig. 1B). While time windows (typically ~4.5 to 6.9 ns) and energy windows (typically 430–645 keV) are employed to filter these events (Spencer *et al* 2021), they are not entirely effective. In a clinical scan, a non-filtered scatter fraction of approximately 37% and a significant fraction of accidental coincidences, often exceeding 25% depending on the administered activity, deteriorate image quality and contrast (Spencer *et al* 2021).

Photons are spin-1 particles characterized by their momentum and transverse polarization (Bass *et al* 2023). When produced via electron-positron annihilation, these photons are quantum entangled in their polarization states. Moving beyond classical intuition, this entanglement describes a system where the photons act as a single, correlated entity regardless of their spatial separation. This is especially significant when photons originate from a pure quantum state -such as the ground state of para-positronium- where all physical properties are defined and the resulting pair exhibits maximal entanglement. Until recently, experimental efforts were centred on proving this maximal state. It was widely believed that if even one photon within an entangled pair were to undergo scattering, the entanglement would be instantly destroyed (Moskal 2024). Unexpectedly, in 2023, it was experimentally demonstrated for the first time that the quantum correlation between photons is not lost when one of them undergoes Compton scattering (Ivashkin *et al* 2023). This observation, confirmed in more detailed experimental and theoretical studies (Bordes *et al* 2024, Caradonna 2024, Parashari *et al* 2024, Tkachev *et al* 2025), sheds new light on the application of quantum entanglement, specifically regarding the potential to improve PET image contrast. Furthermore, in 2025, it was shown for the first time that the degree of quantum entanglement for photons arising from electron-positron annihilation in matter is not maximal (Moskal *et al* 2025); this indicates that the entanglement depends on the annihilation mechanism and, consequently, the tissue type. This observation demonstrated that the degree of quantum entanglement can, in principle, be used not only to reduce noise in PET images but also as a diagnostic biomarker, opening promising perspectives for the non-invasive mapping of tissue oxygenation in the human body (Moskal 2026).

The two above mentioned unexpected discoveries (Ivashkin *et al* 2023, Moskal *et al* 2025) shifted the paradigm in understanding the entanglement of photons from positron-electron annihilation. These breakthroughs have accelerated the development of Quantum Entanglement PET (QE-PET) by opening new perspectives for using the degree of quantum entanglement as a novel diagnostic indicator (Moskal 2026). Recently the first-in-human quantum entanglement imaging was demonstrated using the modular J-PET scanner (Moskal *et al* 2026).

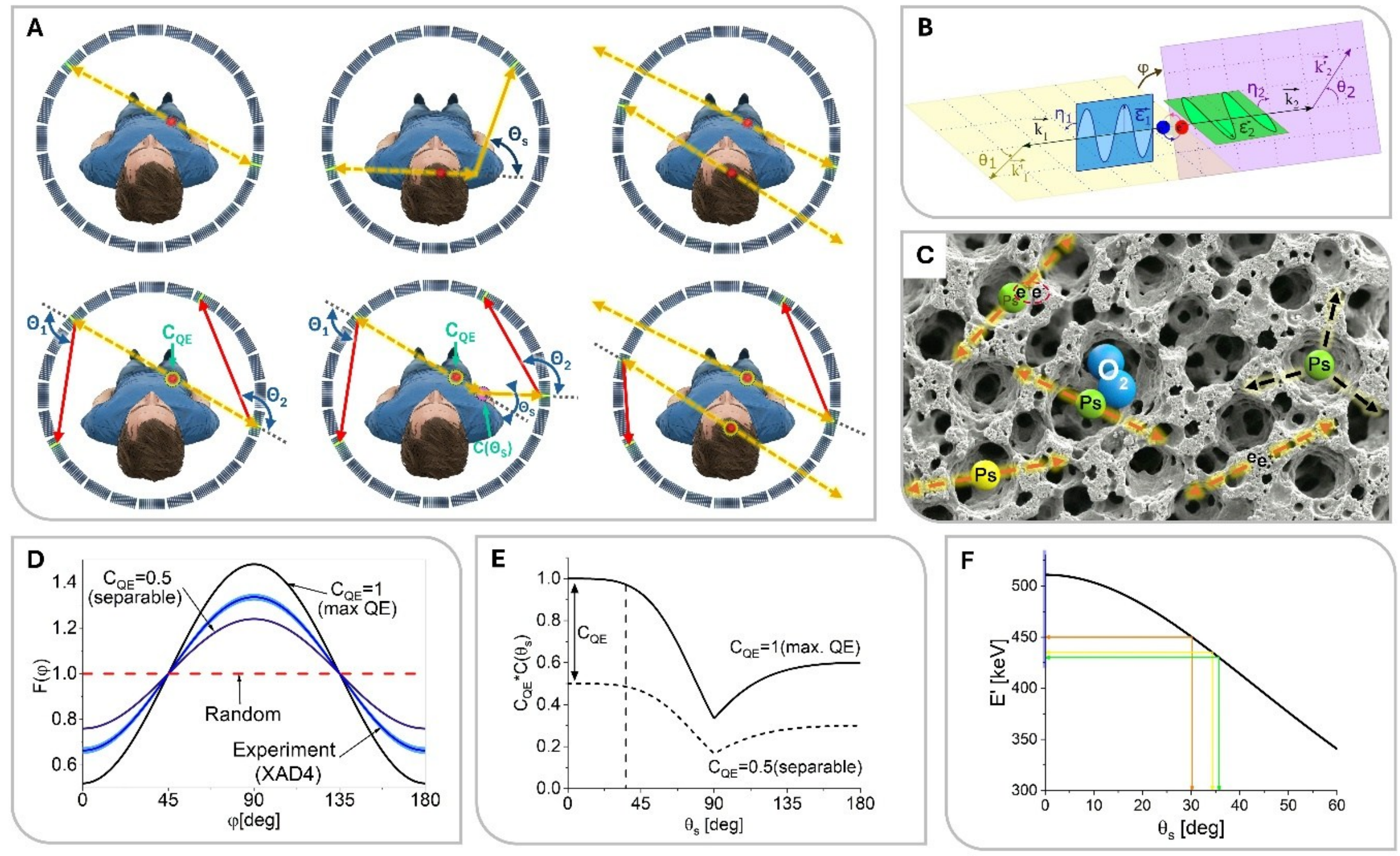

**Figure 1.** Illustration of phenomena and main results relevant for the development of Quantum Entangled PET. **(A)** Cross section for the modular J-PET scanner (Moskal *et al* 2024), the first Quantum Entanglement PET system with superimposed photons from electron-positron annihilation indicated in red-yellow dashed arrows (left), event in which one of the photons underwent Compton scattering in the body at angle $\theta_S$ (middle), and event contributing to random coincidences where registered photons originate from two different annihilations. Lower panel of this figure shows the same events as upper panel with additional scattered photons (red arrows). $C(\theta_S)$ is the measure of correlation after scattering, and $C_{QE}$ is the measure of the degree of quantum entanglement. **(B)** Pictorial illustration of annihilation photons with indicated polarization and Compton scattering planes. $k_1$ and $k_2$ indicate momentum vectors of annihilation photons, $k'_1$ and $k'_2$ denote momentum vectors of scattered photons. $\eta_1$ and $\eta_2$ denote the angle between polarization and scattering planes, $\theta_1$ and $\theta_2$ denote the polar scattering angles, and $\phi$ denotes the angle between the scattering planes of Compton scattered annihilation photons. Figs. B ©2023 American Physical Society. Reprinted with permission from (Bass *et al* 2023). **(C)** Pictorial representation of primary processes leading to the annihilation of positron in the matter. Lower right indicate direct annihilation of positron with the electron, lower left shows annihilation via formation of parapositronium (yellow circle). Also annihilation via formation of ortho-positronium (green circle) is shown for the case of self-annihilation into three photons, and annihilation into two photons via pick-off process and conversion on oxygen. **(D)** Distribution $F(\phi)$ of the relative angle $\phi$ between the scattering planes of Compton scattered annihilation photons for the scattering at angles $\theta_1 = \theta_2 = 81.7°$, where the correlation is the largest. Solid lines indicate theoretical predictions for the maximally quantum entangled photons ($C_{QE} = 1$) and separable state ($C_{QE} = ½$). Blue band indicates experimental results obtained for the annihilation in the porous polymer (Moskal *et al* 2025). **(E)** The effective measure of entanglement of annihilation photons, $C_{QE} * C(\theta_S)$, as a function of the Compton scattering angle of one of the photons ($\theta_S$) for the maximally entangled photons (solid line), and for separable photons (dashed line) (Caradonna 2024, Tkachev *et al* 2025). Vertical dashed line marks the range of $\theta_S$ values between 0 and 36° which are not filtered by the energy criteria in the current PET systems. **(F)** Dependence of the energy of the Compton scattered annihilation photon (E') as a function of the scattering angle $\theta_S$ in the patient. The vertical axis shows the energy E' of a 511 keV annihilation photon after scattering at an angle $\theta_S$. The lines indicate the lower threshold of the energy window (430 keV, 435keV ,450keV) used for scattered photon suppression in the state-of-the-art PET systems (Moskal 2026). Figs. E,F ©2026 Index Copernicus. Reprinted with permission from (Moskal 2026) under a Creative Commons Attribution License 4.0 (CC BY).

At the MeV energy scale, photons interact primarily with individual electrons, rendering traditional optical methods for measuring polarization ineffective. Instead, polarization must be inferred through Compton scattering, which exhibits a non-isotropic distribution relative to the polarization plane, as described by the Klein-Nishina formula (Klein and Nishina 1929). While an individual annihilation photon does not possess an inherent polarization, the pair produced in electron-positron annihilation is characterized by orthogonal polarization planes (In Fig. 1B $|\eta_1-\eta_2|$ = 90°). The quantum state of this two-photon system is physically manifested in the relative angle ϕ between their respective scattering planes (Fig. 1B). Consequently, the information defining the two-photon state is encoded within this angular correlation. As recently postulated, the distribution of angle ϕ may serve as a sensitive probe into the underlying annihilation mechanism (Fig. 1C), potentially varying according to the local environment and tissue characteristics (Moskal 2026).

The relative distribution of the azimuthal angle ϕ, denoted as F(ϕ), depends on the polar scattering angles $\theta_1$ and $\theta_2$. This distribution is described by the following formula (Pryce and Ward 1947, Tkachev *et al* 2025):

$$F(\theta_1, \theta_2, C_{QE}, \varphi) = F_0(1 - C_{QE}\, A(\theta_1)\, A(\theta_2) \cos(2\varphi)).$$

For 511 keV photons, which have an energy equivalent to the electron rest mass, the modulation factor A(θ) is given by (Pryce and Ward 1947):

$$A(\theta) = \frac{\sin^2\theta\,(2-\cos\theta)}{2+(1-\cos\theta)^3}.$$

The $C_{QE}$ coefficient denotes the degree of the quantum entanglement (Tkachev *et al* 2025). It is equal to 1 for a maximally entangled state and to ½ for the separable photons propagating independently of each other. The ϕ distributions for $C_{QE}$ = 1 and for $C_{QE}$ = ½ are shown in Fig. 1D. Notably, $C_{QE}$ is a newly introduced measure of quantum entanglement that remains independent of the scattering angles $\theta_1$ and $\theta_2$ (Tkachev *et al* 2025).

So far, typically the modulation factor $A(\theta_1)A(\theta_2)$ or the correlation factor R were employed as a measure of the entanglement (Bordes *et al* 2024, Ivashkin *et al* 2023, Kožuljević *et al* 2026, Moskal *et al* 2018, Moskal *et al* 2025, Romanchek *et al* 2024, Sharma *et al* 2022). This factor is defined as the ratio of the maximum to the minimum values of the distribution F(φ):

$$R_{QE}(\theta_1, \theta_2, C_{QE}) = \frac{F(\theta_1, \theta_2, C_{QE}, \varphi = 90^\circ)}{F(\theta_1, \theta_2, C_{QE}, \varphi = 0^\circ)}. \quad (11)$$

Unlike $C_{QE}$, the value of R depends on the scattering angles $\theta_1$ and $\theta_2$. At the specific angles of $\theta_1$ = $\theta_2$ = 81.7°, R reaches its maximum value of 2.84 for maximally entangled photons and 1.63 for separable (non-entangled) photons.

Fig. 1D also illustrates the recent discovery that the degree of entanglement is not maximal for photons from electron-positron annihilation in the porous polymer (Moskal 2026). The experimental result is represented by the blue band. This observation can be explained by assuming that photons created via the pick-off effect (see Fig. 1C) are separable, while photons from other processes remain maximally entangled. Although this may be a simplification, it demonstrates that the degree of entanglement, $C_{QE}$ (and equivalently R), varies as a function of the annihilation mechanism. As depicted in Figure 1C, there are four primary mechanisms leading to the emission of two 511 keV photons. These are (i) direct e+e- process, (ii) annihilation via formation of para-positronium, (iii) conversion of ortho-positronium to para-positronium upon interaction with paramagnetic molecules, such as oxygen, (iv) the pick-off process where the positron from the ortho-positronium atom annihilates with an electron from the surrounding molecular environment.

The distribution F(ϕ) is twice as enhanced at 90° for the state of maximally entangled photos compared to the separable state of photons. This enhancement at 90° can be utilized to reduce the ratio of accidental (random) coincidences by selecting events within a window around 90°. This is effective because, as shown by the dashed line in Fig. 1D, the distribution for uncorrelated photons from unrelated annihilation processes is independent of the angle ϕ. Historically, it was also thought that such filtering could reduce the fraction of events where at least one photon undergoes scattering within the patient (McNamara *et al* 2014). This was based on the common assumption that scattering would destroy the entanglement, halving the correlation (as evidenced by comparing the solid curves for $C_{QE}$ = 1 and for $C_{QE}$ = ½ in Fig. 1D). However, recently established dependencies of the entanglement degree as a function of the scattering angle $\theta_s$ (Bordes *et al* 2024, Caradonna 2024, Ivashkin *et al* 2023, Parashari *et al* 2024, Tkachev *et al* 2025), shown as a solid line in Fig. 1E, exclude the possibility of reducing the scatter fraction in PET diagnostics through ϕ angle measurements. This is because the energy windows used in current PET systems already filter out scatterings above $\theta_s$ = 36° (Fig. 1F). For events where $\theta_s$ is less than 36°, the correlation of entangled photons is not lost; the correlation measure $C(\theta_s)$ remains close to unity in this region as shown in Fig. 1E.

The determined dependence of $C(\theta_s)$ is, however, highly beneficial or the application of $C_{QE}$ as an additional diagnostic parameter in PET imaging (Moskal 2026). This is because the $C_{QE}$ value will not be significantly affected for PET events filtered by an energy window, even if one photon scatters. In such cases, the measured value is effectively the product $C_{QE}$ * $C(\theta_S)$, as shown in Fig. 1E. While $C_{QE}$ values have not yet been measured for specific human tissues, initial predictions, based on the assumption that photons from the pick-off process are separable, yield values of $C_{QE} \approx 0.89$ for adipose tissue and $C_{QE} \approx 0.82$ for water (Moskal 2026).

**Current and Future Challenges**

In order to take advantage of the quantum entanglement for suppression of random coincidences and for using the degree of quantum entanglement as a diagnostic indicator, the Quantum Entangled PET (QE-PET) must enable measurement of the angle ϕ between the scattering planes. This requires registration of two additional signals in the PET detector as indicated in the lower row of Fig. 1A. Fig. 1A shows cross section of the first QE-PET constructed from plastic scintillators (Moskal 2026, Moskal *et al* 2024) in which 511 keV photons interact predominantly via Compton scattering with the negligible fraction of the photoelectric effect of 6 10^-5 fraction only (Moskal *et al* 2021). In the case of plastic scintillators, the mean distance between the first scattering of 511 keV photon and the second scattering (the length of the red arrows in Fig. 1 A) amounts to ~7 cm. In the case of the modular J-PET geometry shown in Fig. 1A it amounts to more than 20 cm because some of the back scatter photons may reach longer distances when passing through the air in between the detectors. In the case of the crystal PET detectors Compton scattering constitute about 59% (BGO) to 60% (LYSO) of interactions (Moskal *et al* 2021) and the distance between scatterings (red arrows in Fig. 1A) is on the average equal to about 0.6 cm (Moskal *et al* 2025), with both scatterings in the same crystal block.

The sensitivity for the simultaneous registration of annihilation photons and the Compton scatter photons is at present rather low and for crystal detectors constitutes only about 3-5% of standard PET imaging sensitivity (Kožuljević *et al* 2021). Another challenge is the proper assignment of primary and secondary interaction and accidental coincidences.

In this roadmap we will discuss the current status and challenges of the development of QE-PET, including (i) description of the first full scale QE-PET prototype constructed from plastic scintillators (Moskal 2018, Moskal 2021, Moskal *et al* 2024, Moskal *et al* 2025), (ii) the double module QE-PET systems prototypes from crystal scintillators (Bordes *et al* 2024, Kožuljević *et al* 2026) and from semiconductors (Bharathi *et al* 2026), (iii) the Compton-PET development for QE-PET imaging (Shimazoe *et al* 2020, Tashima and Yamaya 2022, Yoshida *et al* 2020), including electron-tracking Compton imaging (Yoshihara *et al* 2017), combined PET-Compton Camera (Romanchek *et al* 2024) and extension of PET imaging by application of single Compton scattered photons (Das *et al* 2024). The discussion will include also (i) multiple Compton scatterings (Caradonna 2024), (ii) studies of quantum decoherence of annihilation photons (Bordes *et al* 2024, Ivashkin *et al* 2023, Parashari *et al* 2024, Tkachev *et al* 2025), (iii) the dependence of Compton scattering on quantum entanglement (Caradonna *et al* 2024), (iv) quantum entanglement in electron-positron annihilations into three photons (Nowakowski and Bedoya Fierro 2017), (v) sensitivity of total-body QE-PET system, and (vi) perspective of application of the degree of quantum entanglement as a diagnostic biomarker (Moskal 2026), and feasibility of quantum sensing and imaging of pH (Shimazoe *et al* 2024).

**Acknowledgements**
We acknowledge support from the European Union within the Horizon Europe Framework Programme (ERC Advanced Grant POSITRONIUM no. 101199807), the National Science Centre of Poland through grants no. 2021/42/A/ST2/00423, 2021/43/B/ST2/02150, 2022/47/I/NZ7/03112, the Ministry of Science and Higher Education through grant no. IAL/SP/596235/2023 and SPUB/SP/627733/2025, the SciMat and qLife Priority Research Areas budget under the program Excellence Initiative – Research University at Jagiellonian University.

## 2. Development of Plastic Scintillator-Based PET Scanners for Quantum Entanglement Imaging

*Deepak Kumar, Sushil Sharma, and Paweł Moskal*
Institute of Physics, Jagiellonian University, Poland
Center for Theranostics, Jagiellonian University, Poland

deepak.kumar@uj.edu.pl, sushil.sharma@uj.edu.pl, p.moskal@uj.edu.pl

**Status**

Positron Emission Tomography (PET) has come a long way since the first positron imaging performed in the 1950s (Wrenn *et al* 1951), achieving several milestones such as the integration of anatomical correction (PET/CT and PET/MRI) (Antoch *et al* 2009, Beyer *et al* 2000), enhancement of PET images using Time-of-Flight (TOF) information (Ter-Pogossian *et al* 1981), and the development of digital and total-body PET scanners (Cherry *et al* 2017, Surti *et al* 2020). While the major progress in PET imaging has focused on tracer uptake and metabolic activity assessment, it has largely overlooked the quantum information inherently present in the annihilation photons (McNamara *et al* 2014).

The Jagiellonian-PET (J-PET) detector uses plastic scintillators and aims to provide a cost-effective and total-body imaging solution (Moskal *et al* 2021). Developed by J-PET group in Krakow, Poland, this system is designed to overcome limitations of conventional PET scanners that rely on inorganic crystals for photon detection, to achieve superior time-of-flight resolution and an extended axial field of view by employing long strips of plastic scintillators. However, plastic scintillators have a unique advantage for studying quantum properties such as photon polarization and entanglement (Moskal *et al* 2018), owing to their dominant Compton scattering interactions with 511 keV photons (Ziessman *et al* 2013). This makes J-PET especially suited for investigating quantum-entanglement PET imaging (QE-PET), opening the door to a new dimension of diagnostic capabilities (Moskal *et al* 2025, Moskal 2026).

**Current and Future Challenges**

Despite many advancements, several challenges still remain. Traditional PET systems using crystal scintillators lack the necessary timing and angular resolution (Kemp *et al* 2006, Meng *et al* 2005) to differentiate the primary and scattered photon. Furthermore, in crystal-based PET, photons interact substantially via the photoelectric effect (Ziessman *et al* 2013), resulting in the loss of quantum information. This lower Compton interaction limits access to the polarization information crucial for QE-PET. However, a primary challenge for J-PET, which is built from plastic scintillators, stems from their low atomic number (Z) which leads to a relatively low probability of interaction for 511 keV annihilation photons, thus reducing detection efficiency compared to high-Z inorganic crystals.

Understanding and quantifying quantum entanglement in photon pairs is another challenge. Experimental parameter R-value (entanglement witness) or any such parameter (Ivashkin *et al* 2023, Parashari *et al* 2022, Watts *et al* 2021), is a statistical parameter and to use quantum properties to improve SNR is still challenging and requires highly sophisticated algorithms and substantial computational power. Another aspect of QE-PET would be to use it as diagnostic indicator (Moskal 2026). However, the behaviour of polarization correlation in tissue has yet to be determined. The lack of established polarization-correlation behaviour in biological tissue is a significant challenge for achieving a widespread adoption. The multiphoton modular J-PET prototype achieves a coincidence resolving time of ~460 ps FWHM (Kacprzak *et al* 2026), and the total-body design is expected to reach ~3.7 mm/4.9 mm spatial resolution and ~240 ps FWHM (Moskal *et al* 2021). Pushing towards even finer resolution and improving event reconstruction accuracy in challenging scenarios remains an active area of research. Finally, integrating J-PET technology with existing clinical workflows and ensuring its compatibility with various other medical imaging modalities also represent important practical challenges for its future translation into clinical applications (Bayerlein R *et al* 2026). Additionally, translating these quantum measurements into clinically useful biomarkers, such as distinguishing healthy from diseased tissue based on R-values, requires further exploration and validation.

**Advances in Science and Technology to Meet Challenges**

Unlike crystal-based systems, J-PET's plastic modules allow greater spatial separation between multiple Compton scattering events of the same photon, in turn, offering better angular resolution for reconstructing polarization vectors (Moskal *et al* 2025). Combined with high-precision timing and fast signal processing, this system can resolve the angular correlations between the scattering planes of annihilation pair photons.

The Jagiellonian-PET (J-PET) group's recent experimental campaign extends PET capabilities toward developing quantum-entanglement PET (QE-PET) (Moskal *et al* 2025). In the experiment, a $^{22}$Na source was surrounded by a porous XAD-4 medium to enhance positronium formation probability by 68%, inside a vacuum chamber (Jasińska B *et al* 2016). During a 122-day acquisition period, the system recorded over one billion events. Analysis focused on four-hit time-coincidence events, which included two primary 511 keV annihilation photons and their Compton-scattered counterparts (Fig. 1a). Measurements of the angular distribution between scattering planes confirmed the expected $\cos(2\Delta\phi)$ dependence, a signature of orthogonal polarization correlation (Bohm and Aharonov 1957).

The entanglement witness R, measured at scattering angles of 82° and 94° with ±20° windows, yielded values of 2.00 ± 0.03 and 1.93 ± 0.03, respectively. These observed R values fell between the theoretical maximum for maximally entangled photons and the theoretical minimum for fully separable photons, indicating non-maximal entanglement (Fig. 1b) (Moskal *et al* 2025). This partial entanglement is attributed to material-dependent annihilation channels within XAD-4: while direct annihilation (45%) and para-positronium self-annihilation (24%) produce entangled photons, ortho-positronium pick-off annihilation (31%) is hypothesized to create separable pairs through electron interactions. Monte Carlo simulations incorporating this mixture of processes inside the XAD-4 medium predict an R value of 2.08, which matches the experimental results (pink squares in Fig. 1b). The sensitivity of R to different annihilation mechanisms allows it to function as a molecular-environment biomarker, thereby offering potential for detecting tissue types or pathologies and, consequently, enabling tissue-environment characterization via PET imaging (Moskal 2024, Moskal 2026).

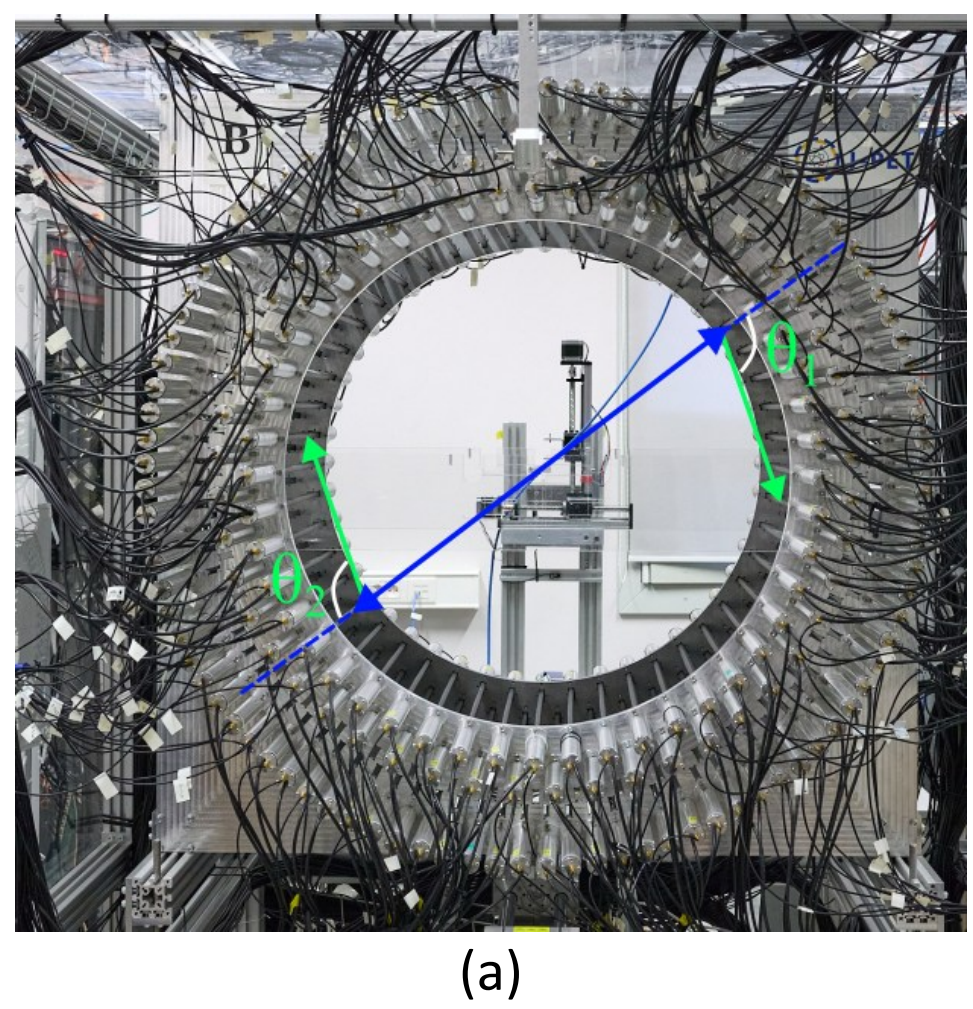

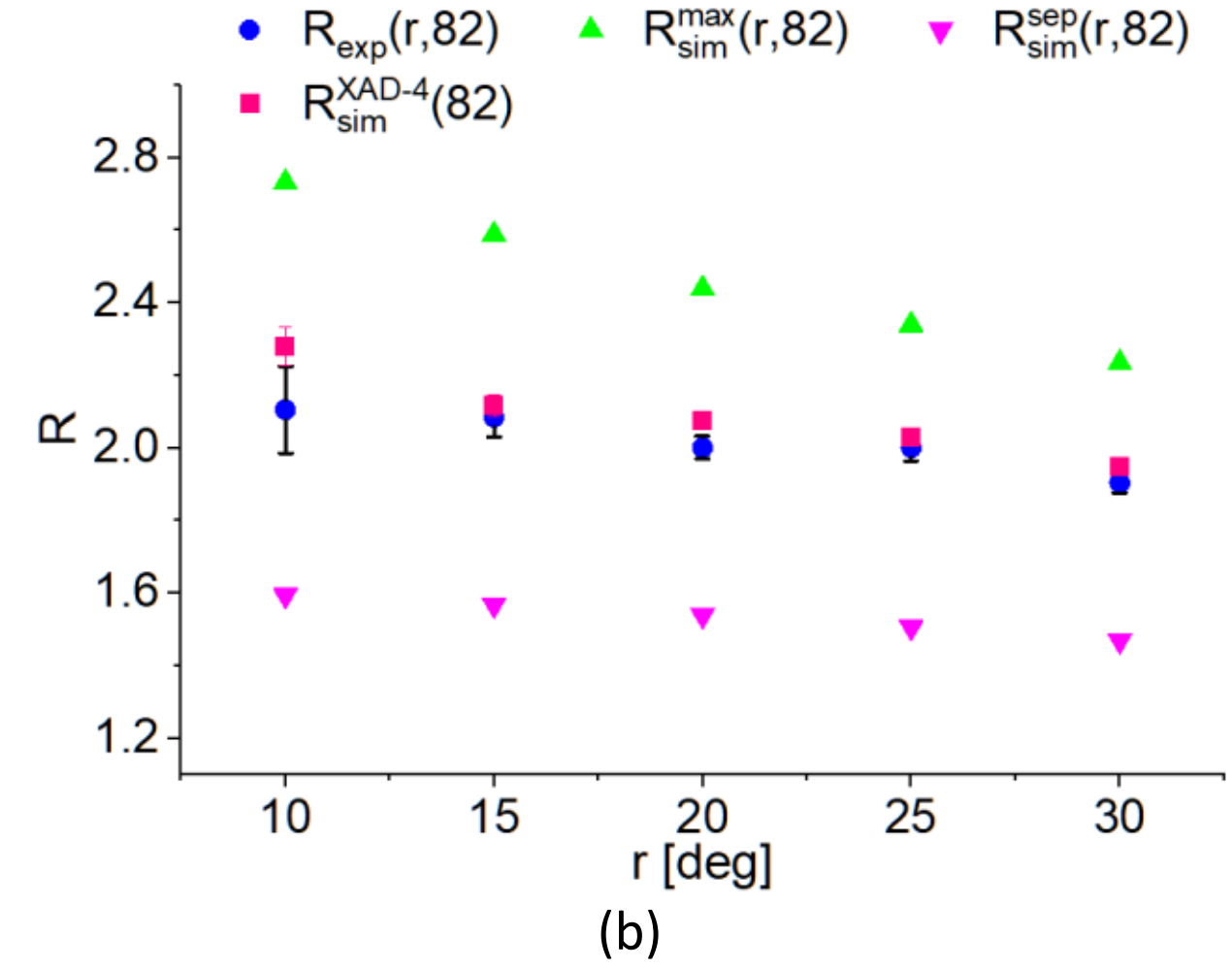

(a) (b)

**Figure 1. (a)** Photograph of the 192-strip J-PET tomograph with a superimposed illustration of the example event with annihilation photons (blue solid arrows) and scattered photons (green arrows) used in the study (Moskal *et al* 2025). **(b)** Plot showing the value of polarization correlation for the scattering angles of the pair of photons around 82 degrees within a radii 'r' as the x-axis of the plot (Moskal *et al* 2025). The plot indicates the value of R, if pair of photons maximally entanglement with green triangle, separable with magenta inverted triangles and experimental result with blue dots. The pink square represents the value of R by assuming the hypothesis in the porous material.

The current progress has not eliminated all existing challenges. The 192 strip J-PET prototype need extended data acquisition times because of the low efficiency. The calculation of entanglement through R requires complex computations while translating R-values for clinical use needs verification across different biological conditions. The study proves that quantum correlations in annihilation photons can be measured using PET based on plastic scintillators.

### Concluding Remarks

Quantum-Entanglement PET (QE-PET) represents a fundamental shift in medical imaging, utilizing the quantum properties of annihilation photons to probe tissue microenvironments in greater detail than conventional PET allows. The R-factor, a novel diagnostic biomarker, experimentally demonstrates the potential for distinguishing tissue types (Moskal 2026). More recently, a scattering angle independent entanglement measure, $C_{QE}$, has been introduced. It provides a single intrinsic measure of entanglement to which the experimentally measured R is directly linked (Moskal 2026), offering a robust, geometry-independent biomarker. Establishing $C_{QE}$ across different tissue types would mark a significant advancement for PET, enabling quantum-entanglement-based tissue characterization in clinical imaging. Future efforts will focus on advanced reconstruction algorithms, AI/ML integration, and standardization within DICOM. The J-PET group has already played a pioneering role, experimentally demonstrating how entanglement is influenced in porous media, notably confirming non-maximal entanglement (Moskal *et al* 2025). The total-body J-PET scanner is critical to QE-PET's clinical realization. With its 243 cm axial field of view and high multi-photon efficiency, it will record more than ten times the coincidences of standard scanners. This capability will facilitate voxel-wise R-mapping in routine scan times, allowing entanglement loss to be analysed alongside metabolic tracers. Furthermore, by leveraging polarization correlation, QE-PET can improve signal-to-noise ratios, which is crucial for whole-body or low-dose imaging.

### Acknowledgement

We acknowledge support from the European Union within the Horizon Europe Framework Programme (ERC Advanced Grant POSITRONIUM no. 101199807), the National Science Centre of Poland through grants no. 2021/42/A/ST2/00423, 2021/43/B/ST2/02150, 2022/47/I/NZ7/03112,

2023/50/E/ST2/00574, 2024/53/N/ST2/04279 the Ministry of Science and Higher Education through grant no. IAL/SP/596235/2023 and SPUB/SP/627733/2025, the SciMat and qLife Priority Research Areas budget under the program Excellence Initiative – Research University at Jagiellonian University.,

# 3. Development of Crystal Scintillator-Based PET Detectors for Imaging with Quantum Entangled Annihilation Photons

*Mihael Makek*[1], *Zdenka Kuncic*[2], *Ana Marija Kožuljević*[3], *Siddharth Parashari*[4]
[1]University of Zagreb Faculty of Science, Department of Physics, Zagreb, Croatia
[2]School of Physics, University of Sydney, NSW 2006, Australia
[3]Institute for Medical Research and Occupational Health, Division of Radiation Protection, Zagreb, 10000, Croatia
[4]Instituto de Fisica Corpuscular (IFIC), CSIC-UV, Catedrático José Beltrán, 2, E-46980 Paterna, Spain

makek@phy.hr, zdenka.kuncic@sydney.edu.au, akozuljevic@imi.hr, siddpara@ific.uv.es

**Status**

Scintillator crystals have established a benchmark as efficient and robust PET detectors. Development of new materials further boosted their potential to improve the energy and timing resolution of the detectors. With the introduction of silicon photomultipliers, the readout of individual small crystals became efficient, reliable and available in a compact form factor, also significantly improving the spatial resolution of PET scanners. Therefore, most state-of-the-art clinical and pre-clinical PET systems are based on scintillation-pixel detector technology (Gonzalez *et al* 2022).

Further improvement to PET imaging may come from utilization of quantum entanglement of the annihilation quanta, reflected in the strong correlation of their polarizations (Kuncic *et al* 2011, McNamara *et al* 2014). This correlation offers another handle, in addition to energy and time information, to discriminate random background where no correlations are present. The polarization of gamma-rays can be measured with appropriate Compton polarimeters. Scintillator pixel-based detectors are suitable for such task, although they typically measure the Compton scattering in dual-layer configurations, which would be very costly if scaled up to clinical systems. Therefore, a more promising approach to PET imaging with quantum entangled annihilation photons is to implement the polarization measurements in single-layer scintillation pixel-based detectors.

The development of the scintillator-pixel based approach for imaging annihilation quanta began with a proof-of-concept study (Makek *et al* 2020). That experimental setup consisted of a pair of 4x4 Lutetium Fine Silicate matrices, with 3.2 mm pitch and SiPM arrays in one-to-one match to the crystals. It proved that measuring the polarization correlations of the entangled annihilation quanta was possible with single-end readout scintillators. A follow-up study assessed the polarimetric performance of different detector configurations (Kožuljević *et al* 2021, Parashari *et al* 2022). They included GAGG crystals of 2.2 mm and 3.2 mm pitch, as well as LYSO crystals of 2.2 mm pitch. The results showed up to 15% higher polarimetric sensitivity could be achieved with more finely segmented pixels (Figure 1). On the other hand, both GAGG and LYSO performed comparably in terms of polarimetric sensitivity, pointing to the fact that the benefit of GAGG's higher light output could not be fully exploited due to the non-ideal matching with SiPM's quantum efficiency.

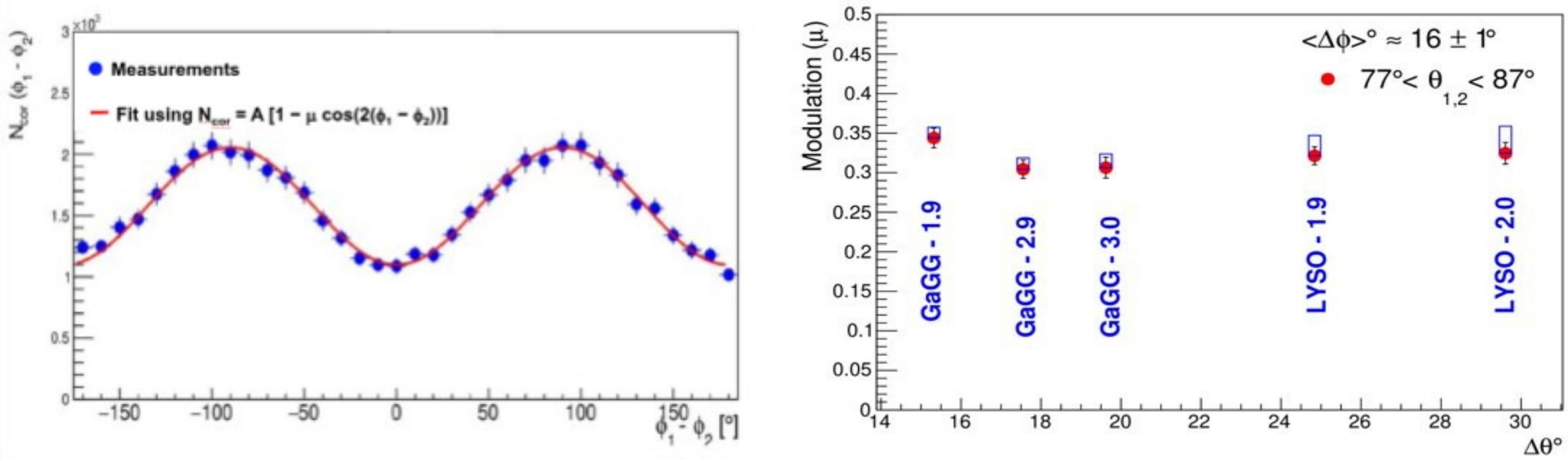

**Figure 1. Reconstructed azimuthal correlation distribution - example (left); Comparison of the polarimetric modulation factors obtained with different detector material and crystal width (right). Adapted from (Parashari *et al* 2022).**

**Current and Future Challenges**

The first PET imaging using quantum-entangled information was reported by Kim et al. (Kim *et al* 2023). They used GAGG matrices with single-end readout in a small-animal PET compatible configuration with ~80 mm diameter to image laboratory sources. These results were encouraging since they showed improvement of the signal to random background and signal-to-noise ratios, compared to standard PET event identification. A major challenge, however, was the very low count rate, only 0.1% -0.2% of standard PET.

The first pixel-based demonstrator comparable with clinical-sized PET systems, capable of exploiting quantum-entangled information, was built by the Zagreb group (Makek *et al* 2022). It encompasses a pair of modules with 3.2 mm pitch, each containing 256 GAGG:Ce crystals of ~3x3x20 mm$^3$, in 16 x 16 configuration, read out by 4 HAMAMATSU S13601-3050 SiPMs in one-to-one match, mounted on a rotating gantry with adjustable diameter (420–700 mm) to emulate a full detector ring by recording events in adjacent positions. Measurements with extended sources (two $^{68}$Ge line sources, 45 MBq each) were performed at the University Hospital Centre Zagreb using two ring diameters, 430 mm and 620 mm (Kožuljević *et al* 2024).

The results obtained by selecting the polarization-correlated Compton events (PCE) were compared to the ones obtained by selecting the events where the total energy of a gamma-quantum was absorbed in a single pixel (SPE). The comparison was made in terms of signal to random background events ratio (SBR), the spatial resolution and the relative sensitivity. The single-pixel events were selected requiring the energy deposition in 400 keV<E<600 keV range. The polarization-correlated events were selected by requiring two fired pixels in each module, with their energy sum in the 400 keV<$E_{sum}$<600 keV range. In all cases the cut on coincidence time of $|t_1-t_2|<2.5$ ns was applied. Further selection was done by the inter-pixel distance (d), the reconstructed Compton scattering angle (θ) and the range of the azimuthal angle difference, $\phi_1$-$\phi_2$. The specific criteria optimized for a high SBR were: 72°<θ<90° (maximum expected visibility) and 70°< $\phi_1$-$\phi_2$<110° (highest polarimetric correlation probability), and no limitation to inter-pixel distance, d.

The SBR=(T-B)/B was assessed from coincidence time spectra, as the ratio of events in 5 ns coincidence window around the peak (T) and events in 5 ns window away from the peak (B). The relative sensitivity was estimated as the ratio of signal S=T-B events (PCE/SPE).

Finally, the spatial resolution was assessed as the full width at half maximum (FWHM) of the profile widths of the two $^{68}$Ge rod-source images at 20 mm distance in the central FOV. The image was reconstructed in OMEGA software (Wettenhovi *et al* 2021) using the OSEM algorithm with 10 iterations and 6 subsets, without any corrections (Figure 2).

With the afore-mentioned selection criteria, an average SBR ~125 was obtained for single-pixel events, while SBR ~155 was reached for polarization-correlated events using the 620 mm diameter ring. The 430 mm ring achieved SBR ~60 (for SPE) and SBR ~75 (for PCE), showing that an average 25% improvement of SBR can be achieved (Kožuljević 2025, Kožuljević *et al* 2026). The relative sensitivity of polarization-correlated events to single-pixel events was moderate 3.2%+-0.1% with 620 mm ring diameter and 2.9%+-0.1% with 430 mm ring diameter (Kožuljević 2025). Finally, the best spatial resolution obtained in case of single pixel events was (2.9+-0.1) mm at 620 mm ring diameter and (2.8+-0.1) mm at 430 mm ring diameter. When imaging exclusively with polarization correlated events the spatial resolution was (6.1+-0.1) mm at both ring diameters, as shown in Figure 2.

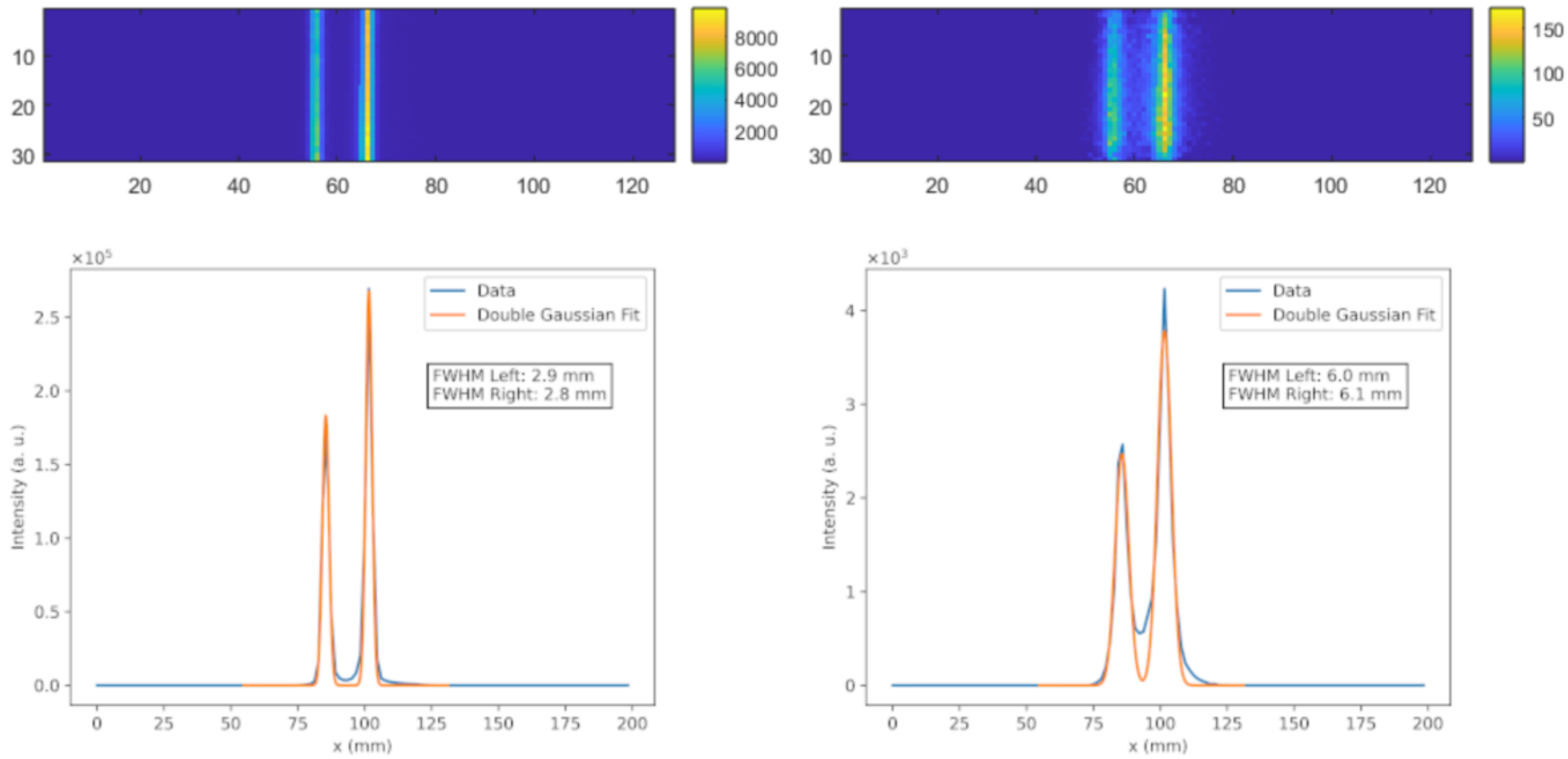

**Figure 2. Coronal images and the corresponding intensity profiles of the Ge-68 Line sources obtained with the SPE (left) and the PCE (right). (Kožuljević 2025)**

A further optimization of the event selection criteria revealed that spatial resolution of (2.5 +- 0.1) mm can be achieved for single-pixel events, while spatial resolutions below 4.5 mm can be reached with polarization-correlated events. At the same time, the relative sensitivity of the latter reaches ~ 10% (Makek *et al* 2025). Combining both events samples for imaging resulted in spatial resolution below 3.0 mm in the central field-of-view, with about 10% sensitivity improvement, demonstrating the potential of imaging with quantum-entangled Compton-scatter events.

**Advances in Science and Technology to Meet Challenges**

While the increase in signal-to-random background ratio is promising, the main challenges in imaging with the polarization-correlated events were the reduced spatial resolution and lower sensitivity compared to the single-pixel events, which are the cornerstone of the conventional PET. The first is inherent to inter-crystal scatter Compton events in general, due to the ambiguity in the interaction order determination. A possible way forward would be using depth-of-interaction (DOI) capable detectors. This would reduce the ambiguity, however it would not eliminate it completely, especially in the most sensitive region for the polarization measurements around $\theta$~90°, where both interactions happen at the same depth. When scaling up laboratory setups to clinical PET devices it is important to keep the module design robust and cost efficient. A promising approach in that sense may be to use the modules based on scintillator pixels with an addition of a light sharing layer (Zatcepin *et al* 2020). They would not be significantly more complex than existing PET technology, however their performance in Compton event detection in terms of energy resolution, timing resolution and depth-of-interaction yet needs to be verified and possibly improved by selecting specific materials and pixel geometries. The second challenge of imaging with the polarization-correlated events is the relatively low coincidence count compared to single-pixel events. One possibility to increase the relative coincidence count is to use broader selection criteria for polarization-correlated events, compromising on SBR and spatial resolution (Kožuljević *et al* 2026), or to use scintillator materials with a more favourable Compton-to-photoelectric interaction probability.

Scientific and technological advances needed to address both challenges must be based on understanding the underlying physics of entangled quanta. This includes understanding the behaviour of the entanglement correlations before and after Compton scattering, a subject that is still under experimental and theoretical study (see Sections 9, 10, 11). Development of efficient simulations (e.g. Žugec *et al* 2026) and implementation of the entanglement physis will also play a role in a more rapid advancement of this technology.

**Concluding Remarks**

It has been shown that scintillator pixel-based detectors can measure the polarization correlations of annihilation quanta. More importantly, is has been demonstrated that single-layer Compton polarimeters are suitable for this task (Makek *et al* 2020, Parashari *et al* 2022), which is of practical importance for scaling up the system to clinical size. Further, it has been demonstrated that PET imaging is possible with single-layer Compton polarimeters (Kim *et al* 2023, Kožuljević *et al* 2024, Kožuljević 2025). These initial results revealed that random background rejection can be improved using the polarization-correlated events compared to standard PET event identification (Kim *et al* 2023, Kožuljević 2025, Kožuljević *et al* 2026), with a moderate increase of the total sensitivity. The main identified challenges of PET imaging with the polarization-correlated events are hence to improve the sensitivity and the address the ambiguity in the first pixel hit in the Compton scattering to avoid degradation of the spatial resolution. Although these improvements may not seem quantitatively high, it is crucial to notice that the polarization-correlated events shall be used as an addition to conventional event selection in PET, thus improving the total sensitivity and signal-to-background ratio. Moreover, this improvement in performance could be achieved by implementing algorithmic modification of the standard pixel-based PET system, without the need for implementation of complex new hardware.

**Acknowledgements**

This work was supported by the Croatian Science Foundation under the project number IP-2022-10-3878 and by the “Research Cooperability” Program of the Croatian Science Foundation, funded by the European Union from the European Social Fund under the Operational Programme Efficient Human Resources 2014–2020, grant number PZS-2019-02-5829, and by the European Union – NextGenerationEU through the National Recovery and Resilience Plan 2021-2026 Institutional grants of University of Zagreb Faculty of Science (ProPubFO).

# 4. Development of Semiconductor-Based PET Systems for Imaging with Quantum Entangled Annihilation Photons

*Gregory R Romanchek*[1,2], *Shiva Abbaszadeh*[2]
[1]Department of Nuclear, Plasma, and Radiological Engineering, Grainger College of Engineering, University of Illinois Urbana-Champaign, Urbana, IL 61801, United States of America
[2]Department of Electrical and Computer Engineering, Baskin School of Engineering, University of California Santa Cruz, Santa Cruz, CA 95064, United States of America
gromanch@ucsc.edu, sabbasza@ucsc.edu

**Status**

Although cadmium zinc telluride (CZT) has been widely implemented in single-photon emission computed tomography (SPECT) and gamma cameras, its adoption in positron emission tomography (PET) has been limited. In conventional PET systems, full-energy absorption events are typically preferred because they help reduce the effects of scatter coincidences and eliminate the need to handle complex multiple event pairing and time ordering. This preference, however, disadvantages CZT-based systems, which have a lower linear attenuation coefficient (~0.567 $cm^{-1}$ at 511 keV (Redus 2002)) and time resolution compared to scintillator crystals like lutetium yttrium orthosilicate (LYSO, ~0.83 $cm^{-1}$ at 511 keV (Kemp *et al* 2001)). As a result, significantly thicker CZT detectors are needed. Yet, building large-volume, high-purity CZT crystals is expensive and technically challenging, which limits the scalability of CZT-based PET systems. In contrast, quantum entanglement PET (QE-PET) represents a paradigm shift because it focuses on scattering kinematics as the key measurable, not full-energy absorption.

When para-positronium decays, or more generally when electron–positron annihilation proceeds through a singlet two-photon state, the resulting annihilation photon pair emerges in an entangled state and consequently has Compton scattering kinematics distinct from non-entangled photon pairs (such as random coincidences) (Bohm and Aharonov 1957). In 2021, Watts et al. drew attention to the practical application of quantum entanglement in PET through their double Compton scatter (DCSc) methodology (Watts et al., 2021). A DCSc coincidence occurs when each photon scatters within the detector and is subsequently absorbed such that scattering kinematics can be extracted. Their simulation study demonstrated scatter signal correction (within the image domain) based on the expected difference in azimuthal scattering angle ($\Delta\phi$) described by the entanglement-informed double differential scattering cross section. Their implementation of these kinematics in GEANT4 was experimentally validated with a CZT-based PET-demonstration apparatus developed by the Kromek Group. This experimental measurement of enhancement ratio – the frequency ratio of the entangled gammas scattering with $\Delta\phi = 90°$ to them scattering with $\Delta\phi = 0°$ – was the first of its kind. While their findings regarding scattering decoherence were eventually contradicted with a refined evaluation (Parashari *et al* 2024), the potential for randoms correction remains.

While non-CZT systems have since demonstrated QE-PET capabilities (Kim *et al* 2023), the greater scatter potential, energy resolution, and hit localization of CZT are immensely desirable for collecting and analyzing DCSc coincidences. Despite this, full system investigations with CZT-based systems have yet to be conducted. This gap is largely attributable to the limited availability of CZT-PET scanners compared with conventional scintillator-based systems.

**Current and Future Challenges**

Rather than focusing on the 511 keV peak of annihilation gamma full energy absorption events, Multiple Interaction Photon Events (MIPEs) which integrate to 511 keV must be considered. While CZT offers superior intrinsic energy resolution ($3.06 \pm 0.39\%$ FWHM at 511 keV (Gu *et al* 2011)) compared to conventional scintillators ($15 \pm 2\%$ FWHM at 511 keV for LYSO (González *et al* 2016)), its charge carrier mobility–lifetime products ($\mu_{\tau_e} \sim 1 \times 10^{-2}\ \mathrm{cm^2/V}$ and $\mu_{\tau_h} \sim 1 \times 10^{-4}\ \mathrm{cm^2/V}$ (Iniewski 2014)) constrain the achievable timing resolution, often exceeding ~10 ns (Iniewski 2014, Meng and He 2005). And so, timing information cannot be used for event ordering. Energy threshold optimization, defined as the minimum deposited energy required to trigger a readout channel while remaining above the system noise floor, is another important consideration. Low-energy electronic noise can confound event pairing algorithms, which have previously been designed to reject exactly these events. Imprecise energy thresholds may bias $\Delta\phi$ in unforeseen ways. The original impetus for Compton scatter event recovery was to enable MIPE recovery to improve signal strength in traditional PET imaging due to the elevated scatter cross-section of CZT (Abbaszadeh *et al* 2018). With attention turned towards DCSc events, we find such endeavors have dual purpose.

The effective pixelation/voxelization of CZT detectors can permit near-mm event positioning, which is ideal for extracting scatter vectors. For a $1 \times 3 \times 1$ mm photon interaction resolution, we have shown a position-derived average polar angular resolution of $9.30°$ for a CZT-based system (Yang *et al* 2019). Using the same system in that report, we find an enhancement ratio of $\mathrm{R} \approx 1.4$ for a simulated dataset of 50% true DCSc ($\gamma_{\mathrm{ann}_1}$ paired with $\gamma_{\mathrm{ann}_2}$) and 50% random DCSc ($\gamma_{\mathrm{ann}_{1||2}}$ paired with $\gamma_{\mathrm{prompt}}$) – well below the optimum of $\mathrm{R_{opt}} = 2.85$ (Romanchek *et al* 2024b). Inherently, longer scatter ranges result in higher accuracy angular measurements. In Watts et al.'s demonstration, long scatters between distant pixels claimed a 2.9° azimuthal angular resolution while proximal scatters floundered at 20.4° (Watts *et al* 2021). These values corroborate our findings for 1 mm pixel size with $\Delta\phi$ resolution range from $\sim 0°$ to $\sim 20°$ with a median of $\sim 4.4°$ (Romanchek *et al* 2024a). While enhancement modulation in $\Delta\phi$ is prominent (ranging from 1 to 2.85), it is not of great magnitude. Such errors in polar and azimuthal scattering angles can significantly impact the efficacy of these measurements. Overcoming this hurdle involves turning to the excellent energy and spatial resolutions CZT can offer.

**Advances in Science and Technology to Meet Challenges**

QE-PET requires large detector volumes, fine interaction and angular resolution, and advanced MIPE recovery. The UCSC 2-Panel PET scanner was originally designed as a high-resolution, large detector-volume CZT-based PET imaging system optimized for head and neck imaging applications (Li *et al* 2020, Enlow *et al* 2024). It consists opposing $200 \times 80\ \mathrm{mm}^2$ detector panels, each incorporating a $5 \times 16$ array of densely packed, edge-on-oriented CZT crystals ($40 \times 5 \times 40\ \mathrm{mm}^3$). The edge-on configuration side-steps the CZT growth limitations as the photon trajectory is incident on 40 mm thickness of the CZT crystal. Each CZT element is instrumented with a cross-strip readout: 39 anode strips at 1 mm pitch along the x-axis and 8 cathode strips at 5 mm pitch along the y-axis. This configuration achieves $\sim 1 \times 2.5\ \mathrm{mm}^2$ in-plane interaction resolution via charge sharing, and sub-millimeter depth-of-interaction precision along the z-axis through cathode-to-anode signal ratio methods. The maximum reported image spatial resolution is 0.58 mm (Li *et al* 2020). More importantly for QE-PET, the maximum energy resolution is $5.35\% \pm 1.08\%$ at 511 keV (Enlow *et al* 2024). Unlike conventional Compton camera architectures, which typically employ separate scatterer and absorber detector layers, the thick edge-on CZT geometry of the UCSC system permits both the initial Compton interaction and subsequent photon absorption to occur within a single detector panel. As stated, the position-derived polar angular error sits at $9.30°$. When using deposited energy (at 4% energy resolution) instead, this average error shrinks to 2.65° (Yang *et al* 2019). For the 50/50 simulated dataset, the enhancement ratio based on position-derived polar angle is $\mathrm{R} \approx 1.4$ while it is $\mathrm{R} \approx 1.8$ for an energy-based approach (Romanchek *et al* 2024b), this can be seen in the red-triangle curve in fig. 1.

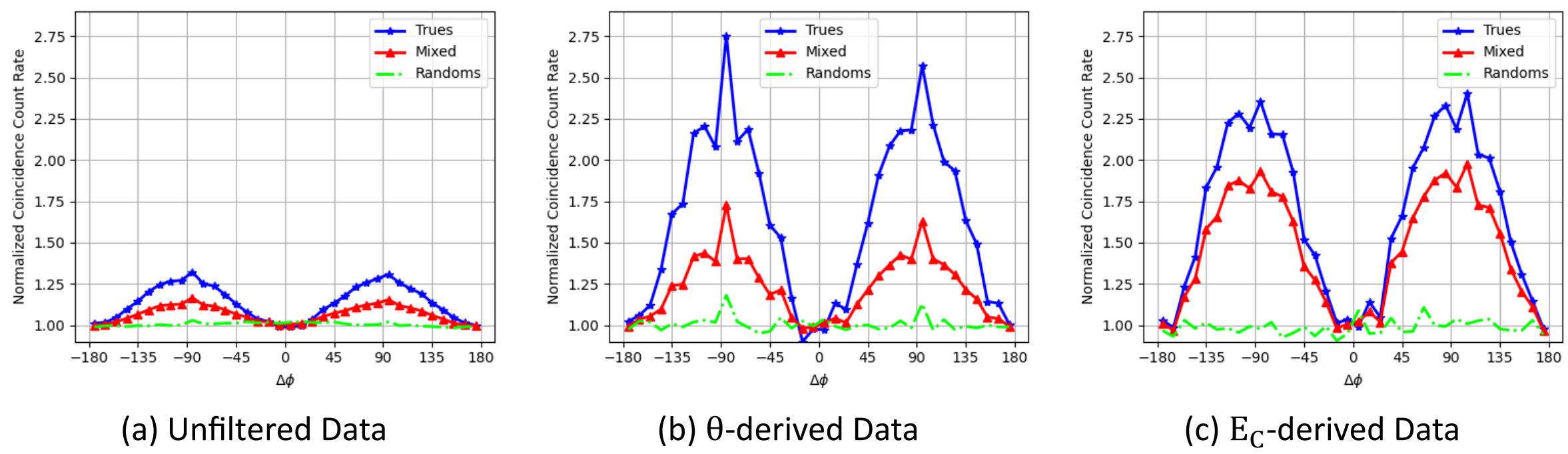

(a) Unfiltered Data (b) θ-derived Data (c) $E_C$-derived Data

**Figure 1.** Azimuthal scattering angle difference (Δϕ) for each portion of the synthetic dataset: total (mixed) in red-triangles, the True portion of the dataset in blue-stars, and the Random portion of the dataset in green-dashes. Each subplot displays the distribution after the given filter: (a) no filter, (b) a position-derived polar filter of $\theta \in [73.12, 91.68]°$, and (c) a Compton energy filter of $E_C \in 235 \pm 10\%$ keV equivalent to the polar range. (Romanchek *et al* 2024b)

Ordering events within the DCSc chain additionally relies on the excellent energy resolution. Using the direction difference angle approach, we can compare the energy-derived and position-derived polar scattering angles to optimally order activation events (Chinn and Levin 2011). Annihilation photons scatter (at least once) in 77.2% of interaction events while full energy absorption occurs with only 22.8% frequency (Abbaszadeh *et al* 2018). For a pair of annihilation gammas in our 2-Panel system, a dual Compton scatter is the dominate coincidence case and accounts for ~31.13% of detected coincidences (Romanchek *et al* 2024b) – a total of 6.81% of all coincidences are DCSc events (Romanchek *et al* 2024a). Such an environment is information dense with known "structure" – following that described by the Klein-Nishina formula. We have demonstrated that a dense neural network trained for Compton event recovery attains superior performance over that of rules-based event positioning; 82% accuracy vs 43% accuracy for DCSc coincidences (Nasiri and Abbaszadeh 2021). We posit that a physics-informed machine learning algorithm can similarly leverage the additional information embedded within the entanglement-based scattering kinematics for event sorting and even coincidence classification.

**Concluding Remarks**

CZT has emerged as a transformative technology in nuclear medicine, offering excellent spatial and energy resolution. Their compact and modular design enables the integration of multiple imaging modalities such as Compton imaging, SPECT, and even PET or CT into a single, versatile platform, facilitating advanced diagnostic workflows and faster, lower-dose scanning. Recent advances have demonstrated that CZT detectors can leverage quantum entanglement in PET imaging, utilizing the angular correlations of annihilation photons. This synergy between CZT's physical detector properties and quantum information processing opens new frontiers for hybrid imaging, multiple PET tracers, and isotopes with prompt gamma enabling more detailed and reliable anatomical and functional assessments. As a result, CZT technology stands at the forefront of next-generation imaging, empowering clinicians and researchers with powerful tools for comprehensive and personalized diagnostics.

**Acknowledgements**

All authors declare that they have no known conflicts of interest in terms of competing financial interests or personal relationships that could have an influence or are relevant to the work reported in this paper. The authors acknowledge the support from the National Institute of Biomedical Imaging and Bioengineering of the National Institutes of Health under Award Numbers R01EB028091 and UG3EB034686.

## 5. Development of Whole-Gamma Imaging for Quantum-Entanglement PET

*Taiga Yamaya, Sodai Takyu, Miwako Takahashi*
Institute for Quantum Medical Science, National Institutes for Quantum Science and Technology (QST), Chiba, Japan
yamaya.taiga@qst.go.jp, takyu.soudai@qst.go.jp, takahashi.miwako@qst.go.jp

**Status**

Sensitivity is the distinctive advantage of PET. Another major modality in nuclear medicine imaging, SPECT, can obtain spatial information only by compromising sensitivity due to the necessity of collimating incoming radiation. In nuclear medicine diagnostics, maximizing sensitivity—i.e., increasing the number of detected-radiation counts— is crucial for improving image quality. Clinically, the superior sensitivity of PET facilitates early lesion detection, reduces scan time, and minimizes radiation exposure.

The 4π geometry of PET detectors maximizes radiation measurement by capturing emissions from all directions. While this approach has been implemented in a small animal PET system (Lovatti *et al* 2023), it remains impractical for clinical applications. Instead, although not perfect, maximizing the solid angle for radiation detection has been effectively achieved in such clinical PET systems as total-body PET and brain-dedicated PET, significantly enhancing sensitivity and imaging quality (Badawi *et al* 2019, Akamatsu *et al* 2022, Takahashi *et al* 2022).

Another important consideration is the use of Compton scattering. When Compton scattering occurs inside a patient's body, the scattered photons should be removed as noise. However, when scattering occurs within the detector, the deposited energy information can be used to determine the direction of incoming radiation on the surface of the Compton cone based on the Klein-Nishina formula (Klein and Nishina 1929). Compton cameras, utilizing this principle, have been actively developed, yet none have been implemented in clinical practice (Tashima and Yamaya 2022). If a Compton camera can be integrated into PET without degrading its performance, such a combined Compton-PET imaging system could utilize a single 511 keV photon for imaging, in addition to annihilation photon pairs, thereby enhancing sensitivity.

Furthermore, if Compton-PET imaging becomes feasible, it would necessitate a paradigm shift in understanding of positron-emitting nuclides that also emit single-gamma rays, such as zirconium-89 ($^{89}$Zr). While these non-pure positron-emitting nuclides have not been widely utilized in clinical practice, a Compton-PET imaging system could employ single-gamma rays for imaging, potentially further enhancing sensitivity and expanding the medical applications of such radionuclides.

**Current and Future Challenges**

One implementation of the combined Compton-PET imaging system is a multi-ring detector geometry, where the inner detector ring functions as the scatterer and the outer detector ring serves as the absorber in Compton imaging (Yoshida *et al* 2020). PET measurements are also possible through various coincidence detection modes, including outer-outer, inner-inner, and inner-outer coincidences. This concept, whole gamma imaging (WGI), was validated using first-generation prototypes (Fig. 1a shows one of them), which successfully produced 909 keV Compton images of a $^{89}$Zr-injected mouse. The resulting images corresponded with PET images obtained from the same $^{89}$Zr distribution; Figure 1b presents example images (Tashima *et al* 2020).

This WGI prototype employed an outer absorber ring composed of four-layer DOI detectors with 2.8 x 2.8 x 7.5 mm$^3$ scintillation crystals (Ce-doped gadolinium oxyorthosilicate co-doped with Zr, GSOZ) (Shimura *et al* 2004) and an inner scatterer ring consisting of 0.9 x 0.9 x 6.0 $mm^3$ scintillation crystals (Ce-doped gadolinium–aluminium–gallium garnet, GAGG) (Kamada *et al* 2012). It was also reported that the GAGG crystals could be optimized to be more suitable for WGI (Kim *et al* 2024).

Triple-gamma imaging, or $\beta^+$-$\gamma$ coincidence imaging, should be considered a potential novel application of WGI. Scandium-44 ($^{44}$Sc), which emits a positron (resulting in a pair of 511 keV annihilation photons) and a 1157 keV γ-ray almost simultaneously, is one of the most suitable radionuclides for utilizing the WGI concept. The source position can be determined at the intersection points between a line-of-response and the surface of the Compton cone (Figure 1c). The WGI prototype achieved a positional accuracy of approximately 6.7 mm for the novel $^{44}$Sc-WGI direct imaging (Yoshida *et al* 2020). This direct imaging method is expected to enable in situ*,* real-time tracking of small activity sources.

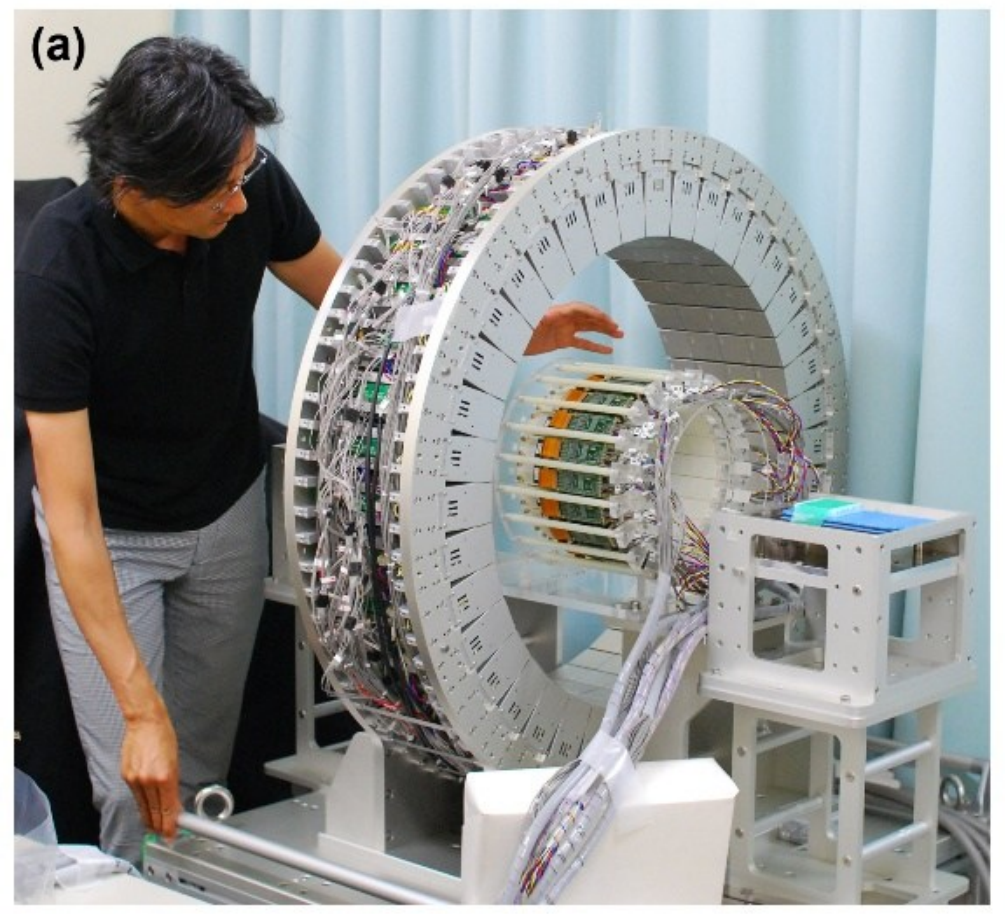

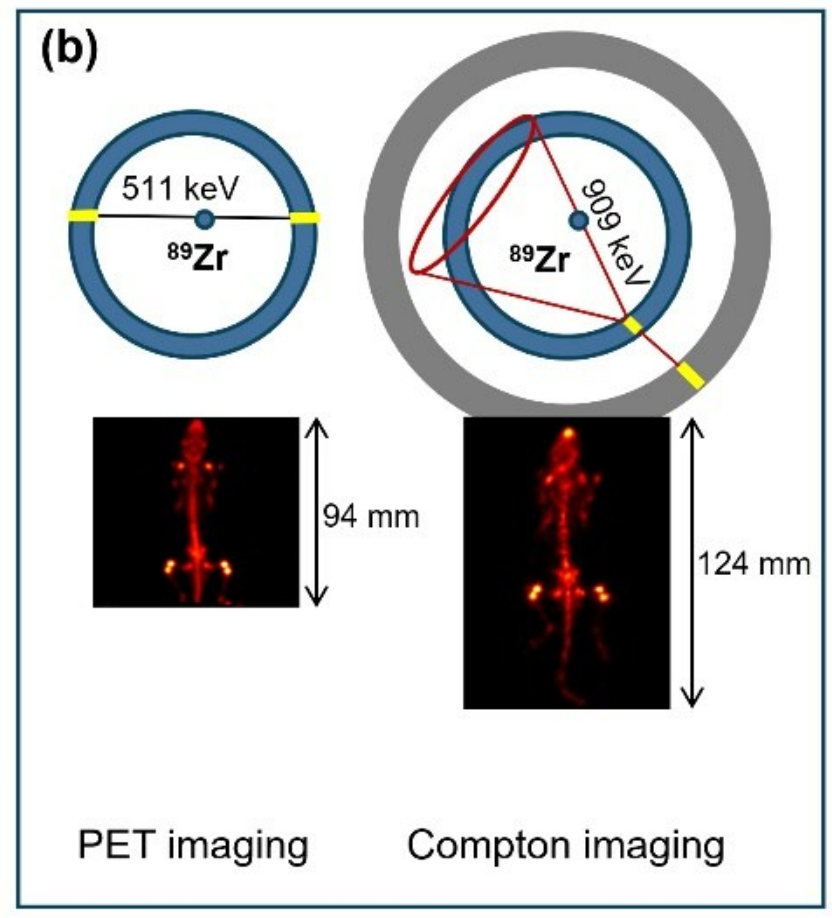

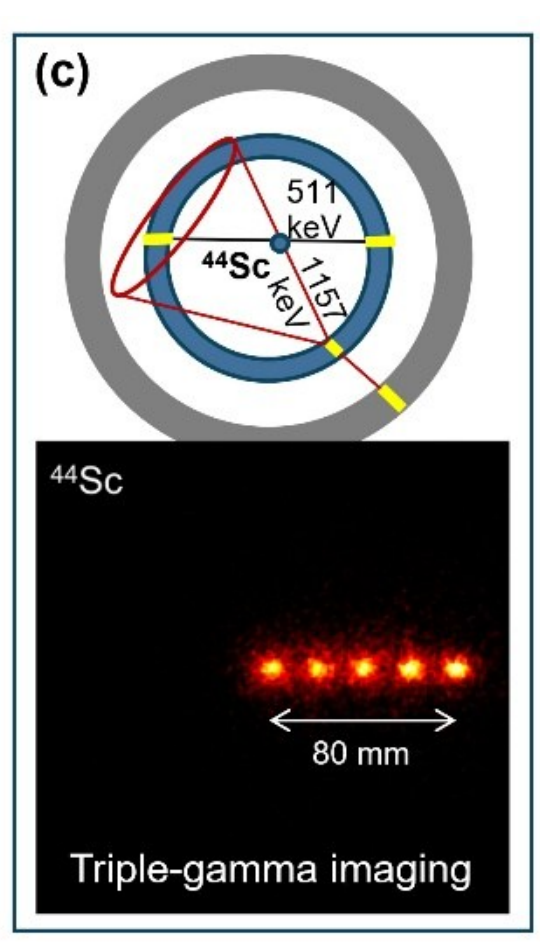

**Figure 1.** A whole gamma imaging (WGI) prototype, which was a PET system combined with a Compton camera (a). The mouse imaging (1-hour measurement started 22 hours after 9.8 MBq $^{89}$Zr-oxalate injection) demonstrates the Compton image obtained using 909 keV gamma rays is comparative in quality to that of the PET image (b). A $^{44}$Sc-containing vial was shifted at 2 cm intervals and visualized with a simple backprojection of the intersection points of the Compton cone and the line of response of PET (c).

**Advances in Science and Technology to Meet Challenges**

Another crucial factor to consider is the time difference between the positron decay and the single-gamma emission (Tashima and Yamaya 2024). The use of radionuclides such as $^{44}$Sc, which emits a single γ-ray almost simultaneously with positron decay, allows the exploration of new nuclear medicine diagnostic methods that focus on positronium (Ps) lifetime, rather than the conventional radioactivity distribution. Ps forms prior to the positron-electron annihilation with a probability of about 30% in the body during PET scans, but its presence has not been utilized in conventional PET imaging. The time until Ps annihilates (Ps lifetime) varies depending on factors such as the surrounding electron density. Recently, several groups have demonstrated in vivo Ps lifetime imaging by PET in the human body (Moskal *at al* 2024, Steinberger *et al* 2024, Mercolli *et al* 2024). Since paramagnetic molecules with unpaired electrons, as well as paramagnetic/oxidized metal ions, influence Ps lifetime in aqueous solutions, it has been suggested that one potential medical application is the sensing of oxygen partial pressure ($pO_2$) in living tissue (Shibuya *et al* 2020, Moskal and Stępień 2021). Following the lab bench experiment by Shibuya et al. (Shibuya *et al* 2020), Takyu et al. (Takyu *et al* 2024b) demonstrated that a clinical PET system could successfully distinguish subtle differences in $pO_2$ values corresponding to hypoxic tumour cells and healthy tissue cells by measuring $pO_2$-adjusted sodium-22 ($^{22}$Na) aqueous solutions (Figures 2a, b) (Takyu *et al* 2024b). Furthermore, it was also demonstrated that Ps lifetime measurements could detect free radicals at millimolar (mM) concentrations (Figure 2c) (Takyu *et al* 2024a). This method could be applied to assessing tumour malignancy (Karimi *et al* 2023) and detecting ionic imbalances, such as those of $K^+$ and $Na^+$ ions, in biological tissues (Zaleski *et al* 2023).

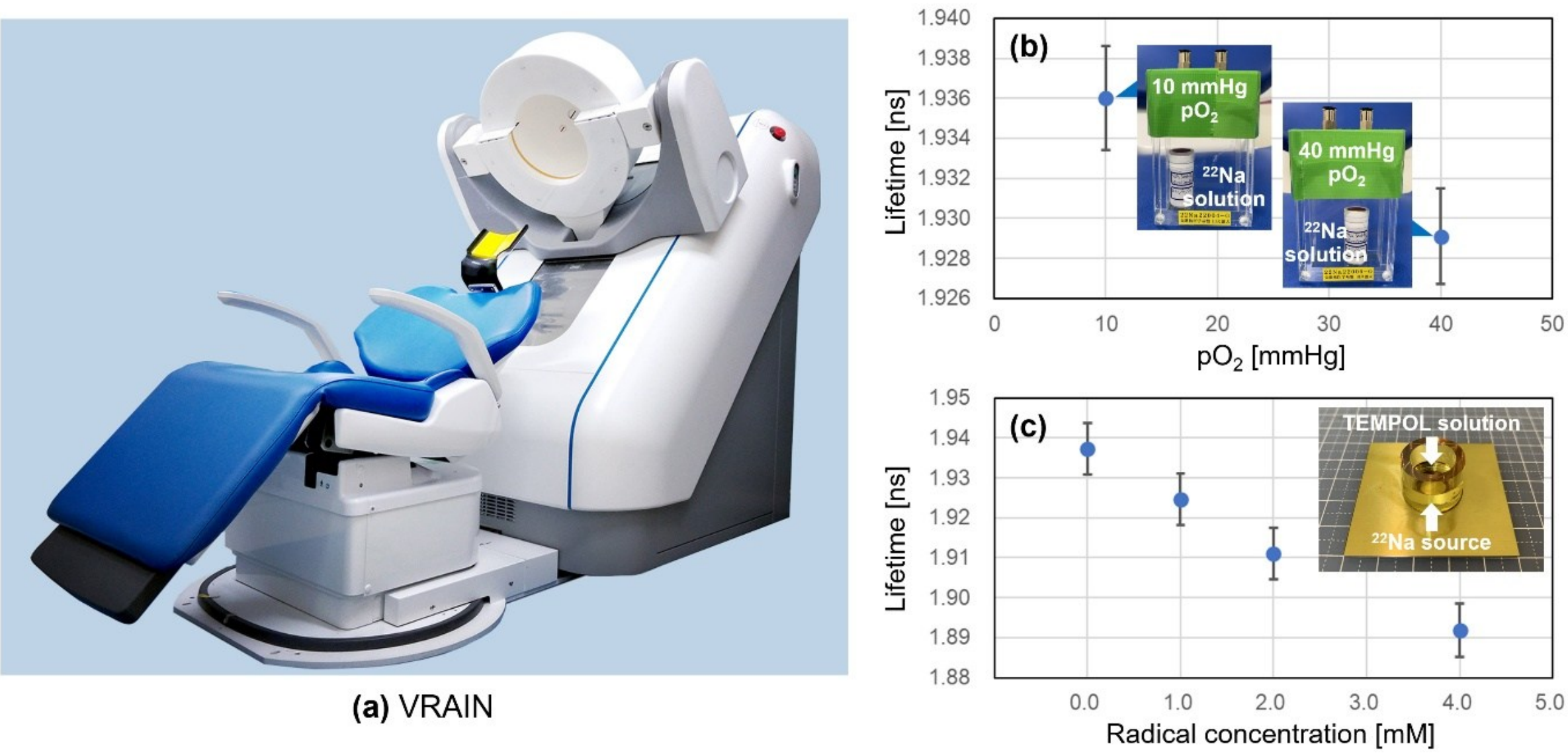

**Figure 2.** Measurement of Ps lifetime by a brain-dedicated PET system, VRAIN (a). Discrimination of oxygen partial pressure ($pO_2$) values of 10 and 40 mmHg, each of which corresponds to hypoxic tumour cells and healthy tissue cells, respectively (b), and correlation between radical (TEMPOL) concentration in water and Ps lifetime (c).

### Concluding Remarks

Obtaining as much patient information as possible to improve diagnostic accuracy is fundamental to realizing better medical practices and treatments, and nuclear medicine imaging has the potential to evolve even further. In addition to the axial extension, known as total-body PET, it is possible to enhance sensitivity more by measuring Compton scattering in detectors, and the combination of Compton-PET imaging, WGI, was invented to realize this concept. The WGI will also enable direct localization through triple-gamma detection or $\beta^+$-$\gamma$ coincidence, and will pave the way for positronium lifetime imaging, which has the potential to reveal previously inaccessible biological information.

### Acknowledgements

The authors were supported in part by the JSPS KAKENHI grant (20H00671, 25K03500 and 26H02613) and the Nakatani Foundation. The authors confirm that any identifiable participants in this study have given their consent for publication.

# 6. Development of Quantum Entanglement Compton-PET imaging

*Kenji Shimazoe*[1], *Mizuki Uenomachi*[2]
[1]The University of Tokyo, Tokyo, Japan
[2]Institute of Science Tokyo, Tokyo, Japan
shimazoe@g.ecc.u-tokyo.ac.jp, uenomachi.m.4403@m.isct.ac.jp

**Status**

With the successful demonstration of quantum entangled PET (QE-PET) (Watts *et al* 2021), subsequent studies (Kim *et al* 2023, Kožuljević *et al* 2024, Moskal 2024) have shown that incorporating quantum information and quantum entanglement can enhance PET scanner capabilities by measuring gamma-ray polarization via Compton scattering, supported by recent theoretical developments (Caradonna *et al* 2024). QE-PET has proven effective in improving the signal-to-background ratio. Since polarization in the MeV gamma-ray regime is measured through Compton scattering, technologies that enable precise Compton event detection are now of great interest.

We proposed and demonstrated the world's first simultaneous imaging system combining a Compton camera and PET – referred to as the Compton-PET hybrid camera (Shimazoe *et al* 2020). This simultaneous acquisition allows for in vivo visualization of multiple molecules, such as $^{18}$F-FDG (PET tracer) and $^{111}$In (SPECT tracer), thereby improving diagnostic accuracy (Uenomachi *et al* 2021), as well as $^{18}$F-FDG and $^{131}$I (therapeutic nuclide) for applications in radio-theranostics (Kim *et al* 2024).

To identify all events in the detector, including both photoelectric and Compton interactions, a one-to-one coupling scheme is preferred. Our detector consists of an 8×8 Ce:GAGG scintillator array coupled with an 8×8 SiPM array (Kamada *et al* 2014). Signals from each pixel are processed by a custom individual readout circuit using dynamic time-over-threshold (dTOT) electronics or custom ASICs (Orita *et al* 2018). TOT-based readout electronics significantly simplify analog-to-digital conversion, enabling multi-channel, high-resolution, and low-power data acquisition. The TOT signals are digitized and recorded by a custom FPGA-based DAQ system (PETnet) with a timing resolution of 62.5 ps (Sato *et al* 2021). The system is capable of detecting energies as low as 20 keV, providing accurate energy information.

Currently, the application of quantum information – such as decoherence effects, quantum entanglement transfer (Bordes *et al* 2024), and multi-photon entanglement – requires more advanced technologies offering fast timing and high energy resolution. While other sections focus on the underlying physics, this section summarizes ongoing efforts in the relevant detector technologies.

As discussed in Section 5.C, quantum entanglement and quantum sensing could enhance not only PET, but also other imaging modalities such as SPECT (Shimazoe and Uenomachi 2022), MRI, and CT. Related technologies were reviewed during the IEEE NSS/MIC 2021 "Quantum Sensing for Biomedical Applications" workshop, as summarized in reference (Shimazoe *et al* 2021).

Compton-PET geometry
CST
PET
Quantum entangled -PET
Compton Imaging
double photon
Compton Imaging

**Figure 1.** Compton-PET geometry which can handle PET, Quantum entangled-PET, Compton Imaging, Compton Scattering Tomography and double photon Compton Imaging. The detector design will play a key role. Right figure shows the ring structure Compton PET prototype.

**Current and Future Challenges**

Despite recent advancements, significant challenges remain in MeV-regime physics, such as accurately characterizing decoherence as a function of scattering angle or electron state (Bordes *et al* 2024). A comprehensive understanding of entanglement loss and decoherence – both experimentally and theoretically – remains one of the fundamental open questions in physics aspect. From an engineering standpoint, the optimal detector configuration for quantum-entanglemnt PET or quantum-entanglementCompton-PET has yet to be fully explored. Investigating system geometries that maximize quantum information extraction is one of the most pressing challenges in biomedical engineering and imaging.

Traditionally, PET system designs have focused on suppressing Compton events; however, quantum-entanglement PET/SPECT imaging requires the opposite approach, as Compton scattering plays a central role. Fully utilizing Compton event information – both within detectors and from within patients – could open new pathways, including the development of novel Compton scattering tomography techniques. This would enable the recovery of valuable information typically discarded in conventional PET/SPECT systems. The double photon Compton detection (Uenomachi *et al* 2020) could provide a new information hidden in the conventional SPECT/PET system.

In our previous work, we demonstrated, for the first time, multi-nuclide imaging using a Compton-PET camera (Llosá and Rafecas 2023, Shimazoe *et al* 2020, Uenomachi *et al* 2021). We also showed that quantum-entangled imaging can improve the signal-to-background ratio, leveraging the strong photon correlations (Kim *et al* 2023, Nakamura *et al* 2025). Nevertheless, several challenges persist in Compton-PET imaging. Notably, spatial resolution in Compton systems is primarily limited by the energy resolution of the detectors, whereas signal-to-noise ratio in modern PET scanners is governed by timing resolution—critical for reducing random coincidences and exploiting time-of-flight (TOF) information.

Unfortunately, no single detector material currently achieves both high energy and high timing resolution, which are required simultaneously for PET and Compton imaging. As is well known, scintillation crystals offer superior timing performance, while semiconductor detectors excel in energy resolution.

In the following section, we will outline several ongoing technical challenges currently under investigation. Moreover, since entanglement detection relies on measuring gamma-ray polarization via

Compton scattering, future work must explore detector geometries optimized for both Compton and photoelectric interactions. Such designs will likely differ significantly from those used in conventional PET scanners. Addressing these challenges will depend heavily on advancements in detector materials, photon sensors, and associated electronics.

**Advances in Science and Technology to Meet Challenges**

As reviewed in the previous section, Compton scattering detector is a key technology, however, there needs something to be compromised to mitigate both energy resolution and timing resolution in the realistic scanner design. Within scintillation materials, CeBr3 can be an interesting option to replace current crystals once the packaging issue to suppress its hygroscopicity is solved (Uenomachi *et al* 2022). One possible solution could be combining an energy resolving scatterer and a time-resolving absorber, for example the combination of Silicon semiconductor and LYSO/LSO scintillation detector. In this configuration, the timing resolution can be defined by LYSO/LSO and angular resolution can be defined by Silicon. Another choice is the combination of fast scatterers and energy-resolving absorber, such as plastic scintillators and TlBr semiconductors. It is also worth to mention that in TlBr semiconductors a few 100 ps time resolution can be acquired via Cherenkov photons as well as an excellent energy resolution better than 2% @ 662 keV (Hamdan *et al* 2024). The recent Perovskite detector development also attracts great interest (Kratochwil *et al* 2025). The time resolution with less than 100 ps, possibly 10 ps with energy resolution 1% @ 662 keV, is the current challenges in PET/SPECT detectors.

One of the key technologies improving Compton-PET scanner could be pixelated silicon detector with the “electron-tracking” capability which enables the tracking of Compton recoil electrons (Yoshihara *et al* 2017). The recoil electron track information greatly improves the signal to noise ratio in Compton images by limiting the Compton cones to Compton arcs. Our group previously reported the improved signal to noise ratio in the reconstructed images. Another interesting aspect of tracking electrons is related to the entanglement measurement depending on the angles of Compton scattering.

Considering the future quantum entanglement Compton PET system and the key role of Compton scattering, the complete reconsideration of geometry instead of using high-density crystal with DOI used in the modern PET system might be necessary. For example, the middle-density individual crystal arrays can be one solution to detect all the interaction including Compton and Photo-electric events. Especially in the future precise QE-Compton-PET, the energy, position detection for determining the Compton scattering angle can be an important key factor to optimize.

As proven by oxygen concentration sensing in previous researches, another interesting direction will be the direct detection of 3-gamma/2-gamma ratio combined with high-energy resolution imaging technique, of which we proved the concept enabling the use of pure positron emitters, such as $^{18}$F (Fujimoto *et al* 2025).

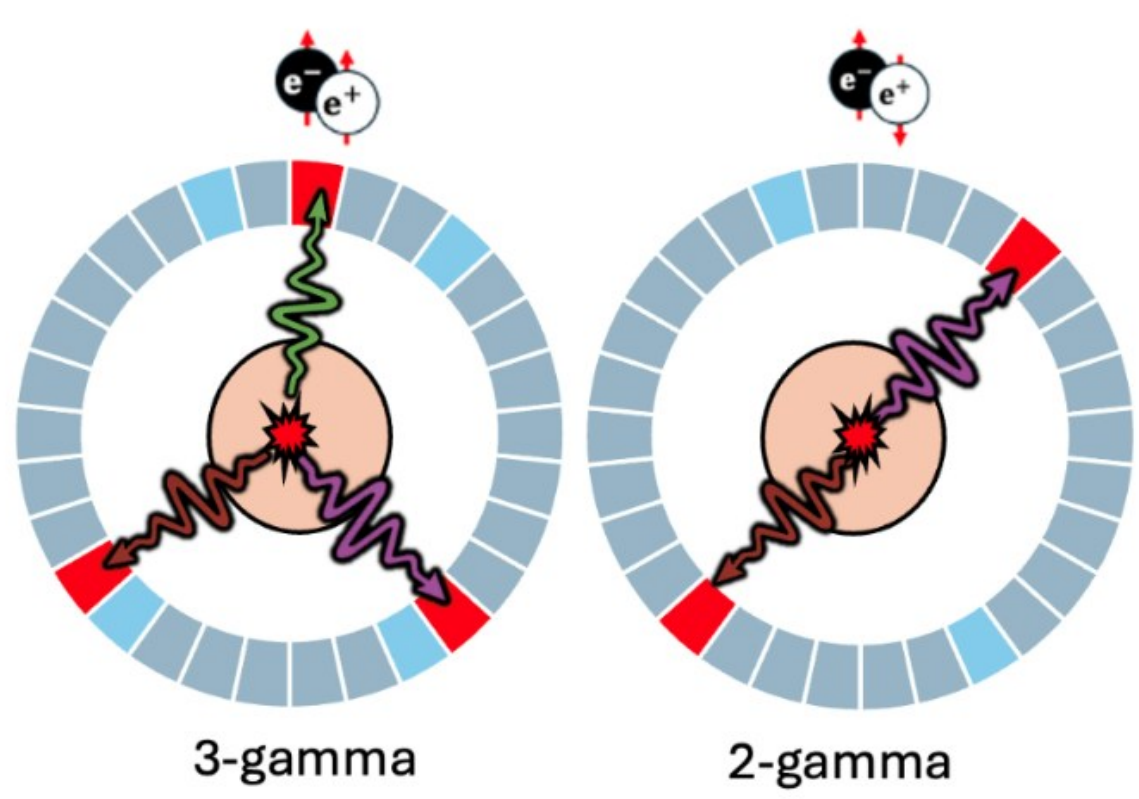

**Figure 2.** 3-gamma/2-gamma imaging concept with high-energy resolution imaging detectors for pure positron emitter.

**Concluding Remarks**

This exciting area using MeV gamma photons with quantum information and quantum entanglement medical imaging is rapidly expanding. For maximizing the quantum information in the future QE-PET and QE-Compton-PET in nuclear medicine, the detector geometry handling both Compton scattering and photo absorption will play a key role. It indicates the completely new geometry, and design can come into consideration for constructing a new PET/Compton/SPECT. In this section, several possible configurations and related pixel detector technologies are reviewed. In terms of medical imaging, increasing the image quality with less dose will be a most important task, and the entanglement detection could play an important role for recovering the neglected events, such as Compton scattering in both patients and detectors. The more basic studies investigating the interaction of entangled gamma-rays with various materials will be also necessary. The work incorporating the quantum effect in Geant4 simulation is also indispensable (Bordes *et al* 2024). The active approach controlling and utilizing nuclear spin and/or electron spin is also a very interesting topic related to two photons/ cascade photons/ multiple photons correlation as it will be discussed in section 15.

**Acknowledgements**

This work was supported by JSPS KAKENHI Grant Numbers JP1193658, JP1193659.

# 7. Development of combined PET and Compton Camera (PET-CC) system for Study of Quantum Entangled Photons

*Praveen Gurunath Bharathi*[1], *G. Romanchek*[2], *S. Abbaszadeh*[1]
[1]Electrical and Computer Engineering, University of California, Santa Cruz, Santa Cruz, California, United States of America
[2]Nuclear, Plasma, and Radiological Engineering, University of Illinois at Urbana-Champaign, Champaign, Illinois, United States of America
pgurunat@ucsc.edu, romanch2@illinois.edu, sabbasza@ucsc.edu

**Status**

Positron Emission Tomography (PET) fundamentally works by capturing and visualizing the metabolic activities within the body using positron-emitting isotopes (Ollinger and Fessler 1997, Muehllehner and Karp 2006). PET imaging's progress was significantly driven by the implementation of radiotracers, which are positron-emitting isotopes which could be injected into the body (Vaquero and Kinahan 2015). The integration of Time-of-Flight (TOF) capabilities into PET systems constituted a major technological leap, enabling the utilization of high-resolution timing of coincident gamma ray detection to achieve superior image quality through enhanced noise reduction and spatial resolution (Surti 2015, Surti and Karp 2020). Spatial resolution in PET is fundamentally limited by the range that positrons travel before annihilation, causing blurring that depends on the energy of the emitted positron and the surrounding tissue properties (Carter *et al* 2020, Levin and Hoffman 1999, Moses 2011). In addition, the intricacies of multi-tracer PET imaging arise largely from the inherent difficulty in disentangling the signals originating from distinct radiotracers, since PET scanners inherently measures the combined activity of all tracers, obtaining precise quantifications and kinetic modelling poses substantial difficulties (Kadrmas and Hoffman 2013, Pan *et al* 2024). **An emerging strategy to mitigate these limitations involves the hybridization of PET with Compton camera technology.** Compton cameras, capable of directly reconstructing the origin of gamma rays through Compton scattering kinematics, offer a complementary imaging pathway that does not rely on collinearity assumptions. By integrating Compton imaging with PET, the reconstructed Compton images can be utilized as prior spatial constraints in PET image reconstruction, helping to correct for positron range-induced blurring and improving the localization of annihilation events. This synergy not only holds potential for enhancing spatial resolution but also facilitates advanced multi-tracer imaging with PET tracers with prompt gamma, using the distinct energies of their gamma photons for clear differentiation. This capability also opens the door for utilizing isotopes that were previously considered impractical-such as those with high positron energies or additional gamma emissions that would interfere with conventional positron-annihilation imaging-thereby expanding the range of tracers available for advanced molecular imaging.

**Current and Future Challenges**

Cadmium zinc telluride (CZT) has been established as a detector of choice for the state-of-the-art single-photon-emission-computed-tomography (SPECT) due to its high energy and spatial resolutions. To have a good sensitivity to gamma-rays of up to 1MeV, a thick CZT detector of up to 4 cm is required. Such thick CZT crystals are not commercially available as its very expensive to manufacture them due to crystal growth complications (Bolotnikov *et al* 2016). The voltage requirement for detectors escalates significantly with thickness, moving from 500V for a 5mm detector to over 5kV for a 40mm one. This steep increase in required voltage poses a substantial problem concerning system scalability. Owing to these identified shortcomings, edge-on CZT configuration is a smart solution to enable high proper stopping power and precise 3D positioning while keeping the applied voltage reasonable. Figure 1 illustrates an edge-on CZT configuration with cross-strip design to make the system compact with minimum number of readout channels.

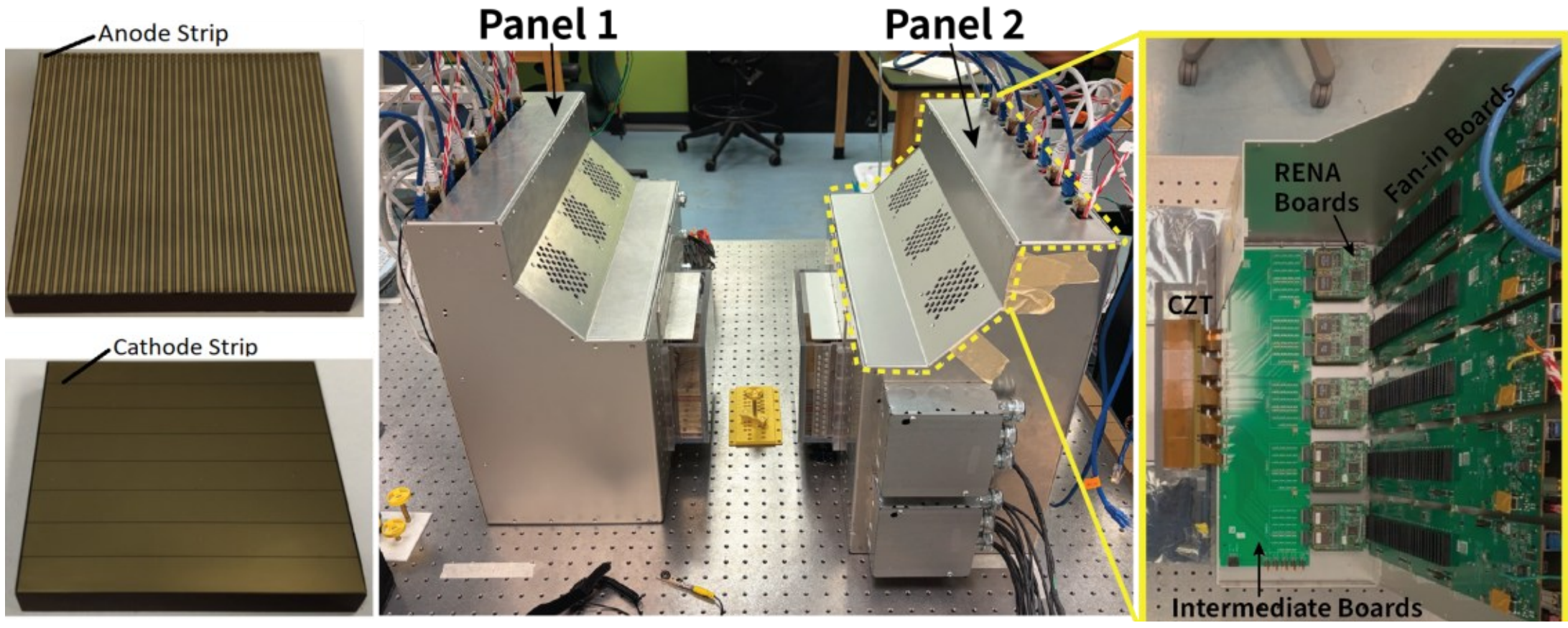

**Figure 1.** (left) Visualization of the cross-strip electrodes spanning opposite faces of a single CZT crystal. (middle) Sub-assembly of the UCSC CZT-based PET system comprising 2 opposing panels and data acquisition electronics. (right) Inset showing the readout electronics housed within one of the panels. One panel can accommodate up to 30 rows of individual CZT detectors (4×4×0.5 cm$^3$).

In PET- Compton Camera (CC) systems, a primary challenge is the sensitivity to multiple Compton scatter events. While edge-on Cadmium Zinc Telluride (CZT) detectors can capture both photoelectric and Compton interactions, accurately reconstructing events involving multiple scatters requires precise event pairing and sophisticated algorithms. The complexity increases as the number of possible interaction sequences rises, leading to ambiguities and reduced event classification efficiency (Llosá and Rafecas 2023, Shoop and Abbaszadeh 2024). High energy resolution is critical for distinguishing between annihilation photons and prompt gammas. CZT detectors offer improved energy resolution (as low as 1 % at 662 keV), but even small degradations in energy resolution can significantly impact the angular resolution and the accuracy of Compton cone reconstruction (Jin *et al* 2023, Parajuli *et al* 2022). In addition, for imaging isotopes with prompt gamma, when the higher energy gamma scatters and fall within the energy window, there is need for smart algorithms to distinguish annihilation photons from randomly pairing a scattered gamma photon with annihilation photons.

**Advances in Science and Technology to Meet Challenges**

Recent advancements explore the quantum entanglement of annihilation photon pairs as a means to enhance discrimination between true coincidence events and scattered prompt gamma rays. Since the two 511 keV annihilation photons are emitted simultaneously and exhibit correlated properties such as polarization and momentum, leveraging these quantum correlations can help differentiate them from uncorrelated or partially scattered photons *By analysing the scatter correlation of detected gamma pairs, it becomes feasible to identify random coincidences from prompt gamma leakage, and offering a road to improve image fidelity and contrast in PET systems.*

In such a dual-modality imaging system, signal contamination works in both ways: prompt gammas interfere with the annihilation gamma coincidence signal as the annihilation gammas interfere with the prompt gamma signal. Energy gating around 511 keV is insufficient to classify a given interaction as annihilation- or prompt-generated, as higher-energy prompt gammas frequently scatter into this energy regime. Prompt gamma scattering similarly complicates Compton recovery strategies, where multiple interaction events are summed into a single energy bin. Examining photons individually leaves much of this ambiguity unresolved; examining coincidence detections, however, opens a new door.

A coincidence measurement in which each photon undergoes at least one Compton scatter prior to absorption, a Double Compton Scatter (DCSc) coincidence, allows the scattering kinematics of both photons to be probed. True coincidences (between two annihilation gammas) differ from random coincidences (between an annihilation gamma and a prompt gamma) in three key ways. First, the prompt gamma carries an energy that may significantly exceed 511 keV, imposing distinct scattering kinematics: its energy-derived polar scattering angle that is computed under the assumption of a 511 keV incident photon will disagree with the angle derived from its interaction positions. Second, and independently, the line-of-response drawn between the initial interactions of a random pair does not pass through the source, introducing a geometric error that further widens the discrepancy between energy- and position-derived angles; true coincidences remain self-consistent on both counts. Finally, and most novel, annihilation gammas are generated in an entangled state with direct implications for their scattering kinematics: entangled pairs (the trues) exhibit a heightened frequency of a 90° difference in azimuthal scattering angle. Collecting DCSc coincidences means all three of these features can be checked, whether by a rules-based algorithm or a machine-learning classifier.

Such a classifier has recently been demonstrated directly (Bharathi *et al* 2026): a permutation-invariant network trained on DCSc coincidences from a dual-emission source uses energy deposition, interaction position, and the azimuthal difference $\Delta\phi$ to identify true annihilation coincidences and reject prompt-gamma randoms. A systematic feature-ablation study found that combining energy with $\Delta\phi$ gave the strongest performance (ROC-AUC 0.87—0.95 across classes, reaching 0.95 for random-coincidence rejection), a roughly two-percentage-point gain over energy alone. Critically, SHAP analysis attributed a marginally higher mean importance to $\Delta\phi$ than to energy within this top-performing model, providing direct evidence that the entanglement-encoded angular correlation contributes discriminative power beyond conventional energy and kinematic features.

**Concluding Remarks**

The feasibility study of quantum entanglement of annihilation photons using a combined PET and Compton Camera system highlights a promising frontier in molecular imaging. Recent advances demonstrate that quantum entanglement, particularly with CZT-based detectors, can address longstanding challenges in multi-isotope PET and dual-emission radiotracer studies, enabling more precise discrimination of overlapping signals and improved image quality. Experimental and simulation results confirm the persistence of strong polarization correlations between annihilation photon pairs, even after Compton scattering, supporting the robustness of entanglement in the MeV regime. Integration of advanced detector technologies-such as edge-on CZT arrays and liquid xenon systems-further enhances spatial resolution, sensitivity, and the ability to utilize a broader range of isotopes, including those previously unsuitable for clinical imaging. These innovations open new avenues for multi-tracer and quantitative imaging, potentially overcoming the spatial resolution limits imposed by positron range and non-collinearity effects. As quantum-entangled PET-CC systems mature, they are poised to significantly expand the capabilities of molecular imaging, fostering new clinical and research applications.

**Acknowledgements**
The authors acknowledge the support from the National Institute of Biomedical Imaging and Bioengineering of the National Institutes of Health under Award Numbers UG3EB034686 and UH3EB034686. The authors also thank Lars Furenlid, Michael King, Matt Kupinski, and Phillip H. Kuo for their valuable discussions, insightful comments, and guidance throughout this research.

# 8. Development of PET imaging with single-scattered photons using polarization characteristics

*Pragya Das, Ritesh Verma, Harmanjeet Singh Bilkhu, Kaustav Prasad*
Physics Department, Indian Institute of Technology Bombay, Mumbai, India
pragya@phy.iitb.ac.in, vermariteshussr@gmail.com, 11harmanjeet@gmail.com, kaustavprasad@gmail.com

**Status**

Positron emission tomography (PET) has made significant advances in achieving high image quality while addressing various challenges, such as high and low count rates, as well as substantial scattering in phantoms. One important consideration in clinical settings is the absorbed dose delivered to the patient. Recent developments in highly sensitive detectors, list-mode data collection, and advanced computational technologies enable us to tackle these challenges effectively. At high count rates, noise can be significantly reduced by utilizing the classical concept of orthogonal polarization of the two annihilation photons (Das *et al* 2024, McNamara *et al* 2014, Toghyani *et al* 2016). In low-count scenarios, it is crucial not to lose any emitted photons (Ghosh and Das 2022, Yoshida *et al* 2020), but to utilize them for imaging.

In some cases, the scattering in phantoms can be as high as 30-60% of all coincidences. In conventional PET, scattered data is typically estimated and discarded (Ollinger 1996). However, there have been attempts (Conti *et al* 2012, Hemmati *et al* 2017) to use this scatter data for imaging rather than rejecting it. Furthermore, images obtained from single-scatter coincidences can serve as an initial estimate in the maximum likelihood expectation maximization (MLEM) reconstruction process (Ghosh and Das 2023). Interestingly, there are inherent geometrical shapes, such as Compton cones, and prolate spheroids defining the scattering locus, and their intersecting surfaces provide valuable insights for determining images (Ghosh and Das 2022). Previous research (Ghosh and Das 2022, 2023, Conti *et al* 2012, Hemmati *et al* 2017) utilized energy information and sometimes timing data for the scattered photons in PET imaging.

Recently, for the first time, a method was proposed (Das *et al* 2024, Verma and Das 2024) to obtain images based solely on timing information without considering energy, suited for the plastic scintillators employed in the J-PET facility (Moskal *et al* 2016, 2024a). Figure 1(a) presents a 2D representation (Das *et al* 2024, Verma and Das 2024), where points **A** and **B** denote the photon detection points in PET ring, and point **C** indicates the scattering point in the phantom for a single-scattered event. The novelty of this approach lies in the circular symmetries: circles centred at **A**, **B**, and their intersection point (green dot) all have the same radius (r). Furthermore, the green circle represents the scattering locus, revealing a clear relationship between r and the scattering angle $\theta_s$. This enables an estimation of the annihilation point by calculating the total path length ACB and the path-length difference of photons based on their travel time. Extending this concept to 3D leads to the annihilation locus — illustrated with red and blue points in Figure 1(b) for a 60° scattering angle, as an example — resulting in a spindle torus shape.

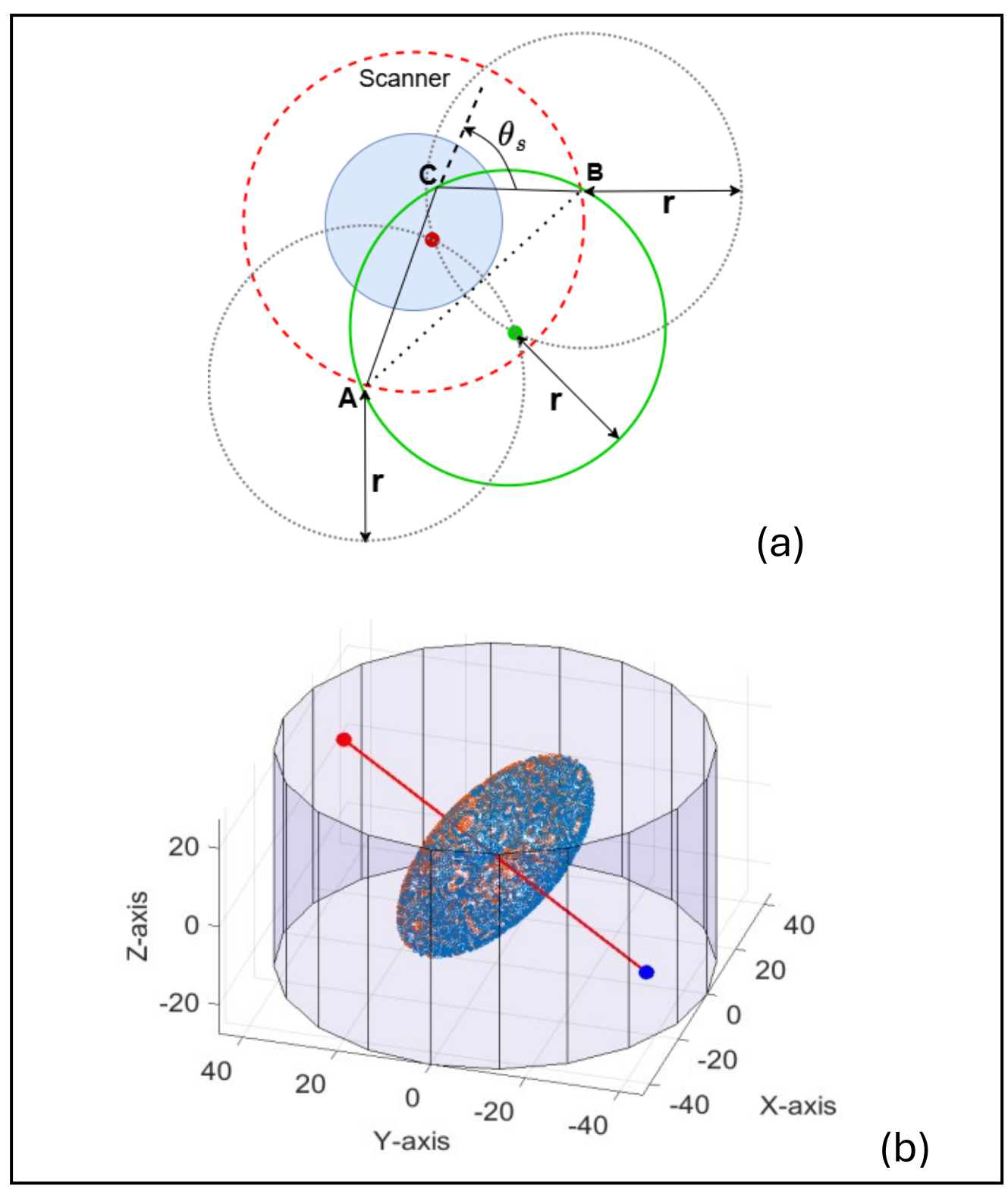

**Figure 1.** (a) A 2D representation of three circular geometries centered at the photon detection points **A** and **B** on the PET ring, also at their intersection point (green dot) describing the locus of the scattering point **C**. (b) A 3D visualization of annihilation-point locus depicted as spindle torus, with blue and red points for 60° scattering angle, as an example.

**Current and Future Challenges**

The proposed model for determining the annihilation volume (spindle torus) for single-scattered photons has been implemented using simulated data. The resulting image was created by combining the annihilation volumes from multiple events across all possible scattering angles, following the application of the Klein-Nishina probability distribution. Figure 2(a) shows the initial image of an off-center line source, generated using precise timing information, with a time-of-flight (TOF) resolution of 1 picosecond. At first glance, the image appears promising, as it correctly indicates the location (1, 1). However, the large width (full width at half maximum, FWHM ~ 6.4 cm) is concerning. The introduction of the Lucy-Richardson (LR) deconvolution technique significantly reduced the width to 2.6 cm, as shown in Figure 2(b). Nevertheless, the challenge remains to further minimize this width. Currently, we are validating the model with experimental data from the J-PET facility at Jagiellonian University in Poland. A primary difficulty we face is distinguishing single-scattered events from true coincidences, given that plastic scintillators provide excellent timing information but lack energy resolution.

One of the research challenges that limits the use of the aforementioned method is the trade-off between computational cost and model accuracy. Increasing the density of scatter locations enhances accuracy but comes at the expense of increased memory usage and processing time. To address this, efficient algorithms for scatter point sampling and GPU acceleration (Zhou and Qi 2011) can be employed. Methods such as Single Scatter Simulation (SSS) (Watson *et al* 1996) offer promising alternatives for utilizing scattered coincidences. Finally, a significant bottleneck in utilizing these scattered events is the difficulty of quantitatively integrating them into standard clinical imaging modalities.

The concept of orthogonal polarization of annihilation photons has been recognized for a long time, but it has not been effectively utilized in PET imaging. A simulation-based study (McNamara *et al* 2014) demonstrated that incorporating photon polarization not only reduces random coincidences but also includes multiple coincidences occurring within the same time window. This results in a significant improvement in the signal-to-noise ratio. Additionally, the number of scattered events is considerably reduced since they do not meet the criteria for orthogonal polarization. While polarization selection is particularly advantageous for environments with high count rates, preliminary results (Das *et al* 2024) indicate that its benefits are more pronounced in low-activity distributions. Further investigation is required to validate this observation. The main research challenge lies in the measurement of polarization, which demands highly sensitive experimental techniques.

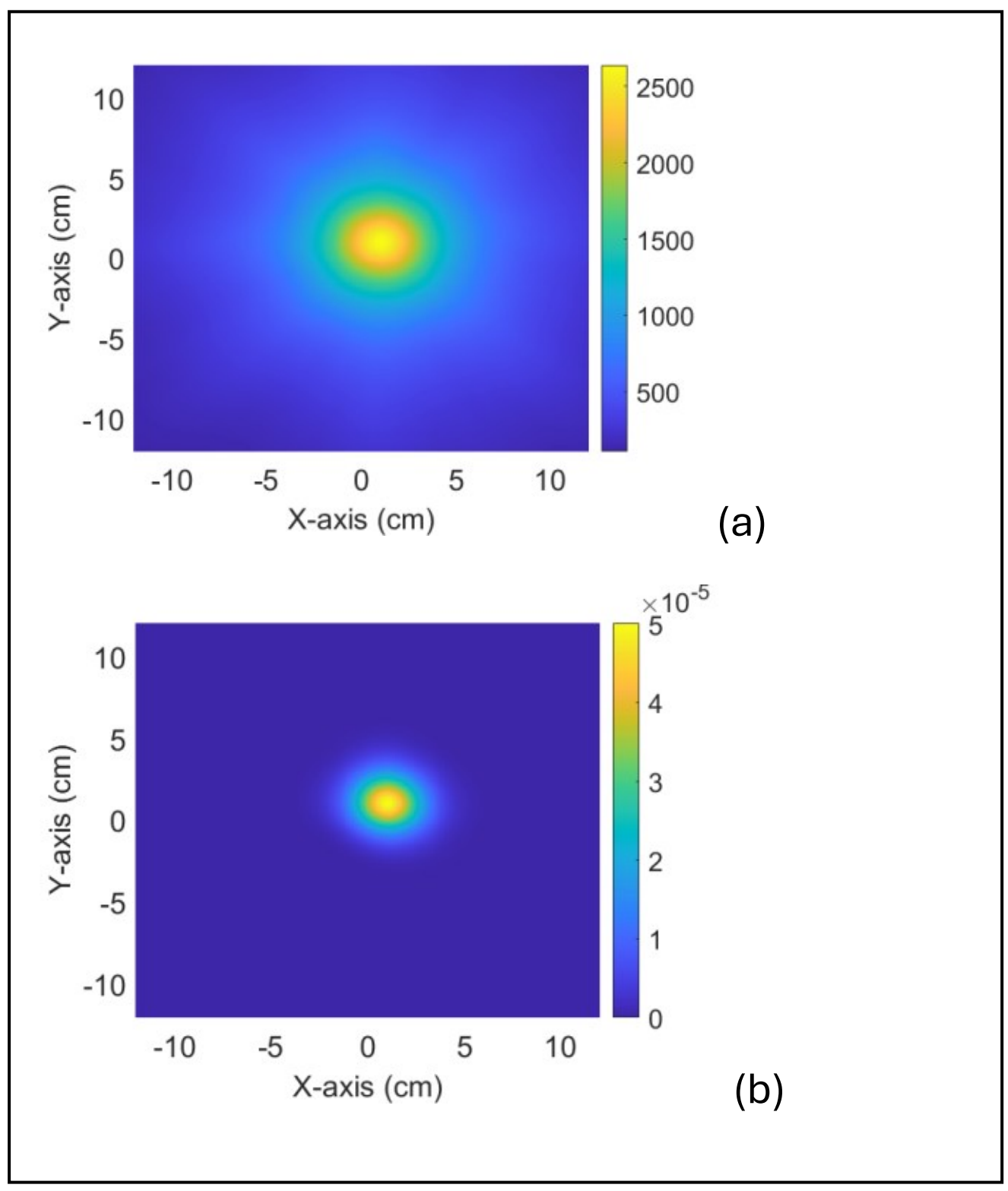

Figure 2. (a) PET image of a line source positioned at (1, 1), obtained from simulated single-scattered events utilizing TOF information. (b) The image processed through the Lucy-deconvolution technique displays a significant improvement in FWHM.

**Advances in Science and Technology to Meet Challenges**

The spatial resolution of PET images, particularly when considering scattered coincidences, is closely related to the time resolution of the detection system. To maximize the excellent timing properties of plastic scintillators, significant advancements in signal processing and electronics are necessary. The development of ultra-fast front-end electronics, along with waveform sampling techniques (such as fast ADCs), is essential to achieve time-of-flight (TOF) resolutions below 100 picoseconds (Lecoq 2017). Additionally, improvements in detector geometry and light collection methods — such as wavelength-shifting (WLS) strips, multi-layered scintillators, and innovative 3D readout designs — can enhance both timing and spatial accuracy (Smyrski *et al* 2017). The use of axial geometry and modular PET ring designs, as demonstrated in J-PET, provides better angular coverage and flexibility (Baran *et al* 2025). Ongoing research into new plastic scintillators with enhanced light yield and decay times, along with hybrid materials that combine plastic and inorganic components, has the potential to improve both timing and spatial performance (Lecoq *et al* 2020).

Innovative reconstruction techniques, such as Scatter-Aware reconstruction algorithms, are being developed to incorporate scattered coincidences. These techniques suggest integrating scatter information into an iterative reconstruction algorithm like OSEM (Ordered Subsets Expectation Maximization) by initially reconstructing the image using only true coincidence events and then including scattered coincidences in subsequent iterations (Verma and Das 2024). The use of Time-of-Flight (TOF) information significantly reduces the uncertainty regarding the origin of scatter, making scattered data valuable for reconstruction. Additionally, emerging designs in modular PET systems will integrate artificial intelligence (AI) at multiple levels, including data acquisition, real-time event classification, scatter modelling, and dynamic reconstruction. These AI-enhanced PET systems are particularly useful for managing the complexities introduced by scatter-aware reconstruction methods (Arabi *et al* 2021).

The conceptual understanding of using polarization criteria in various types of coincidences — true, random, multiple, and scattered — that are either rejected or incorporated in PET imaging is elegantly presented by McNamara *et al.* (McNamara *et al* 2014). Their simulation-based study did not consider photon energy and timing selection; nonetheless, the improvement in image quality can still be significant. Currently, PET ring detectors are not capable of exploiting the polarization correlation of photon pairs. However, Compton-PET systems (Ghosh and Das 2022) are promising candidates for this application. The angular correlation of Compton-scattered annihilation photon pairs can serve as a practical tool for polarization-based discrimination. A Monte Carlo simulation study (McNamara *et al* 2014, Toghyani *et al* 2016) of Compton cameras suggests their potential for coincidence-based imaging, which can be achieved with existing position-sensitive detectors. Moreover, a recent report (Moskal *et al* 2024b) demonstrates the first measurement of polarization correlation using a full-scale tomograph, J-PET.

**Concluding Remarks**

Over the past decade, there have been efforts to emphasize the importance of incorporating data from Compton scattering that occurs within the phantom in positron emission tomography (PET). The event-by-event mode of data collection is crucial for these studies. Recently, a proof-of-concept model was proposed to create PET images using single-scattered photons with time-of-flight (TOF) information, without relying on energy measurements. This innovative method employs symmetrical shapes in its geometric construction — specifically, a prolate shape for the scattering locus and a spindle torus for the annihilation locus in 3D — to generate PET images directly, without the need for reconstruction techniques. While the concept is attractive, it is still far from being applicable in clinical settings. The primary challenge lies in measuring photon travel time with minimal uncertainty (TOF resolution). Currently, plastic detectors used in PET systems, such as J-PET, offer the best potential for achieving this.

The orthogonal polarization of the two annihilation photons enables a strict selection of coincidence events in PET imaging, which helps to reduce noise. However, the difficulty of measuring photon polarization currently limits its practical utility. A proposed technique that utilizes two Compton cameras to measure the angular correlation of the scattered photons shows promise for future research.

# 9. Multiple Compton Scattering in Quantum-Entangled Two-Photon Systems

*Peter Caradonna*
Department of Nuclear Engineering and Management, The University of Tokyo, Tokyo, Japan

**Introduction**

Two key strategies have gained prominence in efforts to improve Positron Emission Tomography (PET) imaging. One approach involves leveraging Compton camera technology, which has the potential to improve image quality by more effectively tracking photon trajectories. This technique has gained world-wide interest due to its potential ability to refine signal detection in PET systems (Abdurashitov *et al* 2022, Frandes *et al* 2010, Uenomachi *et al* 2021, Watanabe *et al* 2008).

Another promising avenue is the investigation of developing methods of using quantum entanglement between annihilation photons to improve image quality. Feasibility studies suggest that entanglement could lead to improvements in image resolution, noise reduction, and novel imaging capabilities. These potential benefits are still theoretical and under investigation (Eslami and Mohamadian 2024, Kim *et al* 2023, McNamara *et al* 2014).

The primary interaction of annihilation photons with both the patient and the Compton camera system is Compton scattering, a process whose scattering distributions are influenced by the degree of entanglement between the annihilation photons (Bohm and Aharonov 1957, Caradonna 2024). In standard PET imaging protocols (Nevo *et al* 2022), approximately 300 billion events involve intermediate Compton scattering, where one of the annihilation photons scatters within the patient before it exits and reaches a detector (Barney *et al* 1991, Ollinger 1996). In a system with a ring of Compton detectors, this process is known as “3-Compton scattering”. The term describes a sequence where an annihilation photon undergoes one Compton scattering event inside the patient, followed by two additional Compton interactions within a pair of detectors. By capturing these multiple scattering events, the system can track photon trajectories with greater precision, offering valuable data for enhanced image reconstruction. In current PET systems, this background scatter degrades image quality by reducing the visibility of cancerous growths (Levin 2008, Zaidi *et al* 2007, Zatcepin and Ziegler 2023). Therefore, accurate modeling of Compton scattering in Compton PET-based imaging systems requires a theoretical framework capable of taking into account how entanglement affects the scattering process.

**Advances in theories**

The theoretical framework for Compton scattering in two-photon systems was first established by Pryce and Ward (Pryce and Ward 1947) and later in that same year by Snyder et al. (Snyder *et al* 1948). Their work examined annihilation photons in the initial Bell state $|\psi_c^-\rangle$, given by:

$$|\psi_c^-\rangle = \frac{1}{\sqrt{2}}|R, R\rangle - \frac{1}{\sqrt{2}}|L, L\rangle,$$

where $|R\rangle$ and $|L\rangle$ denote right- and left-circular polarization states, respectively, as defined in (Caradonna *et al* 2024).

In their analysis, they derived the joint differential Compton scattering cross section in the scenario where two 0.511 MeV entangled photons, initially in the state $|\psi_c^-\rangle$, propagate in opposite directions along a common axis. Each photon undergoes Compton scattering with a stationary electron, where the electron spins are randomly aligned. Additionally, it is assumed that both electrons remain completely uncorrelated with respect to each other. For brevity, this scenario in which each photon undergoes a single Compton scattering event is referred to as 2-Compton scattering.

Recent theoretical advancements have expanded existing frameworks to accommodate a broader class of two- and three-photon systems. In 2019, Hiesmayr and Moskal demonstrated that a Kraus-type structure allows for a straightforward extension of the joint differential Compton scattering cross section to two- and three-photon states, particularly in the context of positronium decay (Hiesmayr and Moskal 2019). That same year, a Stokes-Mueller method (Caradonna *et al* 2019), later refined in (Caradonna *et al* 2024), provided a generalized framework for computing the cross section in Compton scattering of arbitrary two-photon systems. When applied to the scattering scheme originally studied by Pryce and Ward, and later by Snyder et al., both the Kraus and Stokes–Mueller methods successfully reproduce their results.

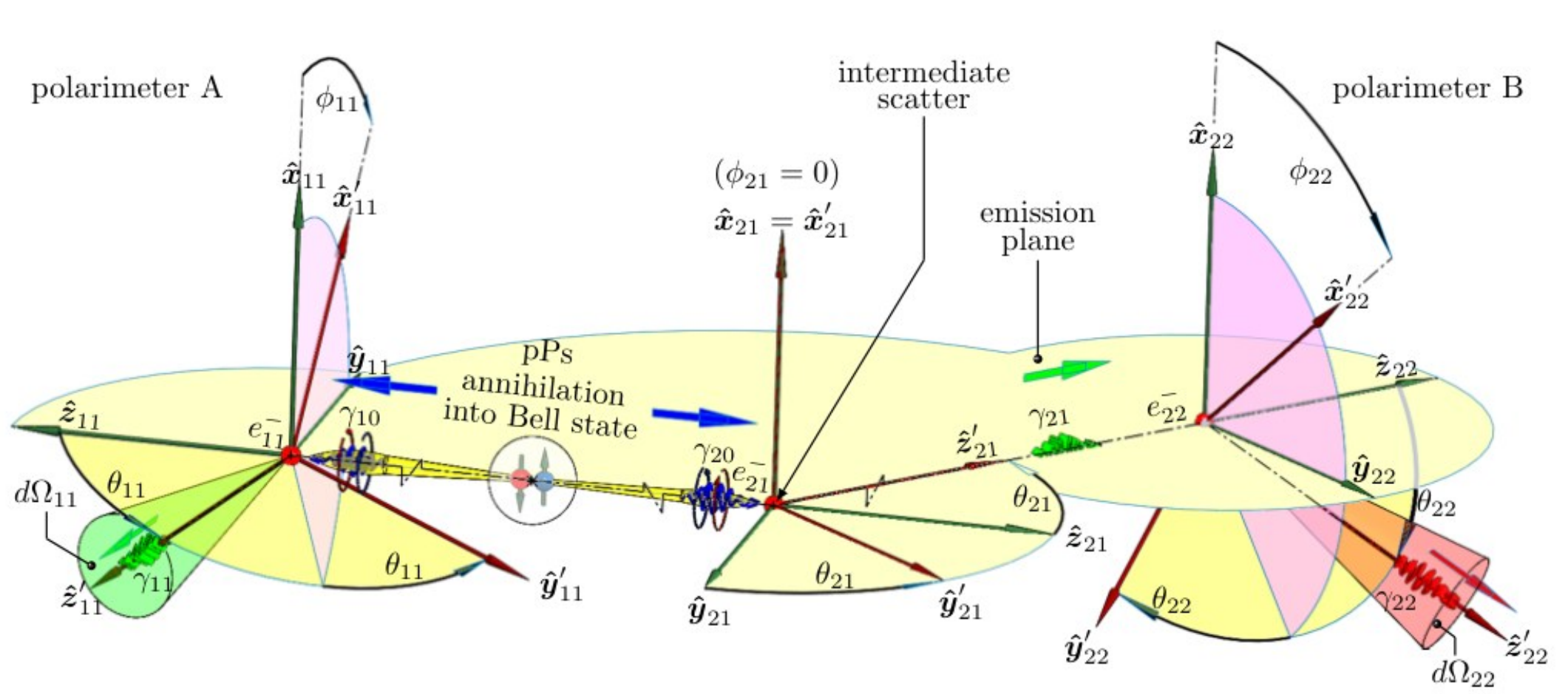

**Figure 1.** The nomenclature for the annihilation of p-Ps, emitting two 0.511 MeV photons, $\gamma_{10}$ and $\gamma_{20}$, in a Bell state, undergoing 3-Compton scattering. The intermediate scattering angle $\theta_{21}$. Photon 21 ($\gamma_{21}$) is the photon that results from the intermediate scattering. We define two hypothetical Compton polarimeters, A and B. The relative angle between the scattering planes of $\gamma_{11}$ and $\gamma_{21}$ is determined by the dot product $\hat{x}'_{11} \cdot \hat{x}'_{22}$.

In 2024, the Stokes–Mueller method was further expanded to incorporate multiple Compton scattering in an arbitrary two-photon system (Caradonna *et al* 2024). This expansion builds on principles that have been shown to accurately reproduce quantum correlations for annihilation photons in scenarios such as 2-Compton scattering with unpolarized electrons (Wightman 1948) and polarized electrons (Schmidt and Simons 1973). As a result, these calculations align precisely with those obtained using relativistic quantum field theory approaches (Pryce and Ward 1947, Snyder *et al* 1948). The robustness of this framework is further evidenced by its ability to account for double Compton scattering effects (McMaster 1961).

This extension specifically focused on analyzing the 3-Compton scattering scenario using the nomenclature illustrated in Fig. 1. The differential cross section for 3-Compton scattering of annihilation photon pairs is given by:

$$\frac{d^2\sigma}{d\Omega_{11}d\Omega_{22}} = \frac{r_0^6}{32}(E_{11}E_{22})^2 P(\theta_{11}, \theta_{22}, \phi_{11}, \phi_{22}; \theta_{21}),$$

where the cross section is parameterized in terms of the intermediate polar angle $\theta_{21}$, which represents the scattering inside a patient. Here, the incident photon energies, each equal to $mc^2$, are expressed in natural units such that $E_{10} = E_{20} = 1$ (see Ref. (Caradonna 2024) for details). The function $P$ is given by:

$$P = \frac{2}{r_0^6}\left(\frac{2}{E_{11}E_{22}}\right)^2 \frac{d\sigma}{d\Omega_{11}}\frac{d\sigma}{d\Omega_{21}}\frac{d\sigma}{d\Omega_{22}} - 2\sin^2\theta_{11}\sin^2\theta_{22}C(\phi_{11},\phi_{22};\theta_{21}) + D(\theta_{11},\theta_{22},\phi_{11},\phi_{21};\theta_{21}),$$

where the first term consists of the product of independent Klein-Nishina differential cross sections $d\sigma/d\Omega$ for Compton scattering events $\gamma_{10} \mapsto \gamma_{11}$, $\gamma_{20} \mapsto \gamma_{21}$, and $\gamma_{21} \mapsto \gamma_{22}$. These expressions apply specifically to the case where the incident photon beams, consisting of $\gamma_{10}$, $\gamma_{20}$, and $\gamma_{21}$, behave as if they are independent and unpolarized. The matrix elements $t_{11}^{(11)}$, $t_{11}^{(21)}$, and $t_{11}^{(22)}$, given in Eq. (31) in Ref. (Caradonna 2024), are replaced by:

$$t_{11}^{(11)} = \left(\frac{1+E_{11}^2}{E_{11}} - \sin^2\theta_{11}\right) = \frac{2}{E_{11}^2 r_0^2}\frac{d\sigma}{d\Omega_{11}} \qquad (\gamma_{10} \mapsto \gamma_{11})$$

$$t_{11}^{(21)} = \left(\frac{1+E_{21}^2}{E_{21}} - \sin^2\theta_{21}\right) = \frac{2}{E_{21}^2 r_0^2}\frac{d\sigma}{d\Omega_{21}} \qquad (\gamma_{20} \mapsto \gamma_{21})$$

$$t_{11}^{(22)} = \left(\frac{E_{21}}{E_{22}} + \frac{E_{22}}{E_{21}} - \sin^2\theta_{22}\right) = \frac{2}{r_0^2}\left(\frac{E_{21}}{E_{22}}\right)^2\frac{d\sigma}{d\Omega_{22}} \qquad (\gamma_{21} \mapsto \gamma_{22})$$

The scattered kinetic energies of the photons labeled $E_{11}$, $E_{21}$, and $E_{22}$ are related to the incident kinetic energies $E_{01} = 1$, $E_{02} = 1$, and $E_{21}$, respectively, through the corresponding Compton scattering angles and are given by:

$$E_{11} = \frac{1}{2-\cos\theta_{11}}, \quad E_{21} = \frac{1}{2-\cos\theta_{21}}, \quad E_{22} = \frac{E_{21}}{1+E_{21}(1-\cos\theta_{22})}.$$

The function $C(\phi_{11},\phi_{22};\theta_{21})$ is given by:

$$C = \cos2\phi_{11}\cos2\phi_{22} - \cos\theta_{21}\sin2\phi_{11}\sin2\phi_{22},$$

and when $\theta_{21} = 0$, $C = \cos2(\phi_{11} + \phi_{22})$. The function $D$ is given by

$$D = \sin^2\theta_{21}[\sin^2\theta_{22}\cos2\phi_{22}\left(\frac{2}{E_{11}^2 r_0^2}\frac{d\sigma}{d\Omega_{11}} + \sin^2\theta_{11}\cos2\phi_{11}\right) - \frac{2}{r_0^2}\left(\frac{E_{21}}{E_{22}}\right)^2\frac{d\sigma}{d\Omega_{22}}\sin^2\theta_{11}\cos2\phi_{11}],$$

and $D$ vanishes for $\theta_{21} = 0$.

When $\theta_{21} = 0$ it can be shown that Eq. (2) reduces to

$$\left(\frac{r_0^2}{2}\right)\left(\frac{r_0^4}{16}\right)(E_{11}E_{22})^2\left[t_{11}^{(11)}t_{11}^{(22)} - t_{12}^{(11)}t_{12}^{(22)}\cos2(\phi_{11}+\phi_{22})\right] = \left(\frac{r_0^2}{2}\right)\frac{\partial^2\sigma}{\partial\Omega_1\,\partial\Omega_2}, \quad \text{for } \theta_{21} = 0,$$

where $\partial^2\sigma/\,\partial\Omega_1\,\partial\Omega_2$ corresponds precisely to the well-known Pryce-Ward cross section for 2-Compton scattering of annihilation photons, following the nomenclature defined in Ref. (Caradonna *et al* 2024). Thus, apart from a factor of $r_0^2/2$, the 3-Compton cross section reduces to the 2-Compton result when $\theta_{21} = 0$.

In 3-Compton scattering, a coincidence count rate $N_{11}^{\perp}$ is considered that represents the number of detected photon pairs where the scattering plane of the transmitted photon $\gamma_{11}$ is perpendicular to the emission plane ($\phi_{11} = 90°$), while the scattering plane of $\gamma_{22}$ lies within the emission plane ($\phi_{22} = 0°$). This configuration is illustrated in the top-left scattering diagram in Fig. 2. The transmitted photons $\gamma_{11}$ and $\gamma_{22}$ are studied under conditions where $N_{11}^{\perp}$ reaches its maximum, occurring at $\theta_{11} = \theta_{22} = \theta = 82.4°$.

Similarly, we define the coincidence count rate $N_{22}^{\perp}$ as the number of detected photon pairs where the scattering plane of $\gamma_{11}$ lies within the emission plane ($\phi_{11} = 0°$), while the scattering plane of $\gamma_{22}$ is perpendicular to the emission plane ($\phi_{22} = 90°$). This configuration is illustrated in the bottom-left scattering diagram in Fig. 2.

In this case, $N_{11}^{\perp}$ reaches its maximum at $\theta_{11} = \theta_{22} = \theta = 81.7°$. These scattering geometries ensure that the scattering planes of the transmitted photons remain perpendicular to each other, regardless of the intermediate scattering angle $\theta_{21}$. The ratio $Z_{(11,22)}^{\perp} = N_{11}^{\perp}/N_{22}^{\perp}$ is then determined based on these proposed measurements.

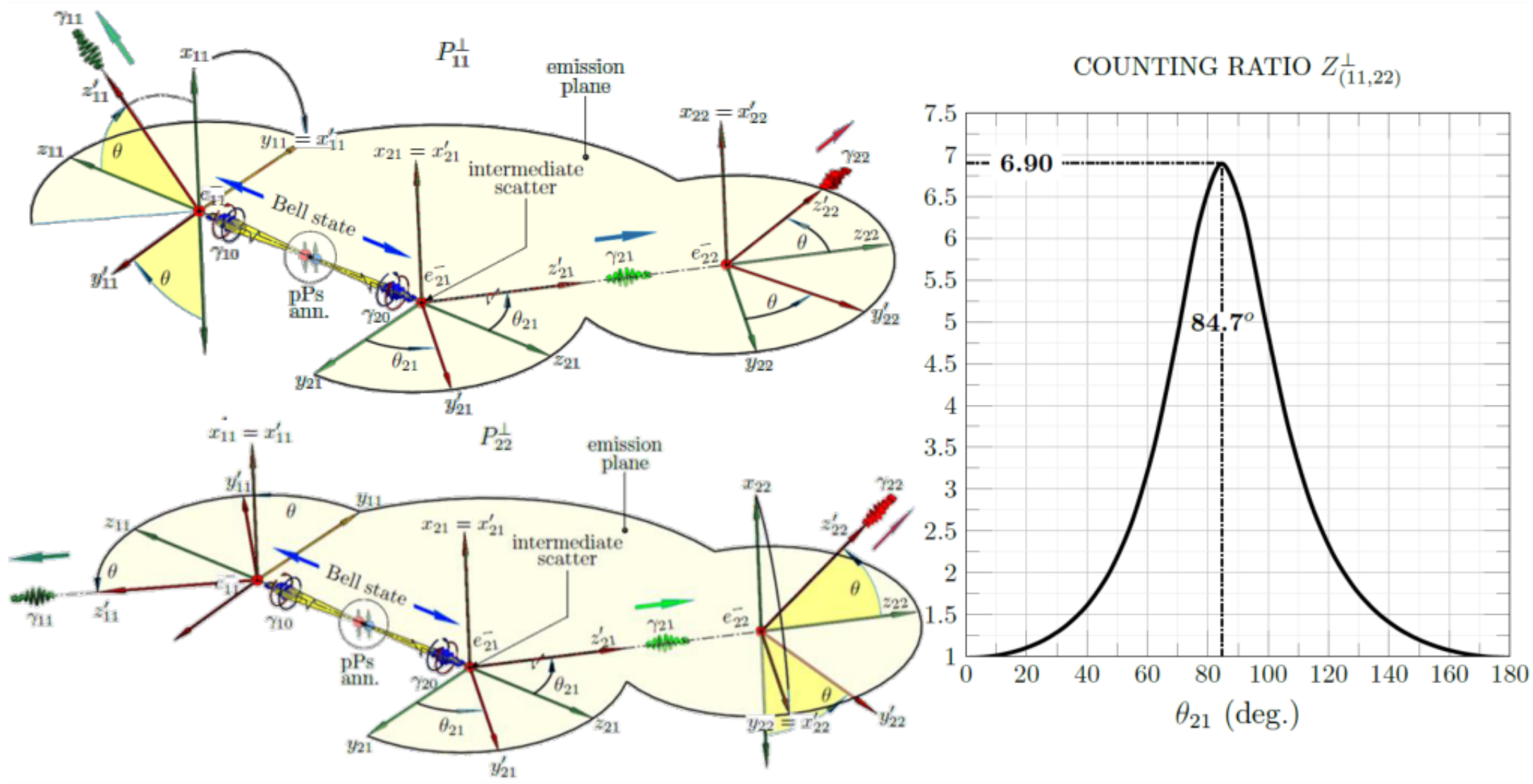

**Figure 2.** The counting ratio $Z_{(11,22)}^{\perp} = P_{11}^{\perp}/P_{22}^{\perp}$ is determined by evaluating $P_{11}^{\perp}$ where it is maximized in terms of $\theta$ at $\theta = 82.4°$ and for $P_{22}^{\perp}$ at $\theta = 81.7°$. At an intermediate scattering angle of $\theta_{21} = 84.7^{o}$, the result indicates that $\gamma_{11}$, which undergoes a single Compton scattering event, is approximately seven times more likely to scatter in a plane perpendicular to the emission plane while $\gamma_{22}$ scatters within the emission plane, compared to the reverse scenario where $\gamma_{22}$ (having undergone double Compton scattering) scatters perpendicular to the emission plane and $\gamma_{11}$ scatters within it.

In an ideal experiment, the count rate $N_{11}^{\perp}$ and $N_{22}^{\perp}$ are theoretically expected to relate to the function $P$ in the following way (see Ref. (Caradonna *et al* 2024)).

$$Z_{(11,22)}^{\perp} = \frac{N_{11}^{\perp}(\theta = 82.4^{o}; \theta_{21})}{N_{22}^{\perp}(\theta = 81.7^{o}; \theta_{21})} = \frac{P(\phi_{11} = 0°, \phi_{22} = 90°, \theta = 82.4^{o}; \theta_{21})}{P(\phi_{11} = 90°, \phi_{22} = 0°, \theta = 81.7^{o}; \theta_{21})} = \frac{P_{11}^{\perp}(\theta_{21})}{P_{22}^{\perp}(\theta_{21})},$$

and the ratio $Z_{(11,22)}^{\perp}$ is depicted on the right in Fig. 2.

**Concluding remarks**

Recent work by Bordes et al. (Bordes *et al* 2024) has incorporated 3-Compton scattering theory into a Monte Carlo simulation to test its predictions for intermediate scattering angles $\theta_{21} < 60°$. Independently, an experimental study by Tkachev et al. (Tkachev *et al* 2025) analyzed correlations at both forward and backward intermediate scattering angles. Both studies show good agreement with theoretical predictions.

Experimental results so far provide evidence that supports the 3-Compton scattering theory. However, further experimental testing is necessary before it can be fully benchmarked to the level of establishing it as a reliable model. If future experiments continue to confirm the predictions of the theory, it could be used to accurately model multiple Compton scattering effects in detection systems. The theory is useful because it accounts for the impact of entanglement on the scattering distributions of annihilation photons undergoing multiple scattering and reduces to the 2-Compton scattering scenario when $\theta_{21} < 0$.

# 10. Studies of the Quantum Decoherence of Quantum Entangled Gamma Photons

*Mihael Makek*[1]*, Siddharth Parashari*[2]
[1]University of Zagreb Faculty of Science, Department of Physics, Zagreb, Croatia
[2]Instituto de Fisica Corpuscular (IFIC), CSIC-UV, Catedrático José Beltrán, 2, E-46980 Paterna, Spain

makek@phy.hr, siddpara@ific.uv.es

**Status**

Recent developments in detector technology, including those of scintillator pixels with high energy resolution and SiPMs for the readout, made possible advancements in gamma-ray Compton polarimetry, which was dormant for a long period due to the lack of appropriate devices. Moreover, the interest in Compton polarimetry in the past decade has been driven by the idea to improve medical imaging in PET using the information about the polarization correlations (Kuncic *et al* 2011, McNamara *et al* 2014, Toghyani *et al* 2016). It was followed by the proof-of-concept in which it was shown that simple single-layer scintillator matrices could be used to measure these correlations (Makek *et al* 2020). The idea to use the polarization correlations to reduce the scatter and the random background in PET also raised the questions about the entanglement or its eventual decoherence after one of the photons had suffered an intermediate Compton scattering (ICS) prior to entering a polarimeter. The first such study gave inconclusive evidence about the decoherence of the scattered quanta, due to scarce statistics (Watts *et al* 2021). More detailed experimental studies quickly followed (Bordes *et al* 2024, Ivashkin *et al* 2023, Parashari *et al* 2024, Tkachev *et al* 2025) along with the theoretical modelling (Caradonna 2025), and hence the question of decoherence of annihilation quanta started to be demystified. The first of these studies made with specific laboratory setup based on large scintillators read out by PMTs (Ivashkin *et al* 2023) showed evidence that no decoherence occurs when one gamma photon scatters at forward (around 0°) angles. This discovery implied that the polarization correlations could not be used to reduce the scatter background in PET, since the correlation was preserved. However, it was important to check this effect with devices compatible to the ones in clinical PET scanners and within a broader kinematic range. The following section describes two experiments on quantum decoherence of entangled gamma photons performed with pixelated scintillator detectors. These experiments showed a way forward to understand the behaviour of polarization of annihilation quanta in scattering processes, which will be crucial for the correct modelling of quantum-entanglement PET devices.

**Current and Future Challenges**

An experiment to test the hypothesis of Compton scattering being a de-cohering process for entangled annihilation quanta was conducted at the University of Zagreb (Parashari *et al* 2024). This experiment used two single-layer gamma ray polarimeters, each comprising an 8x8 matrix of GAGG:Ce crystals with 1.9 mm x 1.9 mm x 20.0 mm dimensions. Detailed performance of these polarimeters was previously evaluated with gamma quanta following $e^+e^-$ annihilation from $^{22}$Na source and it was also shown that the detectors could measure the polarization correlations with high polarimetric sensitivity. A typical value of the polarimetric modulation $\mu=0.28 \pm 0.01$, reaching up to $\mu=0.34 \pm 0.02$, depending on event selection criteria (Parashari *et al* 2022). Translated into asymmetry ratio according to $R=(1+\mu)/(1-\mu)$, these were 1.78 ± 0.04 and 2.03 ± 0.09, respectively. The scatter detector, a 3 mm x 3 mm x 20 mm GAGG crystal read-out by an SiPM was placed geometrically in the center of the coordinate system, 50 mm from the polarimeters, in the path of one of the gamma photons from a point-like $^{22}Na$ source (Figure 1). It measured the energy of the recoil electron in the intermediate Compton scattering process. One of the polarimeters was rotated by $\theta_{scat}$ = 10°, 30°, 50° around the scatter detector measuring the Compton scattering of the pre-scattered gamma photon. One important aspect of the experiment was the selection of the reference frame. For each polarimeter, the reference frame for the Compton scattering was defined by the propagation direction of the incoming gamma ($z_1$, $z_2$) and the common y-axis orthogonal to the plane defined by the three detectors. Hence the Compton scattering angles ($\theta$, $\phi$) of each gamma photon were extracted in their own respective reference frame. In parallel, Geant4 (ver.10) simulation of the setup was performed. Since the modelling of the measured process for the entangled quanta was unknown, initially separable wave functions of the annihilation quanta were assumed. The results are shown in Figure 2 for the back-to-back (no scattering) case, and summarized in the Table 1 for the measured scattering angles. They confirmed that the intermediate scattering does not diminish the correlations up to $\theta_{scat}$ =30°, which agreed with (Ivashkin *et al* 2023). Moreover, the evidence showed that the measured modulation factors were approximately 2 times larger than predicted by the simulation for the separable states. An interesting behaviour was discovered at $\theta_{scat}$=50° where the evidence of reduction of the modulation amplitude was observed.

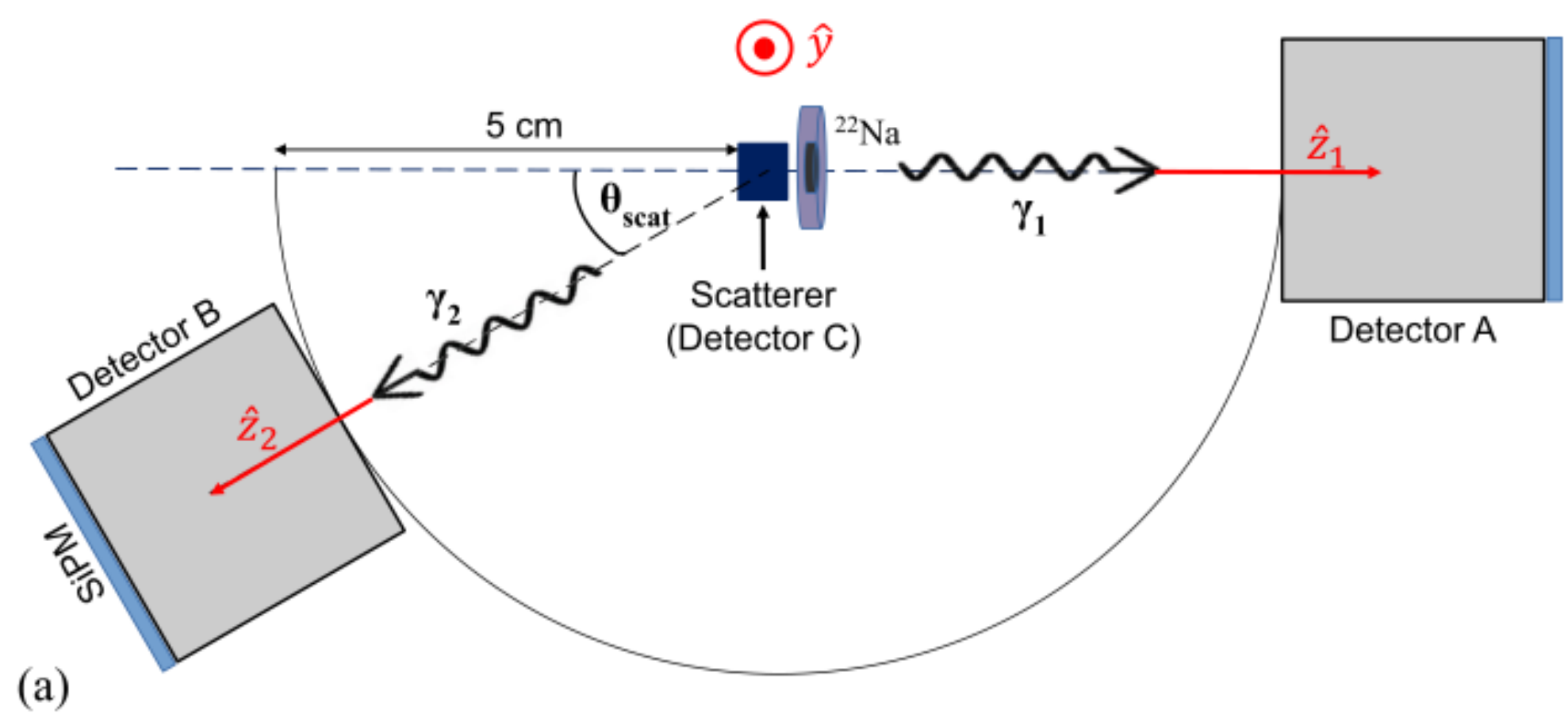

**Figure 1. Schematic view of the experimental setup in Zagreb decoherence experiment (Parashari *et al* 2024).**

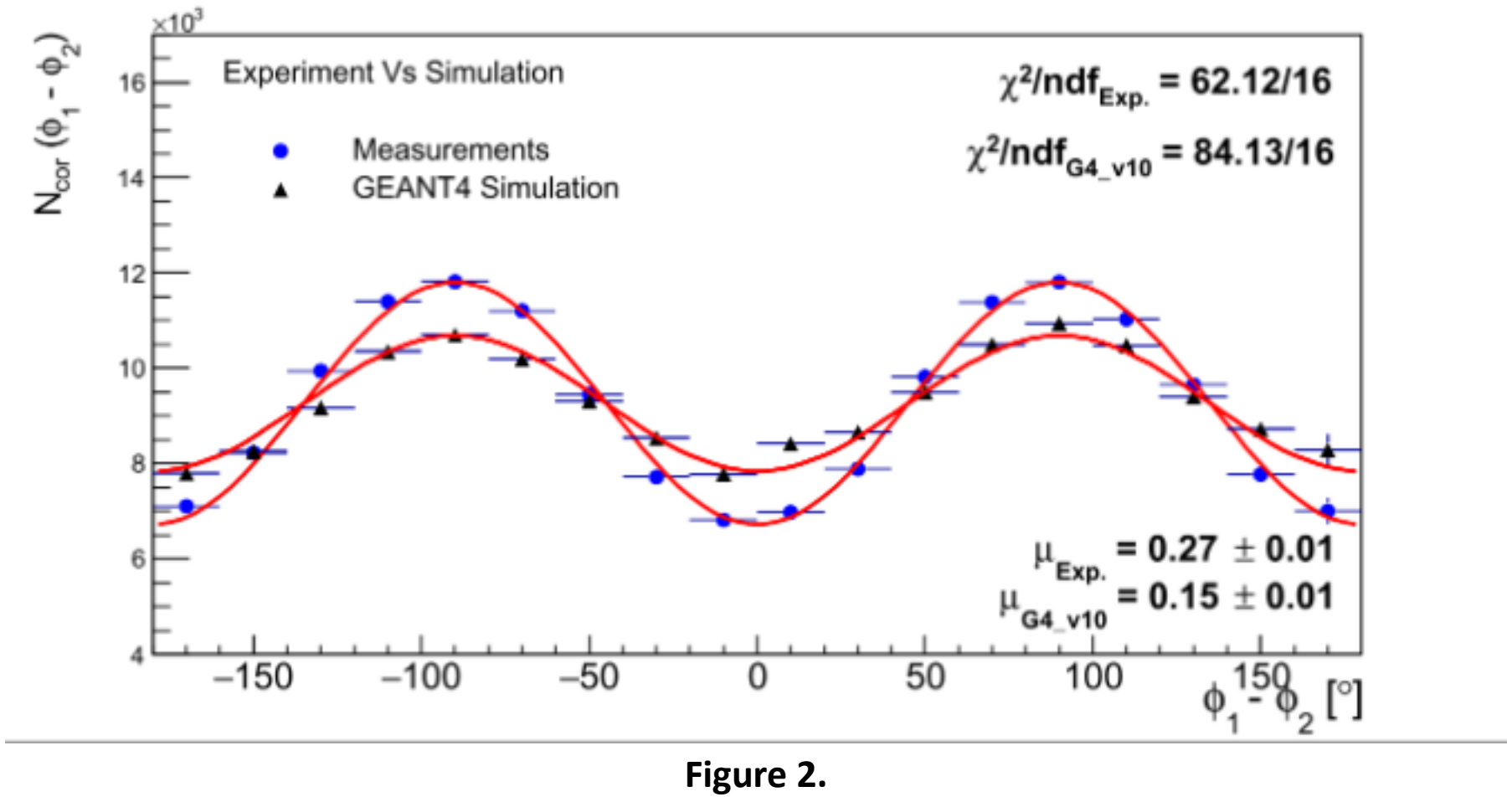

**Figure 2.**

Measured (circles) and simulated (triangles) acceptance-corrected yield, $N_{cor}$, for the back-to-back case. Red lines are fit with the function: $N_{cor} = A [1 - \mu \cos(2(\phi_1 - \phi_2))]$. The selected events are Compton scattered within the polarimeters with $72° < \theta_{1,2} < 90°$.

| $q_{scat}$ | $m_{exp}$ | $m_{sim}$ | $R_{exp}$ | $R_{sim}$ |
|---|---|---|---|---|
| 0° | 0.27 ± 0.01 | 0.15 ± 0.01 | 1.74 ± 0.04 | 1.35 ± 0.03 |
| 30° | 0.30 ± 0.03 | 0.14 ± 0.03 | 1.86 ± 0.12 | 1.33 ± 0.08 |
| 50° | 0.22 ± 0.06 | 0.09 ± 0.04 | 1.56 ± 0.20 | 1.20 ± 0.10 |

**Table 1.** Measured and simulated modulation factors **m** and asymmetry rations **R**. The selected events are Compton scattered within the polarimeters with 72° < $q_{1,2}$ < 90°.

A conceptually similar experiment was conducted by the group from the University of York. Two gamma ray detectors were used, each comprising 16 x 16 LYSO scintillators pixels with 3.0 mm x 3.0 mm x 20.0 mm size, read out by SiPMs (Bordes *et al* 2024). The source was placed in the center of the coordinate system, 38 mm from the polarimeters, and the scatter detector (LYSO crystal 3 mm x 3 mm x 5 mm) was introduced in the path of one of the gammas (see. Figure 3). The measurements were done in two configurations, in one the detector modules were back-to-back, in the other, one module was rotated by $\theta_{scat}$ = 28°. The Compton event analysis was done differently than by the Zagreb group, reflecting the larger solid angles (approx. x 15) subtended by the polarimeters. The first difference was that the intermediate scattering angle, $\theta_{ICS}$, was determined solely from the measured energy in the scatter detector, not relying on the physical rotation of the module. The second difference was the definition of the Compton scattering reference frame, which was fixed to each polarimeter, not varying for each incoming gamma photon. The large detector acceptance enabled measurements for $\theta_{ICS}$ < 70°. The obtained asymmetry ratios are shown in Figure 4. It shows that the azimuthal correlation of the entangled quanta is preserved for $\theta_{ICS}$ < 30° while the asymmetry ratio drops above this angle. In addition to measurements, Geant4 simulations of the setup were performed, implementing different physics scenarios for behaviour of the correlations upon the intermediate scattering. One scenario assumed full decoherence, another assumed keeping the maximum correlation of the entangled quanta and the third implemented theoretical modelling according to (Caradonna 2025). The experimental results clearly discarded the full decoherence scenario. On the other hand, the other two scenarios agreed with the measurements within their respective precisions. It is worth noting that the implemented theory provides the consistent modelling of the azimuthal correlations after the intermediate Compton scattering.

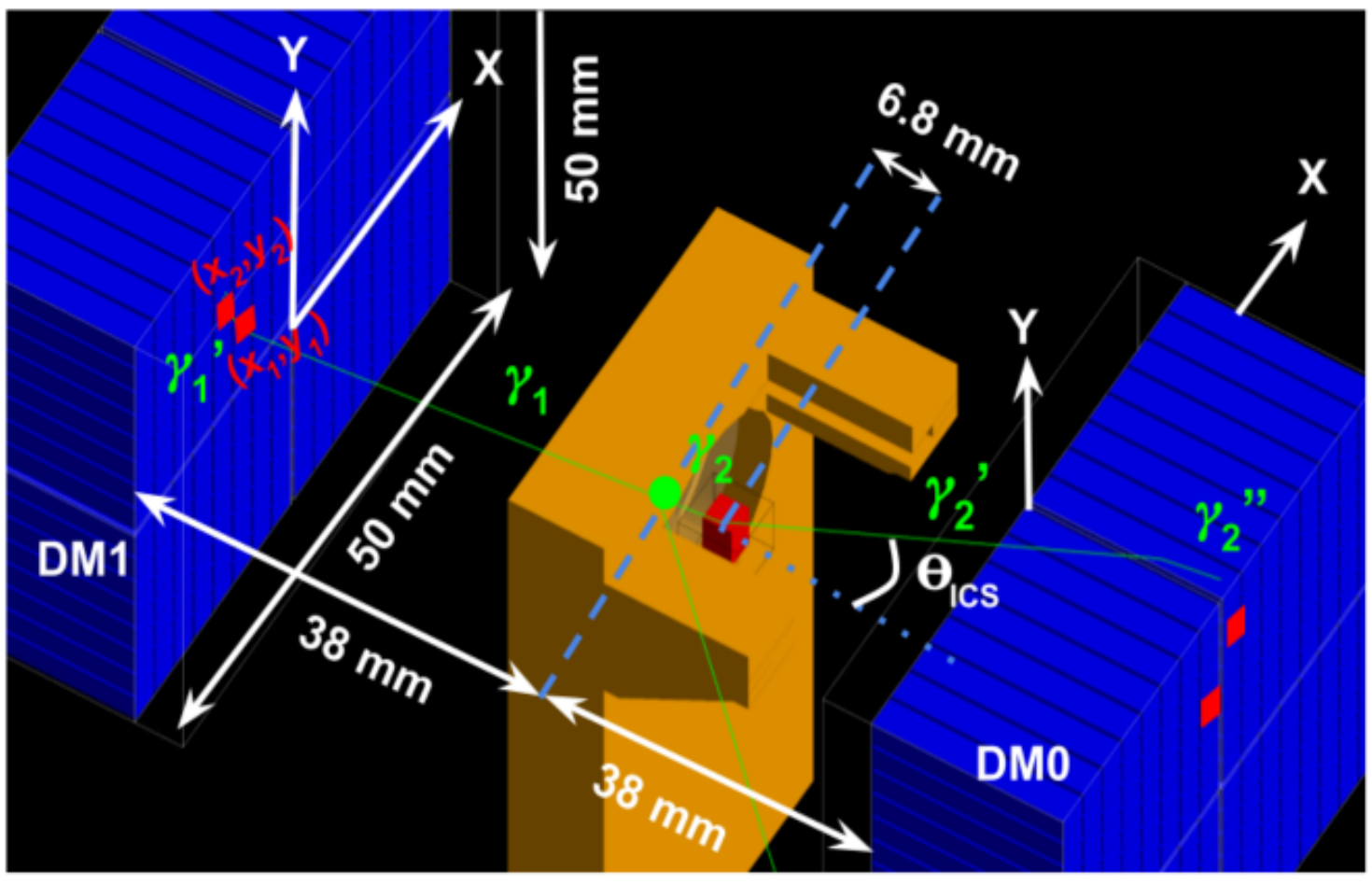

**Figure 3.** Schematic of the experimental setup of the experiment conducted at the University of York (Bordes *et al* 2024).

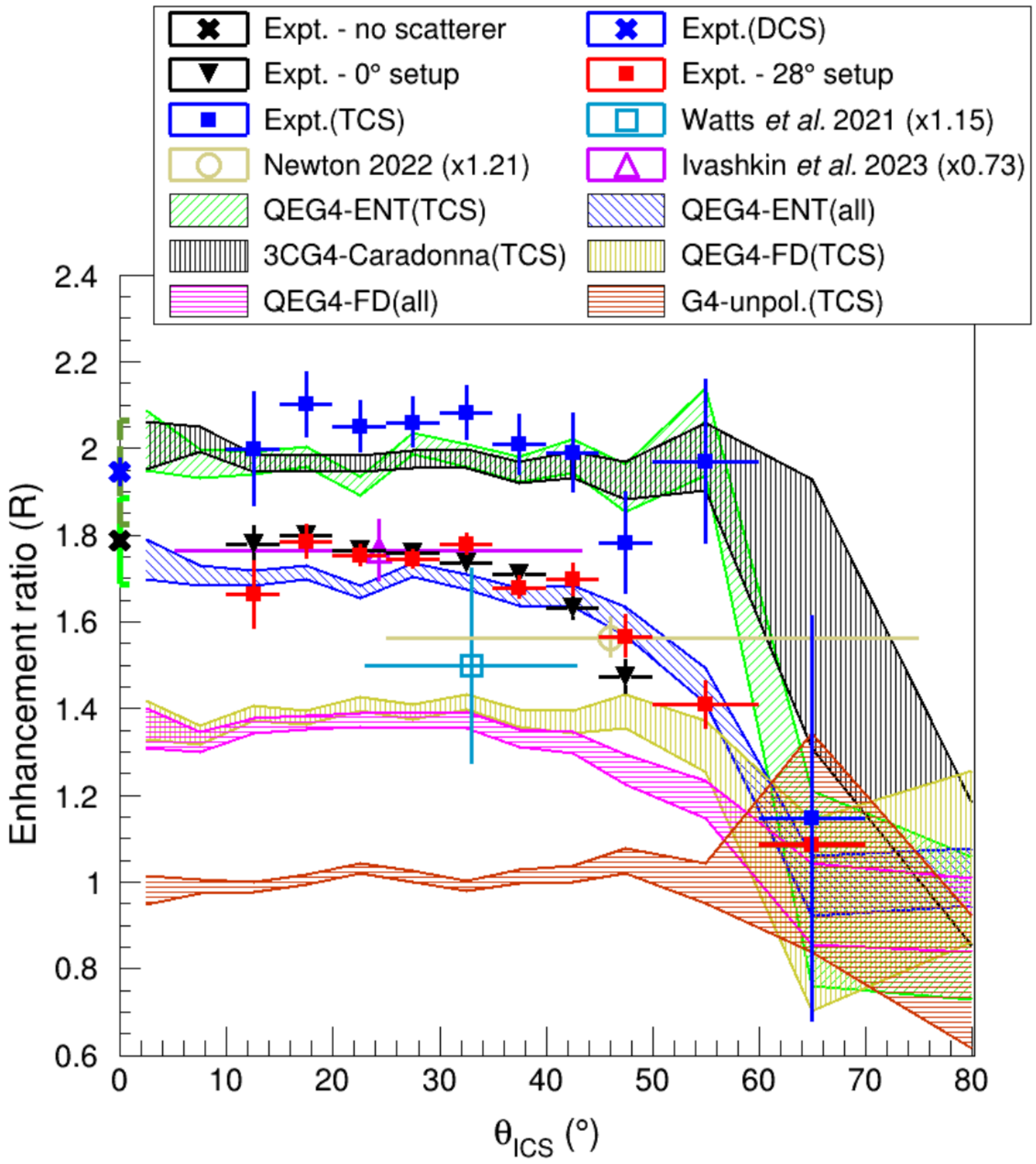

**Figure 4.** Enhancement ratio (asymmetry ratio) **R** for different intermediate scattering angles (Bordes *et al* 2024). Black triangles (red squares) represent measurements with modules back-to-back (rotated by $\theta_{scat}$=28°).

The two described experiments were conceptually very similar, but the differences in the experimental setups (the detector acceptance and resolution), as well as the ones in the analysis (reference frame definitions) could lead to different measured values of the parameter R and pose a challenge for direct comparison. Since the direct comparison between the results would not be reliable, here we compare the relative trends observed in both experiments. Table 2 gives the ratio of the R at given $\theta_{scat}$ with $R_0$ for the back-to-back case obtained with each setup, as the reference. These trends are consistent for the two described experiments within their experimental precision.

| $\theta_{scat}$ | $R(\theta_{scat})/R_0$ (Parashari *et al* 2024) | $R(\theta_{scat})/R_0$ (Bordes *et al* 2024) |
|---|---|---|
| **27.5°** | | 0.98 ± 0.01 |
| **30.0°** | 1.07 ± 0.07 | |
| **32.5°** | | 0.99 ± 0.02 |
| **47.5°** | | 0.88 ± 0.03 |
| **50.0°** | 0.90 ± 0.12 | |
| **55.0°** | | 0.84 ± 0.03 |

**Table 2.** Ratios of the asymmetry parameter R obtained in the experiments conducted by the Zagreb group (middle column) and the York group (right column). The values of the York data are graphically extracted from the Figure 4.

**Advances in Science and Technology to Meet Challenges**

To advance our understanding of the entanglement of the annihilation gamma quanta, challenges in both, experiment and theory must be met. Experimentally, it would be crucial to measure the angular correlation throughout the full angular range of $\theta_{ICS}$, since the current data suggest a different behaviour at larger angles. This, in principle can be achieved with existing polarimeters, like the ones described in this section. Further, a comparison between different experiments is also crucial to check their consistency and influence of systematic uncertainties. To achieve that, the results should ideally be corrected for experimental effects such as detector acceptance and efficiency. Finally, there had been indications that the measured correlation depends on the selected reference frame. Hence, this question should be taken under serious consideration in future measurements and corresponding analyses.

In the theoretical aspect, the modelling of the Compton scatter of the entangled gamma quanta had long been neglected. This is evident from the fact that only recently, the modelling of the Compton scattering of the polarized annihilation quanta had been implemented in Geant4 simulations (ver. 11.1 released in 2023). Therefore, it is encouraging that the first calculations of the triple Compton scattering (Caradonna 2025) give consistent description of the experimental results (Bordes *et al* 2024).

**Concluding Remarks**

The investigations of quantum decoherence of entangled gamma photons are of great interest for fundamental understanding of quantum entanglement at MeV energies, but they are also crucial for correct modelling of applications relying of this phenomenon such as quantum-entanglement medical imaging with PET. In the past few years, a significant progress has been achieved in both experimental part and theoretical modelling, however further investigations are needed to create consensual understanding of the underlying physics.

**Acknowledgements**

This work was supported by the Croatian Science Foundation under the project number IP-2022-10-3878 and by the “Research Cooperability” Program of the Croatian Science Foundation, funded by the European Union from the European Social Fund under the Operational Programme Efficient Human Resources 2014–2020, grant number PZS-2019-02-5829, and by the European Union – NextGenerationEU through the National Recovery and Resilience Plan 2021-2026 Institutional grants of University of Zagreb Faculty of Science (ProPubFO).

# 11. The Compton Scattering Dependence on Quantum Entanglement

*Peter Caradonna*
Department of Nuclear Engineering and Management, The University of Tokyo, Tokyo, Japan

**Introduction**

Quantum mechanics predicts that entangled annihilation photons exist in an indefinite polarization state before measurement. Each photon has an equal probability of being measured in either the right-circularly polarized or left-circularly polarized state. When the polarization of one photon is measured, its state collapses to a definite value. To conserve angular momentum, the polarization of the other photon must instantaneously collapse to the same state, regardless of the distance between them. This instantaneous collapse is what Einstein famously called "spooky action at a distance" (Einstein *et al* 1971), referring to the nonlocal nature of quantum entanglement.

In contrast, if the photons were not entangled, their polarization correlations would be fixed at the moment of creation. Any measurement on one photon could only affect the other after a time delay, constrained by the speed of light, as required by classical physics—a consequence of the local nature of classical particle interactions.

Although Bell's theorem provides a theoretical framework for distinguishing quantum entangled correlations from classical correlations, its application to annihilation photons is hindered by experimental limitations. Specifically, standard Bell tests rely on polarization measurements, but ideal linear polarization analyzers are unavailable for annihilation photons due to their high energy (511 keV). Given this constraint, this discussion instead emphasizes an experimental approach – specifically, how observed Compton scattering distributions can serve as evidence for entanglement.

Without an ideal polarization analyzer, any discrepancies between quantum predictions and experimental results in Bell test experiments could be attributed to instrumental limitations rather than a fundamental challenge to classical interpretations (Kasday 1971, Selleri 1988).

Although standard Bell tests cannot be performed on annihilation photons, multiple lines of evidence suggest that the measured correlations in their scattering distributions cannot be explained by local hidden variable models. First, Bell tests with maximally entangled optical photons, which have five orders of magnitude lower kinetic energy than annihilation photons, have consistently violated Bell's inequalities (Aspect *et al* 1982, Handsteiner *et al* 2017), largely ruling out local hidden variable models in those systems. If these results extend to gamma-ray energies, annihilation photons should also exhibit similar violations. Second, quantum electrodynamics (QED) predicts that electron-positron annihilation produces maximally entangled photon pairs, as described by Eq.(1). Moreover, the predicted scattering distributions from QED are in agreement with experimental results (Bertolini *et al* 1981, Bruno *et al* 1977, Kasday *et al* 1975, Langhoff 1960, Watts *et al* 2021, Wilson *et al* 1976, Wu and Shaknov 1950). Finally, the Klein-Nishina formula establishes a theoretical connection between scattering angles and polarization, reinforcing the expectation that polarization correlations should manifest in scattering distributions.

**Quantum Entanglement in Observed Scattering Distributions**

The transformation of state $|\psi_c^-\rangle$, given in Eq. (1), to the linear basis is particularly useful for delineating between quantum and classical correlations in entangled photon pairs. Using the nomenclature in Ref. (Caradonna *et al* 2024), $|\psi_c^-\rangle$ can be expressed as

$$|\psi_c^-\rangle = |\psi_l^+\rangle = \frac{1}{\sqrt{2}}|V,H\rangle + \frac{1}{\sqrt{2}}|H,V\rangle,$$

where the subscript $l$ in $|\psi_l^+\rangle$ denotes the linear basis spanned by $|V,H\rangle$ and $|H,V\rangle$, where $|V\rangle$ and $|H\rangle$ represent vertical and horizontal polarization states, respectively. The superscript $+$ indicates the relative phase between the eigenstates. If the annihilation photons remain in the entangled state $|\psi_l^+\rangle$, the ratio of coincidence count rates for Compton-scattered photons with perpendicular vs. parallel scattering planes reaches a maximum of 2.84 when both photons scatter at $\theta_1 = \theta_2 = \theta = 81.7°$.

However, Bohm and Aharonov proposed that if $|\psi_l^+\rangle$ breaks down at large separations, it should be replaced by a mixture of factorizable states that still preserves the fundamental conservation laws obeyed by the entangled state. While linear momentum conservation is satisfied pair by pair, angular momentum conservation cannot be maintained for individual photon pairs in a fixed reference frame but holds in the statistical average. By formulating this mixed state, Bohm and Aharonov effectively removed entanglement from the system while leaving nearly all other physical properties intact. Under these conditions, the coincidence count ratio due to this break down would decrease to approximately 1.6 (Bohm and Aharonov 1957).

At the time, this calculation could not be directly compared with the experiment of Wu (Wu and Shaknov 1950), as the detected photons had an angular spread around the ideal scattering angle of $\theta = 81.7°$. By accounting for this spread, an appropriate angular averaging yielded a revised theoretical prediction of $R = 2.00$, which closely matched the experimental result of $R = 2.04 \pm 0.08$.

In hindsight, by introducing the mixture of factorizable states, Bohm and Aharonov effectively transformed Wu's experiment from one that initially served to test the predictions of Pryce and Ward into one that also provided a test for evidence of action at a distance. This reframing situates Wu's experiment within a broader historical framework, highlighting its role as an early indication of nonlocality in high-energy interactions.

Bohm and Aharonov initially analyzed azimuthal correlations at a single scattering angle ($\theta = 81.7°$). In 2009, an analytical solution for the joint probability distribution of their mixture of factorizable states was derived for all scattering angles (Caradonna *et al* 2024).

To gain further insight, it is instructive to compare the joint differential cross section for Compton scattering of both photons, given in Eq.(8) and denoted as $\partial^2\sigma_{(\psi_l^+)}$, with the corresponding cross section for the mixture of factorizable states proposed by Bohm and Aharonov, $\partial^2\sigma_{(BA)}$ (Caradonna *et al* 2024), given by

$$\partial^2\sigma_{(\psi_l^+)} = \frac{r_0^4}{16}[H(\theta_1)H(\theta_2) - \mu(\theta_1)\mu(\theta_2)\cos 2\eta]\,\partial\Omega_1\,\partial\Omega_2,$$

and

$$\partial^2\sigma_{(BA)} = \frac{r_0^4}{16}[H(\theta_1)H(\theta_2) - \frac{1}{2}\mu(\theta_1)\mu(\theta_2)\cos 2\eta]\,\partial\Omega_1\,\partial\Omega_2.$$

respectively, where $\eta = \phi_1 + \phi_2$ and

$$H(\theta_i) = \frac{(1-\cos\theta_i)^3 + 2}{(\cos\theta_i - 2)^3}, \quad \mu(\theta_i) = \frac{\sin^2\theta_i}{(\cos\theta_i - 2)^2}, \quad i = 1,2.$$

Although the two cross sections are structurally similar, $\partial^2\sigma_{(\psi_l^+)}$ is not simply a scaled version of $\partial^2\sigma_{(BA)}$. The key difference lies in the second term of $\partial^2\sigma_{(BA)}$, which is reduced by a factor of $1/2$ compared to $\partial^2\sigma_{(\psi_l^+)}$. This reduction directly weakens the azimuthal correlations between the scattering planes of the final photons, making them generally less pronounced in the Bohm-Aharonov mixture than in the entangled state $|\psi_l^+\rangle$.

Using the differential cross sections $\partial^2\sigma_{(\psi_l^+)}$ and $\partial^2\sigma_{(BA)}$, the asymmetric ratio $Z(\theta)$ – as defined by Bohm and Aharonov (Bohm and Aharonov 1957) – can be obtained for all angles where $\theta_1 = \theta_2 = \theta$. This ratio is given by

$$Z(\theta) = \frac{\partial^2\sigma(\theta, \eta = 90°)}{\partial^2\sigma(\theta, \eta = 0°)}.$$

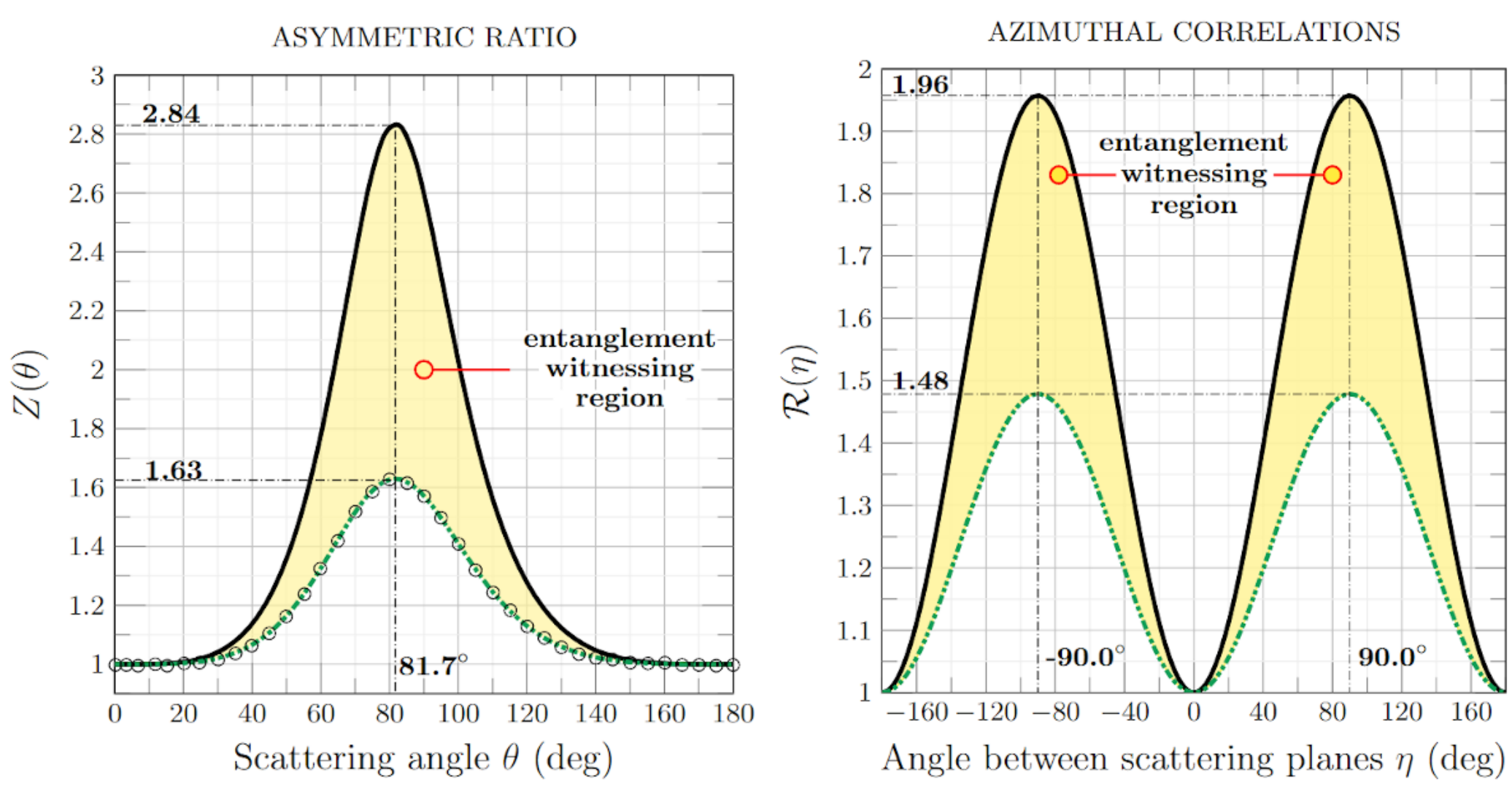

**Figure 1.** Left figure: The graph compares the relative dominance of photons scattering expressed in terms of the ratio of perpendicular divided by parallel scattering planes when entanglement is maximum (solid line) and zero (dash-dotted line), as a function of the Compton scattering angle $\theta_1 = \theta_2 = \theta$. Measurements within the shaded region indicate action at a distance. The analytical solution for the Bohm and Aharonov mixed state is compared with Geant4 simulation results (open circles) from Watts et al. (Watts *et al* 2021). Right figure: Azimuthal correlations for the case Annihilation photons (solid line), and the Bohm and Aharonov mixed state (dash-dotted line) in the case *($\theta_1 = \theta_2 = \theta = 81.7°$).*

Here, $Z(\theta)$ compares the differential cross section at two azimuthal angles, $\eta = 90°$ and $\eta = 0°$, providing insight into the relative dominance of photon scattering with perpendicular scattering planes versus parallel scattering planes. The results for $Z(\theta)$ for both the entangled state $|\psi_c^-\rangle$ and the Bohm-Aharonov mixture of factorizable states are presented on the left side of Fig. (1).

An alternative approach to studying correlations focuses on the configuration where the contrast between perpendicular and parallel scattering is maximized at $\theta = 81.7°$. Instead of analyzing only coincidence count ratios as a function of the scattering angle $\theta$, this method examines the coincidence count rate as a function of the relative angle between the scattering planes, characterized by the quantity $R(\theta, \eta)$, defined as

$$R(\theta, \eta) = 1 + \frac{1 - Z(\theta)}{1 + Z(\theta)}\cos 2\eta.$$

The function $R(\theta, \eta)$ possesses several useful properties. Deviations from unity indicate momentum correlations between the scattered photons (Kasday *et al* 1975). If the momenta are uncorrelated, $R(\theta, \eta)$ remains equal to one for all values of $\eta$. To distinguish between azimuthal correlations influenced by entanglement and those that are not, we define a modified function $\mathcal{R}(\theta, \eta)$, which is constructed so that its minimum value is unity (Refer to appendix in Ref. (Caradonna *et al* 2024)). In an ideal geometry where $\theta_1 = \theta_2 = \theta = 81.7°$, $\mathcal{R}(\theta, \eta)$ evaluates to

$$\mathcal{R}(\theta = 81.7°, \eta)_{(\psi_c^-)} = 1.479 - 0.479\cos 2\eta,$$

and

$$\mathcal{R}(\theta = 81.7°, \eta)_{(BA)} = 1.240 - 0.240\cos 2\eta.$$

and the results are presented on the right side of Fig. (3).

**Concluding remarks**

Directly measuring the polarization state of both photons to access their strong polarization correlations is currently beyond experimental reach. For now, the best alternative is to exploit scattering distributions as an indirect means of extracting information. By analyzing these correlations, quantum entanglement may be inferred without direct polarization measurements. In doing so, quantum theory provides a detailed framework to guide the research and development of quantum-enabled PET systems.

# 12. Quantum Entanglement in Three-Photon Positron-Electron Annihilation

*Marek Nowakowski*
ICTP-South American Institute for Fundamental Research, São Paulo, Brazil
Universidade Federal de São Paulo, São Paulo, Brazil

**Status**

Medical Physics and Fundamental Physics, albeit both an integral part of physical sciences, seem at the first glance to be far apart having different goals and methods. It can come as a surprise how entwined they are in reality. The medical diagnostic method of PET uses several fundamental physical discoveries (Bass *et al* 2023). The first one is the existence of anti-matter (positron), a prediction of relativistic quantum mechanics, confirmed experimentally in 1932 by C. D. Anderson. Even though we are sure that it is there, the universe consists mostly of matter and anti-matter is indeed rare. To have it ready for applications one needs a rare nuclear decay, a variant of beta decay where the proton decays into a neutron, positron and a neutrino inside the nucleus. In free space this decay is not possible due to energy- momentum conservation, i.e., the proton mass is smaller than the mass of a neutron. But the proton in a nucleus is bound, and the decay happens albeit rarely in proton rich nuclei. The PET machine detects photons which are produced by the electron-positron annihilation into two photons, an example of transmutation of matter (here into radiation) allowed and predicted by relativistic quantum field theory. A variant of the prompt annihilation is the formation of positronium; the proton in the hydrogen atom is replaced by the positron and the positronium is just a “sibling” of the hydrogen albeit unstable. The positronium in turn decays into two or three photons depending on the spin alignment of the electron and positron. The three-body decay has an interesting quantum mechanical property: the three photons are entangled (Acín *et al* 2001, Nowakowski and Bedoya Fierro 2017). Very roughly speaking and explained in more detail below, the three photons do not behave independently. Could it be that we can also use this entanglement in medical applications (Moskal 2021, Moskal 2024, Moskal 2026). Or vice versa, can we use the PET machine designed for medical purposes to probe into fundamental issues of physics. This indeed has been done already for a couple of problems addressing fundamental questions about discrete symmetries (Moskal P *et al* 2021, 2024). Is this also possible in the case of a three-body entanglement of photons? Some first steps towards this program have been already proposed and done (Hiesmayr and Moskal 2019).

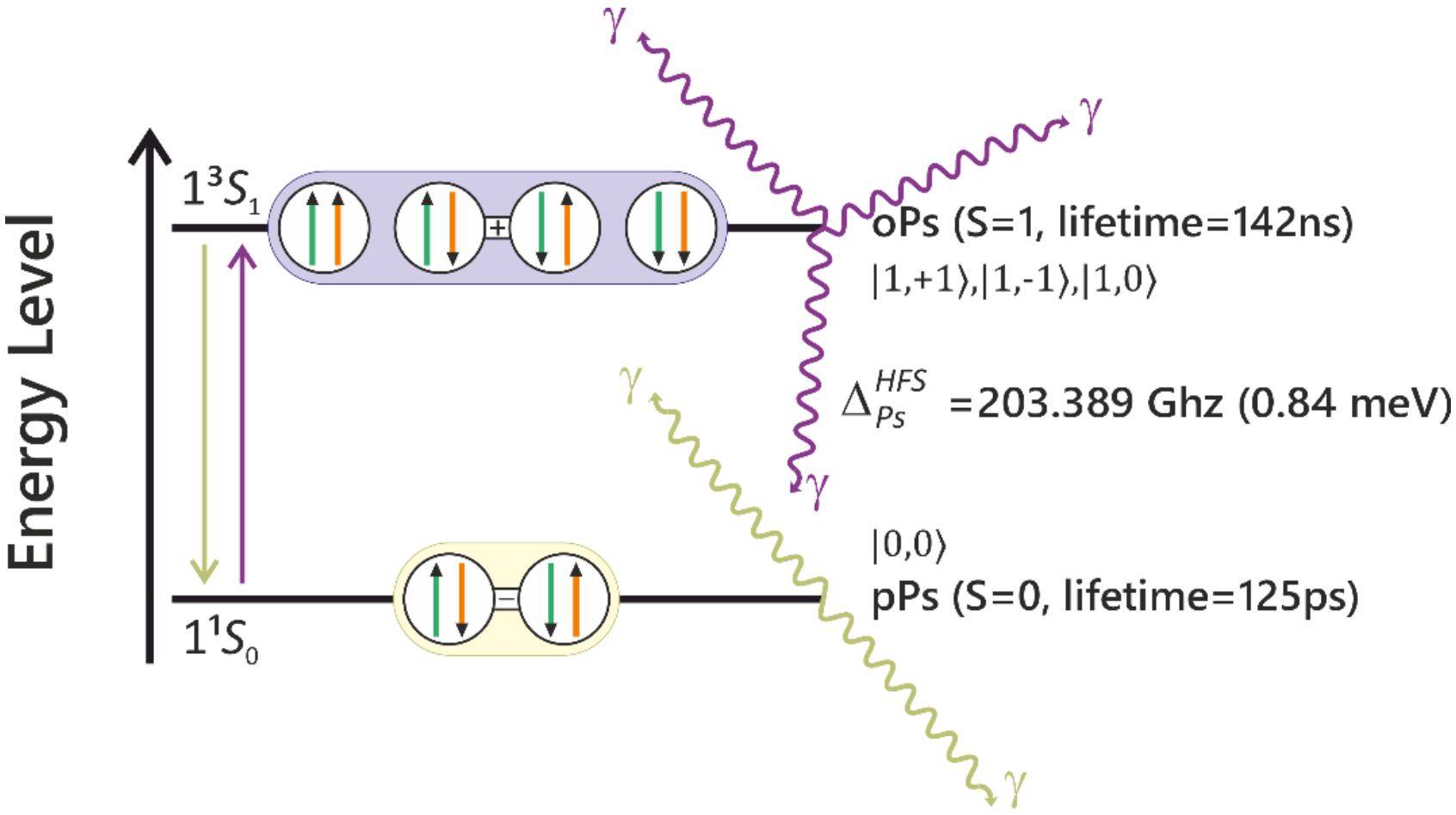

**Figure 1.** Schematic representation of positronium in a compact graphical form.

**Current and Future Challenges**

Quantum mechanics describes the physical reality in terms of a linear space called Hilbert space. The elements of this space are referred to as vectors or states, i.e., $|\mathrm{x}\rangle$. Because the space is linear, we can have a superposition of states. The state can be, e.g., $|\mathrm{S}\uparrow\rangle$ (spin up) or $|\mathrm{S}\downarrow\rangle$ (spin down), but also $|\mathrm{S}\rangle = |\mathrm{S}\uparrow\rangle + |\mathrm{S}\downarrow\rangle$, in which case we talk of an undetermined state. Only after a measurement the state becomes determined. One speaks of wave function collapse: $|\mathrm{S}\rangle \rightarrow |\mathrm{S}\uparrow\rangle$ if we have measured up. Such an indeterminism is classically not possible. We can say a particle is either at a position x or y, but we cannot write this as a superposition. If we have two particles, everyone is described by its proper Hilbert space and therefore the two-particle state is a product state, e.g., $|\mathrm{Product\ Spins}>= |\mathrm{S}\uparrow\rangle|\mathrm{S}\downarrow\rangle$. An undetermined state in the product space is, for instance

$$|\mathrm{Entangled\ Spins}>= |\mathrm{S}\uparrow\rangle|\mathrm{S}\downarrow\rangle + |\mathrm{S}\downarrow\rangle|\mathrm{S}\uparrow\rangle.$$

It is customary to imagine that the left-hand side of the entangled product state is in the hands of Alice whereas Bob has the right-hand side. If Alice measures spin up, the wave function $\mathrm{Entangled\ Spins}$ collapses to $|\mathrm{S}\uparrow\rangle|\mathrm{S}\downarrow\rangle$. One talks about quantum entanglement. Of course, the quantum entanglement can also happen in different variables. It is known since the seminal paper by Einstein, Rosen and Podolsky in 1935 (Einstein *et al* 1935). What is interesting about it, is its clear features of non-locality. Inequalities (Bell (Bell 1964) and Clauser-Horne (Clauser and Horne 1974) inequalities) have been derived on the basis of locality (for instance, a statistical independence of two events A and B is expressed by the fact that the probability that A and B occurs is simply a product P(A, B)=P(A)P(B)) which are not satisfied by the observables of a quantum entanglement which, essentially, is a proof for its inherent non-locality.

There are two kinds of electron-positron bound states (positronium): the para-positronium with total spin 0 and the ortho-positronium with spin one. Because of the conservation of a discrete symmetry (charge conjugation) the ortho-positronium in a PET machine decays into three photons which turn out to be entangled in the photon spins (called polarizations). The three-particle entanglement between the three participants is simultaneous. Because of the big number of events, a PET machine seems ideal to study this peculiar nature of quantum mechanics and it could well be, like with other fundamental discoveries in physics, that it will be useful in medical applications.

**Advances in Science and Technology to Meet Challenges**

A reasonable amount of work to understand the nature of three genuinely entangled photons has already been done by the JPET group in Cracow (Hiesmayr and Moskal 2019). Therefore here, we suggest yet another approach which bears a strong similarity to the two-particle entanglement where the quantum non-locality was revealed because the quantum probabilities in the entanglement did not satisfy certain inequalities like the Bell or Clauser-Horne inequality. Could we proceed in a similar way in the three-particle case? In principle, the answer is affirmative. The Clauser-Horne and Bell inequalities have been derived for the first time by the mathematician George Boole (Boole 1862, 1958). His derivation relies on the classical consistency of the interpretation of the probabilities and therefore it is a priori not clear where he used the classical locality. Nevertheless, the Bell and the Clauser-Horne inequalities appear in his list of consistent inequalities (Pitowsky and Svozil 2001). One can generalize Boole's results for three events (A, B, C) with two or three possible outcomes called Bell-Boole inequalities (Pitowsky and Svozil 2001). Here A means that Alice measures her photon with an outcome $A_1$ or $A_2$ . Similarly, B stands for Bob and C for, say, Charlie. Many of these new inequalities in the list are trivial or can be related to others. For a flavor of their complexity we just cite one of them (Pitowsky and Svozil 2001)

$$\begin{aligned}2 \geq & -P(A_1) + 2P(A_2) + P(B_1) + P(B_2) - P(C_1) + 2P(C_2) - P(A_1B_1) + P(A_1C_1) + 2P(A_1B_2) \\ & + P(A_1C_2) - P(A_2B_1) + P(A_2C_1) - 2P(A_2B_2) - 3P(A_2C_2) + P(B_1C_1) - P(B_2C_1) \\ & - P(B_1C_2) - 2P(B_2C_2) + 2P(A_1B_1C_1) - 2P(A_2B_1C_1) - 2P(A_1B_2C_1) \\ & - 2P(A_1B_1C_2) + 2P(A_2B_2C_1) + 2P(A_2B_1C_2) - P(A_1B_2C_2) + 3P(A_2B_2C_2).\end{aligned}$$

It would be a wonderful undertaking to check which one of the new and non-trivial Boole-Bell inequalities is violated by a three-photon entanglement on ortho-positronium. In case of a positive experimental outcome, i.e., we find a violation, we would have to dig deeper into this inequality to find the reason of its violation, which, like in the two-particle entanglement, lies most likely in the non-locality of quantum mechanics. Seen from yet another perspective, we notice that Boole himself could not have known about quantum mechanics. This fact alone tells us that classical probability is inherently different from its version emerging in quantum mechanics.

**Concluding Remarks**

Quantum entanglement became a resource in quantum computation and quantum communication, but it is always gratifying to study a particular example which occurs naturally in nature. More so, our example lies between two areas of physics as already mentioned above. I outlined the basic features of a three-photon entanglement appearing in the PET machines through the decay of ortho-positronium. I suggested to link the idea of Boole (classical inequalities) to these quantum events to probe their non-classical nature. At the same time, I had to leave out some technical complications caused by the fact that the photons states will not be always pure. This would require then another tool like the density matrix. It remains to be seen whether quantum entanglement can be useful in medical diagnostic applications as it is already in computer science. At the same time, it is already now clear that the Positron-Electron Tomography will contribute in understanding the entanglement issue in fundamental physics (Bramon and Nowakowski 1999).

**Acknowledgements**

Cheerful discussions with Prof. Paweł Moskal are gratefully acknowledged. I would like to thank Aleksander Khreptak for the kind help with the figure.

## 13. Sensitivity of Quantum Entanglement Imaging with Total-Body PET scanners

*Sushil Sharma, Deepak Kumar, Paweł Moskal*
Institute of Physics, Jagiellonian University, Poland
Center for Theranostics, Jagiellonian University, Poland

sushil.sharma@uj.edu.pl, deepak.kumar@uj.edu.pl, p.moskal@uj.edu.pl

**Status**

PET scanners achieve diagnostic performance through their ability to detect a high fraction of 511-keV photons. High acceptance rate enhances the signal-to-noise ratio (SNR) which enables equivalent image quality through shorter acquisitions and reduced injected doses, thereby improving lesion detectability and quantitative accuracy in clinical practice (Badawi *et al* 2019, Cherry *et al* 2018, Karp *et al* 2020). The fundamental requirement for Quantum-Entanglement PET (QE-PET) depends on sensitivity, as it utilizes the polarization correlation of annihilation photons, which necessitates their registration before and after scattering inside the detectors. The Klein-Nishina cross-section defines how photon polarization relates to scattering angles (Klein and Nishina 1929). Entanglement manifests itself as azimuthal correlation (Δϕ) between their scattering planes (Moskal *et al* 2018, Ivashkin *et al* 2023, Bordes *et al* 2024, Parashari *et al* 2024, Moskal et al 2025). Quantum electrodynamics predicts that joint cross-section modulation reaches its peak when both photons scatter near 82° angles (Caradonna 2024, Pryce and Ward 1947). The utilization of entanglement in QE-PET depends on detector-scattered events which are typically discarded in the conventional PET (Moskal *et al* 2018). The measurement of entanglement must precisely register four interactions (two primary interactions and their respective Compton-scattered secondary interactions) occurring within a single coincidence timing window that constitutes an event.

**Current and Future Challenges**

The transition of QE-PET from laboratory demonstration to clinical application faces substantial obstacles despite its promising theoretical foundation. One of the main challenges arises from detector limitations. Conventional LYSO/LSO crystal arrays favour photoelectric absorption, a consequence of their high atomic number, rather than the Compton scattering essential to QE-PET. Because the interaction site is effectively confined to the pixel centre, positional uncertainty propagates into the reconstructed scattering angle, giving state-of-the-art clinical PET systems an angular uncertainty on the order of 10°–15° (mean absolute error, extrapolated from CZT-based position-resolution estimates (Yang et al. 20)), making reliable sequencing of Compton interactions and thus reconstruction of incident-photon directions difficult.The ambiguity of ordering increases when time-of-flight (TOF) resolution exceeds about 100 ps. Therefore, the main detector-related limitations to clinical QE-PET stem from restricted angular resolution, sub-optimal TOF performance and limited axial field-of-view (AFOV). The imprecise measurement of scattering angles makes it difficult to detect entanglement signatures in QE-PET while also making event classification harder because of noisy data and reducing the quantum correlations in coincidences which decreases system sensitivity. Furthermore, scattering within the patient may introduce additional decoherence (Sharma 2022). However, the recent experimental studies show that the polarization correlation is essentially preserved for scattering angles below about 30° and is reduced, though not destroyed, by around 50° (Bordes et al 2024, Parashari et al 2024). A further limitation, and the one that is decisive for sensitivity, is the axial field-of-view (AFOV). Short AFOV of current prototype removes a large fraction of photon pairs and undermines the statistical reliability of quantum-correlation measurements. These limitations motivate two complementary responses. The first combines plastic scatterers with crystal absorbers, or employs Compton cameras based on CZT semiconductor detectors, which offer sub-millimetre positional accuracy and high energy resolution at the cost of greater complexity and more difficult scale-up (Moskal and Stępień 2020, Romanchek et al 2024). The second enlarges the axial coverage itself. Total-body geometries recover the pairs lost to a short AFOV, and their meter-scale extent raises the coincidence yield to the level required for quantum-correlation imaging, as developed in the remainder of this chapter.

**Advances in Science and Technology to Meet Challenges**

**(a) Total-body PET as a route to high sensitivity**: The EXPLORER (Badawi *et al* 2019) and PennPET (Karp *et al* 2020) total-body scanners demonstrate how a long axial field of view raises sensitivity, the property on which QE-PET ultimately depends. Their 1-2 m axial fields of view increase detection sensitivity by up to forty times compared to conventional PET scanners (Badawi et al 2019, Cherry et al 2018). The expanded acceptance increases quantum-event statistics, which improves the detection of weak polarization correlation. The recovery of scattered coincidences further enhances Δϕ reconstruction. Total-body imaging also supports dynamic imaging of fast tracer kinetics and the measurement of positronium lifetime, a parameter sensitive to tissue oxygenation and a candidate hypoxia biomarker (Moskal and Stępień 2021, Takyu et al 2024, Moskal et al 2024). The J-PET prototype demonstrates how plastic scintillators serve as a cost-effective, Compton-favouring platform for QE-PET: their high Compton yield preserves the polarization information, although their low density restricts overall sensitivity and makes event reconstruction more demanding (Niedźwiecki et al 2017, Moskal and Stępień 2020). The total-body J-PET design described in the following sections aims to retain this Compton advantage while recovering the sensitivity lost to the low density of plastic.

**(b) Quantum-Optimized Scanner Design**: For a Quantum-entanglement enabled PET scanner, it is important to keep a good trade-off between the sensitivity of the detector and the amount of solid angle that is covered by Compton-scattering, to preserve the polarization information. Multi-ring detector geometries with a scatterer in combination with a high-Z absorber (LYSO/LSO/CZT) are a feasible approach to achieve these goals. In such hybrid detectors, the scatterer preserves the Compton interaction and thus the polarization information, while the absorber stops the scattered photon and allows for precise event reconstruction. There are several detector concepts that have been designed along these lines (Yoshida et al 2020, Shimazoe et al 2020, Romanchek et al 2024) and complementary single-layer polarimeter studies show that the azimuthal correlation (Δϕ) can be sampled to suppress random background (Kožuljević et al 2026). However, such detectors are more demanding to build than a single-material design, requiring more intricate mechanics, costlier detector materials and elaborate electronics for combining the individual detector signals. Also, the timing and energy information of the scatterer and absorber has to be carefully reconciled before the corresponding events can be matched and sequenced (Tashima and Yamaya 2022). A complementary route keeps a single low-Z material throughout and recovers sensitivity through axial scale; the total-body J-PET (TB-J-PET) concept follows this approach and is described in the following section (Moskal et al 2024, Baran et al 2025).

**(c) Total-Body J-PET: Preliminary Design and Feasibility:** The Total-Body Jagiellonian-PET (TB J-PET) represents an extension of the novel plastic-scintillator-based concept from a research-size device to a whole-body imager of a meter-size. The initial design of 85 cm-diameter cylinder consisted of three layers of 50 cm long scintillator strips in the axial direction to provide whole-body coverage at reduced cost (Niedźwiecki *et al* 2017, Moskal *et al* 2025). The modular J-PET is a portable design of 24 independent detection modules, configured as a single-layer ring of 50 cm axial length. It produced the first in vivo positronium-lifetime image of the human brain, the first clinical application based on multiphoton registration (Moskal et al 2024). However, the single-layer design reached only 0.06 cps

kBq$^{-1}$ triple-gamma sensitivity, which confined quantum-entanglement studies to long acquisition times.

TB J-PET replaces the single ring with seven coaxial rings of independent detection modules (24 modules per ring) that span 2.43 m axially and define a 41.4 cm inner radius. Each module comprises two parallel layers of 16 plastic scintillators (330 mm × 30 mm × 6 mm) read out by SiPM arrays at both ends of each strip (Baran *et al* 2025). The two layers nearly double the interaction cross-section, partly compensating for the lower density of plastic relative to LYSO. Each module also carries a wavelength-shifter (WLS) layer between the two scintillator layers, which provides an axial resolution of about a few mm FWHM and thereby sharpens the angular resolution available for Δϕ reconstruction. The full geometry is shown in Fig. 1(a).

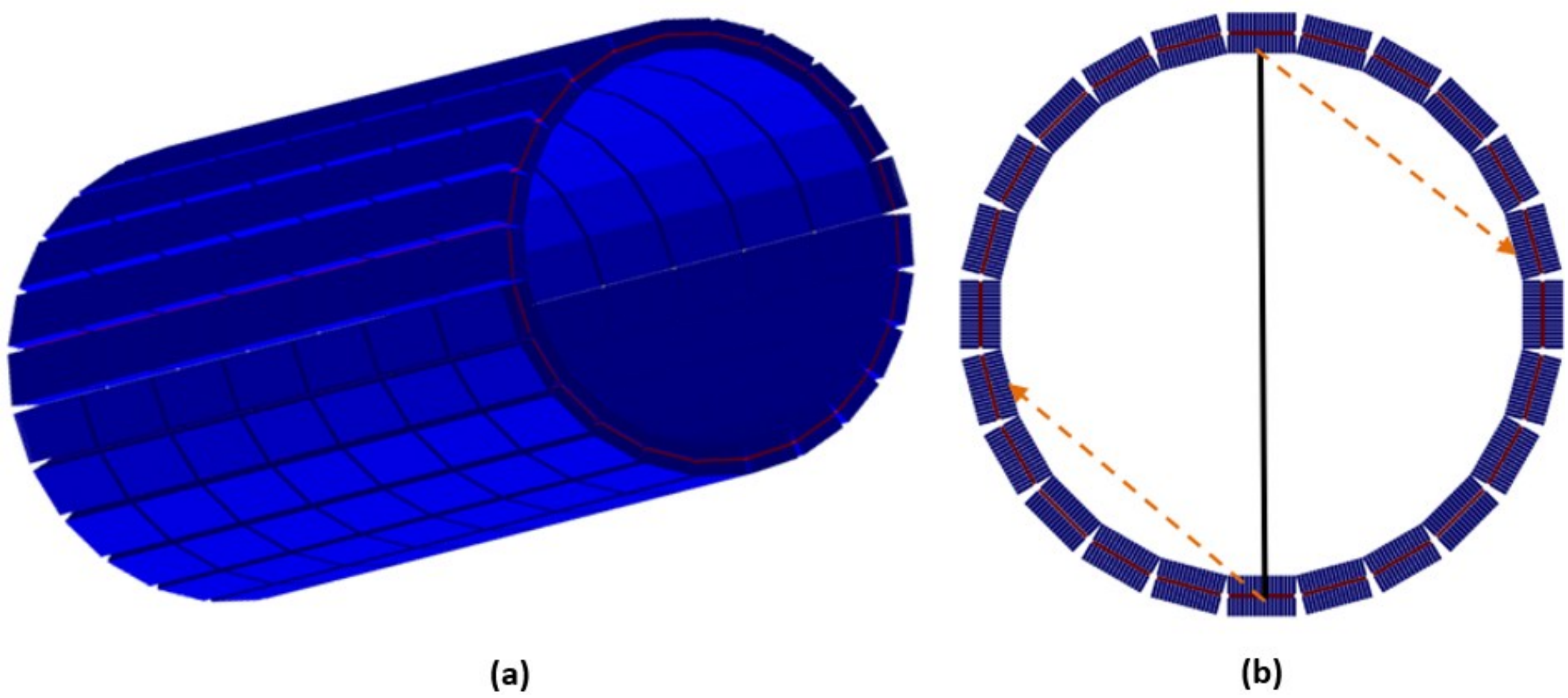

**Figure 1**. Geant4 rendering of the total-body J-PET scanner. (a) A 2.43 m-long cylinder of seven concentric rings with an inner radius of approximately 41.4 cm. The detection modules in each ring consist of two 330 mm EJ-230 plastic-scintillator layers separated by wavelength-shifter plates (red). (b) Transaxial view of a single ring. Blue rectangles denote individual scintillator strips and red lines the interleaved WLS plates. The black solid arrow indicates a back-to-back 511 keV annihilation-photon pair, and the dashed orange arrows their first Compton scatters. Registering all four interactions within one coincidence window allows the azimuthal correlation Δϕ to be obtained, as required for quantum-entanglement PET.

The two-layer geometry together with the extended axial field-of-view raises the two-gamma sensitivity to 34 cps kBq$^{-1}$ while the (2 + 1) triple-gamma sensitivity reaches 1.66 cps kBq$^{-1}$, about a 28-fold improvement over the 0.06 cps kBq$^{-1}$ of the single-layer modular prototype. To estimate the performance of TB-J-PET scanner for QE-PET imaging, a Monte Carlo study using the Geant4 toolkit was carried out, generating two back-to-back 511 keV photons. Events were classified according to the interactions of the annihilation photons and their subsequent scatters within the TB-J-PET volume. The quantity reported here is the geometric acceptance, defined as the fraction of generated back-to-back photon pairs that produce four true hits in the active volume, namely the two primary interactions and their two Compton-scattered secondaries. This acceptance was obtained from true-hit information alone, without energy smearing and without a minimum energy deposition criterion for photon registration and therefore represents an idealized upper limit on the four-hit yield. Since these idealized conditions act in the same manner at every axial position, the acceptance scales with the system sensitivity along the scanner axis. Consequently, the acceptance distribution shown in Figure 2 also describes the axial dependence of the four-hit sensitivity. The acceptance remains above 50% of its central value across a ±85 cm range (**Fig. 2**), which supports whole-organ quantum imaging with approximately uniform statistics. At the centre of the scanner the idealized acceptance is about 2.5%. When realistic energy-window selection, registration thresholds, and intrinsic detection efficiency are taken into account, this value corresponds to a four-hit QE-PET sensitivity of about 3 cps kBq$^{-1}$.

The plastic strips promote Compton interactions, which keep both annihilation photons and their first scatters within the active volume, while the WLS-based axial resolution preserves the scattering-angle precision needed to extract the azimuthal correlation Δϕ. The combination of long axial coverage, dual layers for higher interaction probability, WLS-assisted axial resolution, and triggerless multi-gamma acquisition together make TB-J-PET a clinical platform capable of recording the four-hit signatures required for polarization-based QE-PET, while the use of plastic scintillators keeps the material cost comparatively low.

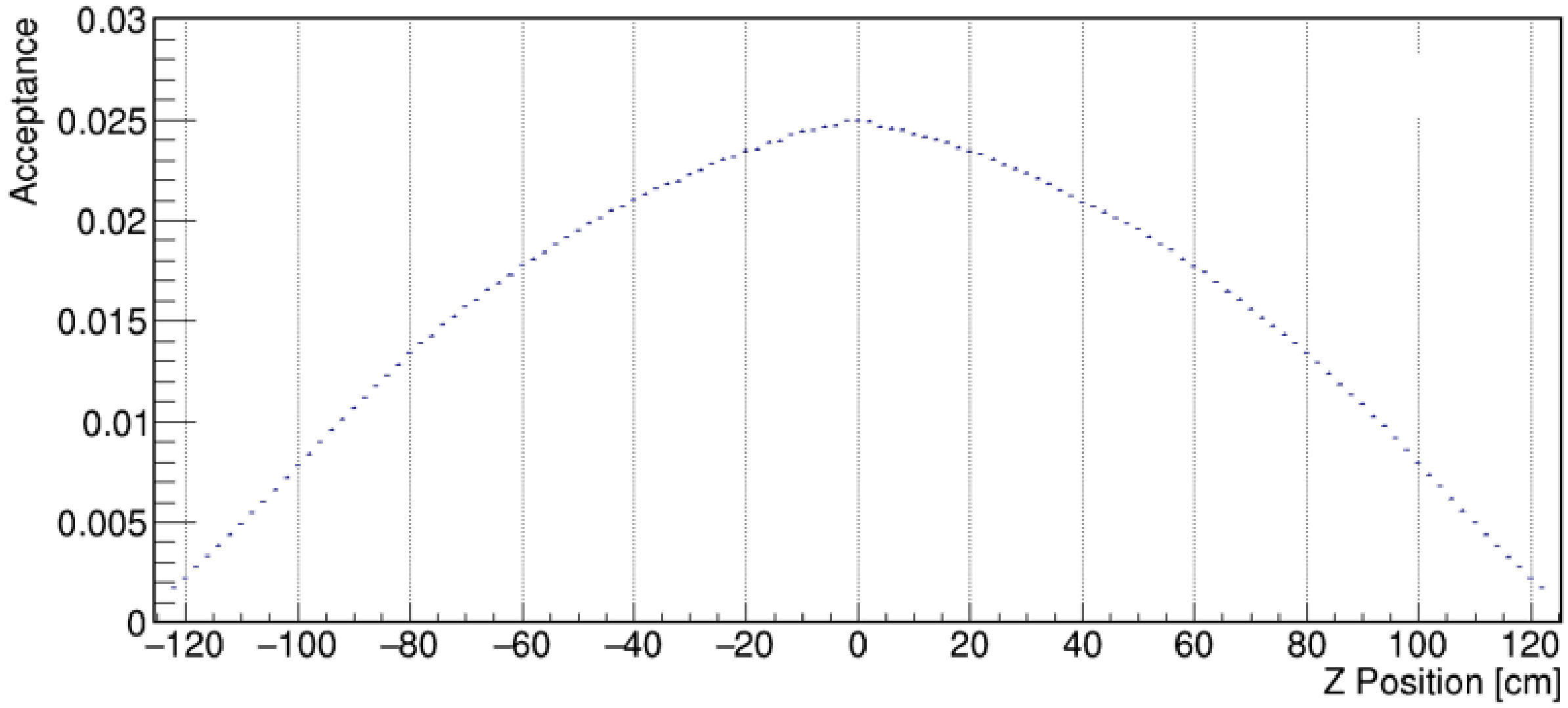

**Figure 2**. The simulated axial acceptance of TB-J-PET for QE-PET four-hit events is shown. The acceptance is defined as the fraction of back-to-back 511 keV photon pairs that generate four true hits i.e. two primary interactions and two Compton interactions for the given four hits, without energy smearing and without any threshold cuts. The seven-ring detector spanning 2.43 m axially reaches a peak acceptance of about 2.5% at z = 0, and the acceptances of more than 50% of this maximum value are maintained over a long axial length of ±85 cm.

**Concluding remarks**

Sensitivity in QE-PET is critical to translate it into clinical application. A QE-PET event consists of four interactions: two 511 keV photons and their first Compton scatters within the detector material. Hence, it is a much rarer signal than the two interactions registered by standard PET. Four interactions demand high yield of starting events and high angular resolution of scattering interaction. Total-body coverage is one of the ways to increase the yield of starting QE-PET events, i.e., to recover lost photon pairs discarded by short axial field-of-view of conventional PETs, while keeping the detector acceptance nearly uniform along the body of object under study. The Compton-dominated interactions registered by plastic scintillators, improved by WLS readout providing high axial resolution, represent the second ingredient crucial for the scattering angle resolution necessary for the reconstruction of Δϕ. Among various designs of detector under study, seven-ring configuration of TB-J-PET represents a good compromise between the plastic detector with its low cost and whole organ imaging instead of only central slice of interest. The four-hit sensitivity of seven-ring (preliminary design) TB-J-PET amounts to about 3 cps $kBq^{-1}$ at the center of detector, keeping its half-value over ±85 cm.

**Acknowledgement**

We acknowledge support from the European Union within the Horizon Europe Framework Programme (ERC Advanced Grant POSITRONIUM no. 101199807), the National Science Centre of Poland through grants no. 2021/42/A/ST2/00423, 2021/43/B/ST2/02150, 2022/47/I/NZ7/03112, 2023/50/E/ST2/00574, 2024/53/N/ST2/04279, the Ministry of Science and Higher Education through grant no. IAL/SP/596235/2023 and SPUB/SP/627733/2025, the SciMat and qLife Priority Research Areas budget under the program Excellence Initiative – Research University at Jagiellonian University.

# 14. Degree of Quantum Entanglement as a Novel Diagnostic Biomarker

*Paweł Moskal, Ewa Łucja Stępień*
Marian Smoluchowski Institute of Physics, Jagiellonian University, Kraków, Poland
Center for Theranostics, Jagiellonian University, Kraków, Poland
p.moskal@uj.edu.pl, e.stepien@uj.edu.pl

**Status**

Positron emission tomography (PET) is a well-established diagnostic method delivering both static and dynamic images of pharmaceutical uptake (e.g., FDG) and receptor expression on cell membranes (e.g., PSMA) (Bharathi *et al* 2025). PET images are reconstructed based on the registration of the 511 keV photons created in the annihilation of electrons from the patient with positrons emitted by radioactive isotopes attached to the administered pharmaceuticals. So far in the routine PET diagnostics only information about the density distribution of annihilation sites is used. The standard PET scanners are insensitive to the annihilation mechanisms and to the degree of quantum entanglement $C_{QE}$ (eq. 1 in Introduction section), that may in principle deliver additional information about the tissue molecular architecture and degree of oxygenation (Moskal 2026).

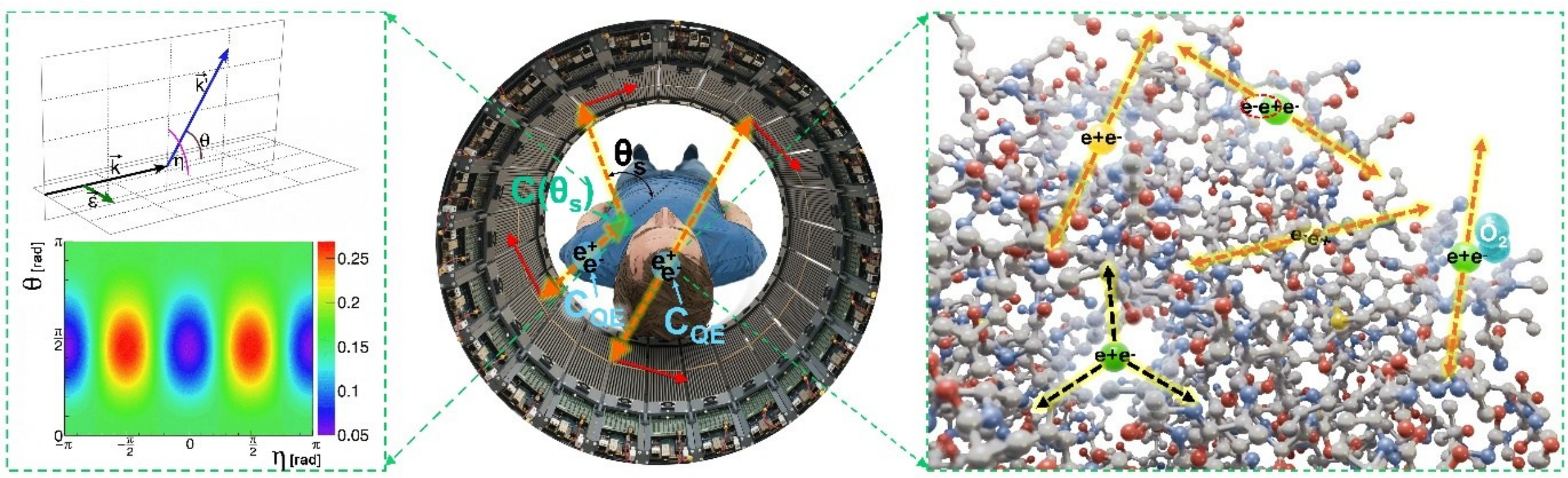

**Figure 1.** Illustration of the main processes underlying the imaging of quantum entanglement degree. The middle panel shows a photograph of the J-PET modular scanner (Moskal *et al* 2024) -the first QE-PET scanner- with a superimposed patient and electron-positron annihilation events. Yellow-red dashed arrows indicate annihilation photons, while red arrows indicate photons after Compton scattering in the detector. The angle θs denotes the scattering angle if the annihilation photon undergoes Compton interaction within the patient. The zoomed image on the right illustrates the primary processes of electron-positron annihilation, including direct annihilation (center) and annihilation via the formation of para-positronium (yellow circle) and ortho-positronium (green circles). Ortho-positronium may self-annihilate into three photons, or it can annihilate into two photons via pick-off (upper right) and conversion into para-positronium via oxygen (right). Observation of non-maximal entanglement in porous material (Moskal *et al* 2025a) indicates that the degree of quantum entanglement $C_{QE}$ depends on the annihilation mechanism. The zoomed image on the left illustrates the Compton scattering of annihilation photon on electron in the patient body. Figure indicates an angle η between polarization and Compton scattering planes. Photon momentum before and after scattering is indicated as k and k', respectively. Lower figure shows the normalized dependence of the scattering angle θ as a function of the angle η (Moskal *et al* 2018), indicating that scattering is at most probable in the plane orthogonal to polarization plane (η = +- π/2), and that the correlation is at best visible at scattering angles close to θ ≈ 82°. The degree of correlation after scattering is described by $C(\theta_s)$, that was determined to be close to unity for the θs < 36° (Caradonna 2024, Tkachev *et al* 2025) relevant for PET imaging (Moskal 2026).

To determine $C_{QE}$, the registration of Compton-scattered photons is required (red arrows, Fig. 1), as explained in detail in the Introduction. In standard PET imaging, all processes leading to two 511 keV photons are indistinguishable. However, the various fractional contributions of these processes, which depend on the tissue's molecular architecture and degree of oxygenation (Moskal 2026), influence positronium properties. As recently observed, these contributions also influence the degree of quantum entanglement (Moskal *et al* 2025a).

Positrons emitted by the radionuclide may annihilate in the patient's body directly with an electron or via the intermediate formation of a positronium atom. Possibilities for electron-positron annihilations in tissue into two or three photons are detailed in Fig. 3 of reference (Moskal 2026). Only the main mechanisms are shown in Fig. 1 below.

Information about the relative fraction of the various annihilation mechanisms and about properties of positronium atoms such as mean lifetime of ortho-positronium and the ratio of 3γ/2γ decays rates delivers information about the tissue nano-architecture and may become useful in clinics (Moskal *et al* 2025b). Recently introduced method of positronium imaging (Moskal 2020) enables imaging of positronium lifetime in addition to the uptake value. The first clinical positronium images, demonstrated with the J-PET scanner, show differences in positronium lifetime between healthy brain tissue and glioma, proving that positronium lifetime depends on tissue type under *in vivo* conditions (Moskal *et al* 2024).

Recent observation of non-maximal photon entanglement ($C_{QE} < 1$) originating from porous polymers (demonstrated using J-PET, the first QE-PET scanner (Moskal *et al* 2025a)) indicates that the degree of quantum entanglement, $C_{QE}$, may also be sensitive to the chemical environment of the annihilation site. This opens perspectives for developing $C_{QE}$ as a diagnostic biomarker, particularly for hypoxia (Moskal 2026).

Importantly, the information carried by $C_{QE}$ is not lost even if one of the photons undergoes Compton scattering within the body, provided the scattering angle, $\theta_s$, is less than ~36°. This is typical for events accepted by standard PET systems (see Fig. 1F in the Introduction). This follows recent observations showing that the degree of quantum correlation, $C(\theta_s)$ (see Fig. 1E in the Introduction), is preserved after a photon undergoes Compton scattering at angles $\theta_s <$ ~36° (Bordes *et al* 2024, Caradonna 2024, Ivashkin *et al* 2023, Parashari *et al* 2024, Tkachev *et al* 2025).

**Current and Future Challenges**

Imaging diagnostic parameters sensitive to the annihilation mechanism, and thus to the molecular environment at the annihilation site, requires the registration of more than two photons. In case of the positronium lifetime imaging, $\tau_{oPs}$, in addition to two annihilation photons a registration of prompt gamma is needed (Moskal 2020, Moskal *et al* 2019, 2020, 2024). In case of the 3γ/2γ decay ratio for positron annihilation, three annihilation photons have to be registered (Moskal *et al* 2021, Shimazoe and Uenomachi 2026). In case of 3γ/2γ decay rate ratio for ortho-positronium, four signals are required: three annihilation photons and prompt gamma. And for the case of the degree of quantum entanglement, $C_{QE}$, two annihilation photons and two Compton scattered photons have to be registered (Bordes *et al* 2024, Kožuljević *et al* 2026, Moskal *et al* 2025a, Romanchek *et al* 2024, Tkachev *et al* 2025). Registration of such events (illustrated in Fig. 3) requires high sensitivity PET systems with the capability of multi-photon data acquisition. In case of the degree of quantum entanglement, $C_{QE}$, it is at present difficult to precisely estimate the precision required for the useful medical applications. The first estimations of $C_{QE}$ were only recently reported (Moskal 2026). Calculations were performed assuming that photons are not entangled during the pick-off process (where positron annihilation with an environmental electron occurs in a non-pure quantum state), while they remain maximally entangled in all other processes (Moskal 2026). The estimated $C_{QE}$ values are 0.890 for adipose, 0.886 for isopropanol, 0.867 for water, 0.818 for cyclohexane, and 0.784 for isooctane (Moskal 2026). These results indicate that distinguishing between various tissue types requires a precision of $\sigma(C_{QE}) \approx 0.01$, as the measurement resolution must be significantly finer than the observed differences between substances, such as adipose and water (Moskal 2026). Furthermore, the changes in $C_{QE}$ as a function of oxygen partial pressure were estimated (Fig. 2). In the case of adipose tissue, the estimated change in $C_{QE}$ between physoxia ($pO_2 \approx 50$ mmHg) and hypoxia ($pO_2 = 10$ mmHg) amounts to 0.0019 (Moskal 2026). Therefore, to distinguish between physoxic and hypoxic conditions *in vivo*, the precision must be several times higher, on the order of $\sigma(C_{QE}) \approx 0.0005$. Fig. 2 shows the predictions for $C_{QE}$ compared to other possible diagnostic parameters as $\tau_{oPs}$, $R_{oPs\text{-}3\gamma/2\gamma}$, and $R_{3\gamma/2\gamma}$ (Moskal 2026).

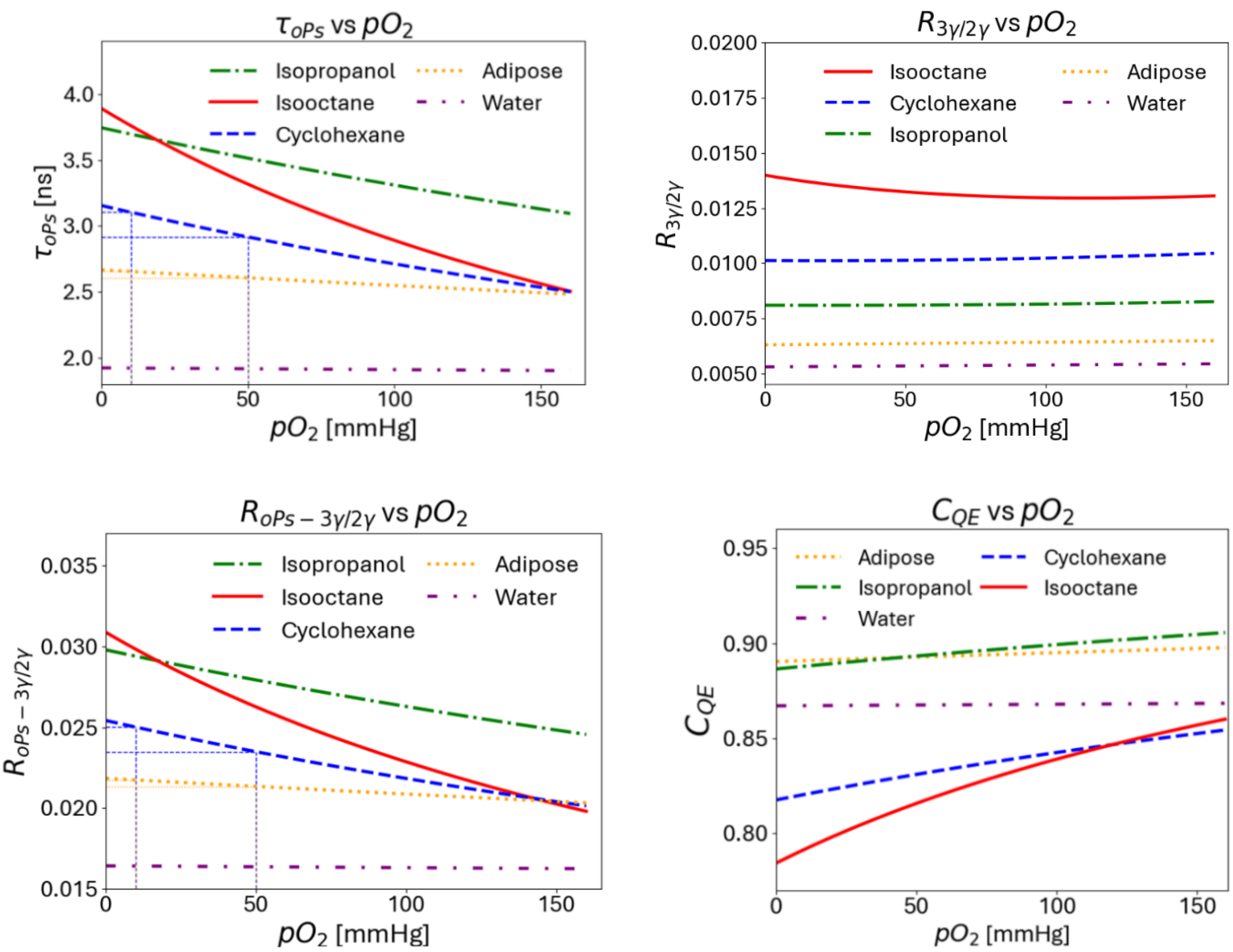

**Figure 2.** Predictions for the dependence of novel diagnostic parameters on oxygen partial pressure in adipose tissue, water, and organic liquids (isopropanol, isooctane, and cyclohexane) (Moskal 2026). Figures indicate mean ortho-positronium lifetime, $\tau_{oPs}$ (upper left), the 3γ/2γ decay rate ratio for ortho-positronium, $R_{oPs\text{-}3\gamma/2\gamma}$ (lower left), the overall 3γ/2γ rate ratio for positron annihilation, $R_{3\gamma/2\gamma}$ (upper right), and the degree of quantum entanglement, $C_{QE}$ (lower right), as a function of oxygen pressure, $pO_2$. In the left panel, vertical and horizontal lines indicate the change in $\tau_{oPs}$ and $R_{oPs\text{-}3\gamma/2\gamma}$ between physoxic (50 mmHg) and hypoxic (10 mmHg) conditions. The calculations of $C_{QE}$ are performed assuming the hypothesis that annihilation photons are not quantum entangled for the pick-off process but are maximally quantum entangled for all other processes. 

### Advances in Science and Technology to Meet Challenges

The sensitivity for quantum entanglement imaging is much lower than that of standard PET systems. In crystal-based PET detectors, double Compton scattering followed by the registration of scattered photons constitutes only a few percent of all detected 2γ events (Bharathi *et al* 2026, Kožuljević *et al* 2026). Although this is a small fraction, if total-body PET scanners with high sensitivity, such as 174 cps/kBq (Spencer *et al* 2021), were upgraded to register such events, they could enable effective quantum entanglement imaging with a sensitivity of approximately 3 cps/kBq. Crystal detectors capable of resolving double interactions are currently under rapid development (Bharathi *et al* 2026, Bordes *et al* 2024, Kožuljević *et al* 2026, Shimazoe *et al* 2026, Yoshida *et al* 2020).

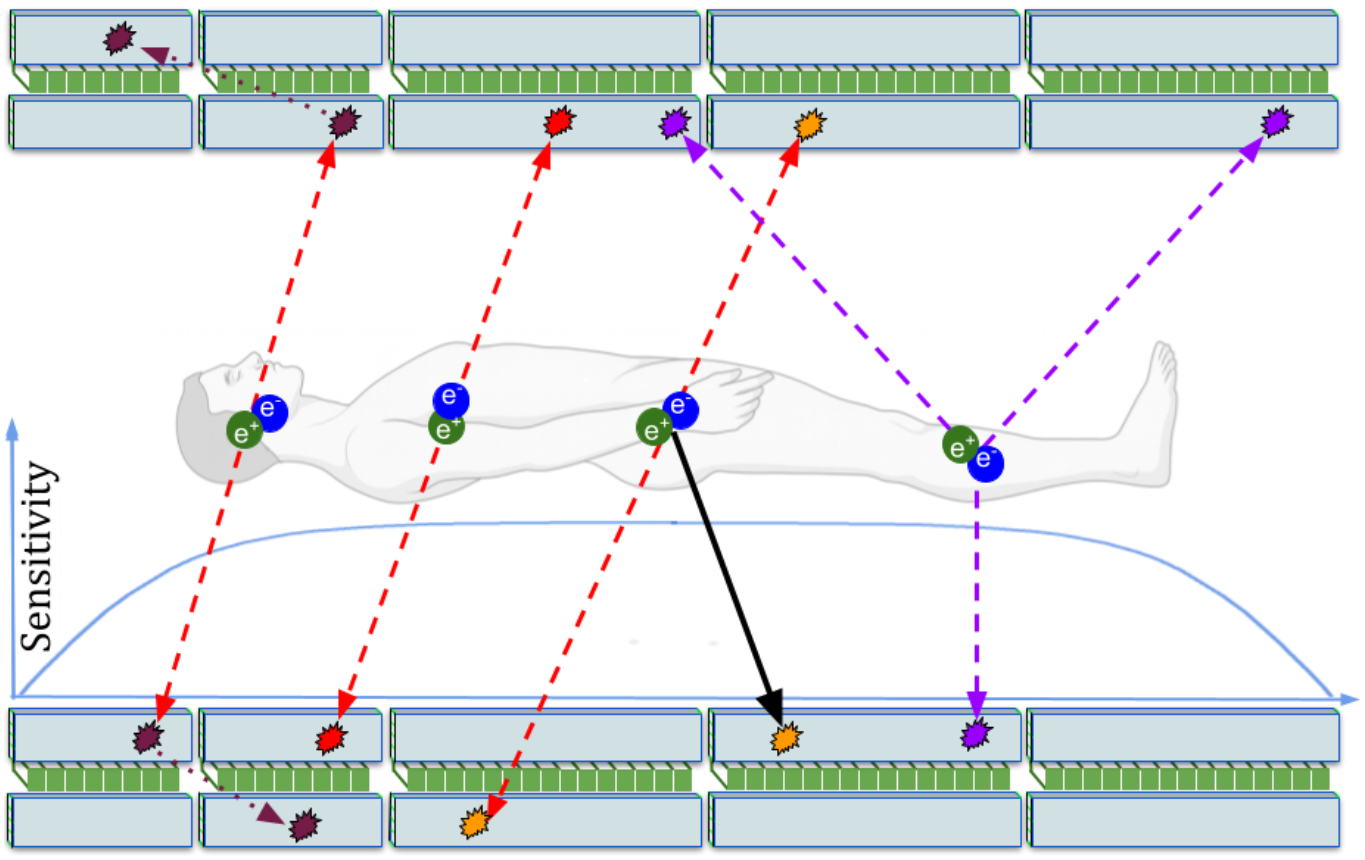

**Figure 3.** Schematic cross-section of the multi-photon total-body J-PET scanner, currently under construction at Jagiellonian University in Poland (Moskal *et al* 2021a). The system is designed for positronium and quantum entanglement imaging (Moskal 2018, Moskal 2021) and is capable of registering annihilation photons from e+e- annihilations into 2γ (red dashed arrows) and 3γ (blue dashed arrows). It also records events with 2γ and prompt γ from radionuclide de-excitation (solid black arrow), as well as Compton-scattered photons (dotted maroon arrows) for quantum entanglement imaging.

In the case of the plastic scintillator-based J-PET, the first QE-J-PET prototypes with a 50 cm axial field-of-view, capable of quantum entanglement imaging, have been built and used to study the degree of quantum entanglement (Moskal *et al* 2025a). The total-body J-PET shown in Fig. 3 is currently under construction, and its estimated sensitivity for quantum entanglement imaging reaches about 3 cps/kBq (Moskal 2026). Such sensitivity will enable to collect few million events useful for quantum entanglement imaging per organ using a standard PET protocol. This will be sufficient to achieve precision of $\sigma(C_{QE})$ = 0.0005 requiring about 4 000 000 events.

Quantum entanglement imaging (reconstructing a map of $C_{QE}$ parameter) will require development of the iterative image reconstruction methods that in addition to density distribution will enable reconstruction of the image of parameter $C_{QE}$. The challenge is analogous as in the case of the development of methods for positronium lifetime imaging (Moskal *et al* 2025). Hence in principle the methods developed so far (Chen *et al* 2023, Huang B *et al* 2024, Huang H *et al* 2025, Qi and Huang 2022) should be adaptable also for the degree of quantum entanglement, $C_{QE}$, imaging.

Finally, development of the total body J-PET will also open perspectives for the investigations of multi-partite entanglement in 3γ annihilations (Nowakowski and Bedoya Fierro 2017) as discussed in Section "Quantum Entanglement in Three-Photon Positron-Electron Annihilation".

### Concluding Remarks

Recent observations of non-maximal entanglement of photons originating from electron-positron annihilation in matter (Moskal *et al* 2025a) indicate that the degree of quantum entanglement may depend on the annihilation mechanism. Consequently, it can serve as a hallmark of tissue type and oxygenation levels (Moskal 2026). This opens perspectives for using the degree of quantum entanglement of photons recorded during positron emission tomography as a diagnostic biomarker. High-sensitivity total-body PET scanners, when upgraded for multi-photon acquisition, will enable researchers to conduct studies toward the clinical translation of QE-PET. The first-in-human quantum entanglement imaging was demonstrated recently using the modular J-PET scanner (Moskal *et al* 2026), constituting an important step toward translating quantum entanglement imaging into the clinic.

### Acknowledgements

We are grateful to Mgr. Atharv Dalvi, Dr. Bartosz Leszczyński, Dr. Aleksander Khreptak, and Mgr. Keyvan

Tayefi Ardebili, for the help in preparing figures and editorial corrections. We acknowledge support from the European Union via ERC Advanced Grant POSITRONIUM no. 101199807, the National Science Centre of Poland through grants no. 2021/42/A/ST2/00423, 2021/43/B/ST2/02150, 2022/47/I/NZ7/03112, the Ministry of Science and Higher Education through grant no. IAL/SP/596235/2023 and SPUB/SP/627733/2025, the SciMat and qLife Priority Research Areas budget under the program Excellence Initiative – Research University at Jagiellonian University.

# 15. Feasibility of Imaging and pH Sensing for Nuclear Medicine Using Perturbed Angular Correlation Spectroscopy

*Mizuki Uenomachi*[1], *Kenji Shimazoe*[2]
[1]Institute of Science Tokyo, Tokyo, Japan
[2]The University of Tokyo, Tokyo, Japan
uenomachi.m.4403@m.isct.ac.jp, shimazoe@g.ecc.u-tokyo.ac.jp

**Status**

Molecular imaging is an essential technology in the fields of biology, medicine and pharmaceutical research. For example, fluorescence imaging not only visualizes cells but also can provide information such as intermolecular interactions (Demchenko 2006) or intracellular drug uptake (Motamarry *et al* 2019). Moreover, the emerging technology of nanodiamond quantum sensors can provide local biological information such as pH (Fujisaku *et al* 2019), temperature (Fujiwara *et al* 2020), and viscosity (Gu *et al* 2023). However, these technologies are restricted to superficial tissues and cannot be used to access information deep within the body, as they rely on optical light, which has poor tissue penetration.

In contrast, nuclear medicine imaging techniques, such as positron emission tomography (PET) and single photon emission computed tomography (SPECT), can visualize the accumulation of radiopharmaceuticals deep inside the body. Diagnostic images are generated by estimating the positions of radionuclides through detection of emitted X-rays or gamma rays, which exhibit strong tissue penetration. PET constrains the position of a positron to a line between two detectors by detecting a pair of 511 keV annihilation gamma rays produced by electron-position annihilation. SPECT estimated the position of a single photon emitting nuclide by limiting the direction of incoming gamma rays. However, these conventional imaging methods cannot offer local biological information.

We have proposed and demonstrated double-photon coincidence imaging using cascade nuclides that emit multiple gamma rays sequentially (Orita *et al* 2021, Shimazoe *et al* 2017, Uenomachi *et al* 2020, 2021, Yan *et al* 2023). Coincidence detection of cascade gamma rays using directional detectors, which can estimate the direction of incoming gamma rays, enables precise localization of a radionuclide to a single point. Furthermore, a more detailed analysis of the physical characteristics of cascade nuclides indicates that we can extract information on external fields acting upon the nucleus and their influencing factors. A cascade nuclide in an excited state emits gamma rays sequentially as it transitions through inter mediate energy levels, eventually reaching its stable state. If the second gamma ray is detected after the first, the angle between the directions of the emitted gamma rays is correlated, known as “angular correlation”. In addition, the angular correlation of emitted gamma rays is perturbed when the nucleus is subjected to an external electromagnetic field (Bauer 1985, Rinneberg 1979). We have proposed a novel nuclear medicine imaging method combining double-photon coincidence imaging and perturbed angular correlation (PAC) spectroscopy (Figure.1) (Shimazoe *et al* 2022). This new technology can provide both positional and biological information and is anticipated to constitute the quantum diagnostic imaging modality capable of extracting localized physiological information from deep within the body.

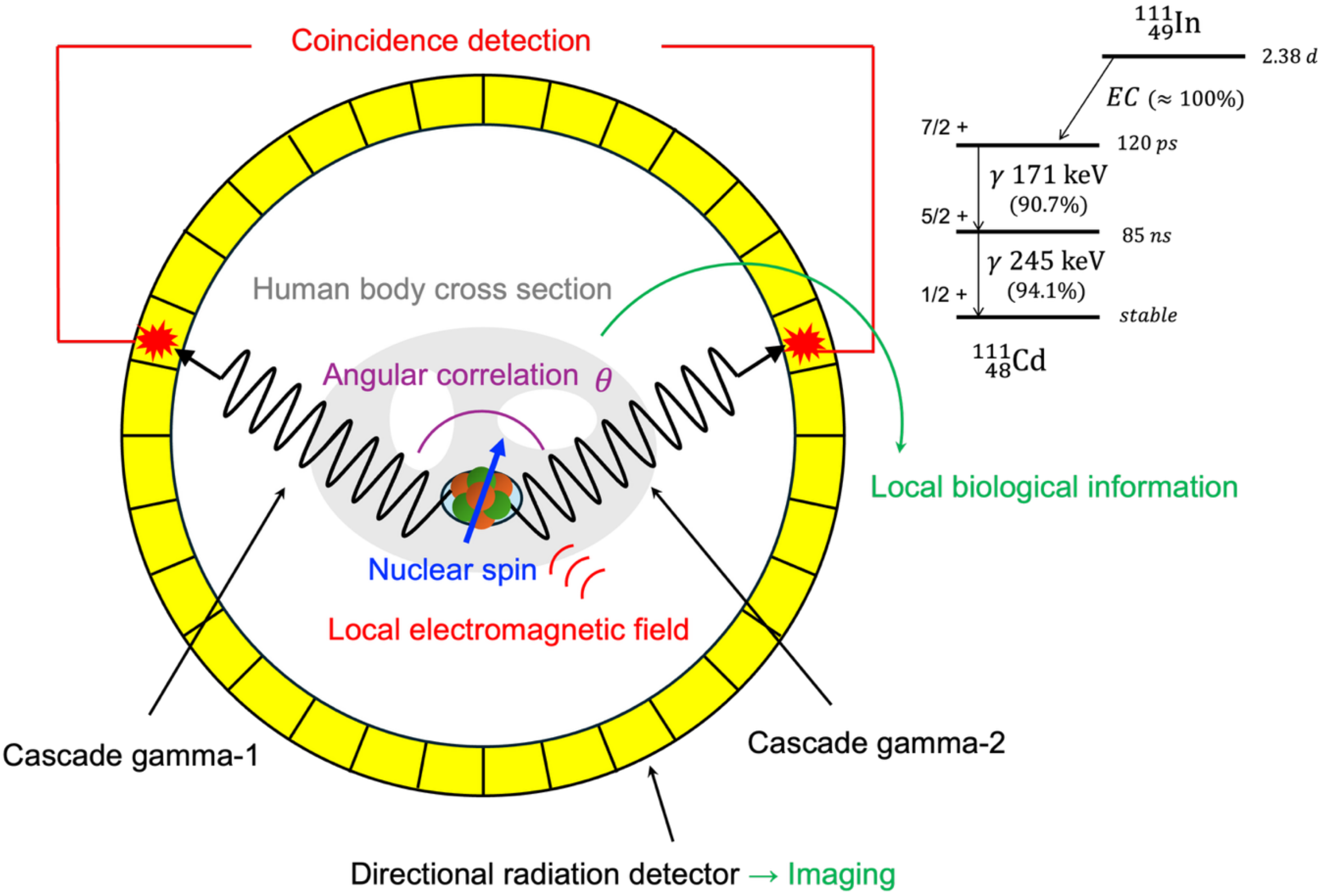

**Figure 1.** Concept of the proposed quantum diagnostic imaging method that combines double-photon coincidence imaging with perturbed angular correlation (PAC) spectroscopy, and $^{111}$In decay scheme. This technique enables simultaneous acquisition of spatial and biological information by detecting angular correlations influenced by local electromagnetic fields, allowing extraction of localized physiological information from deep within the body.

### Current and Future Challenges

We have demonstrated a novel method of imaging and pH sensing in nuclear medicine using $^{111}InCl_3$ solutions (Shimazoe *et al* 2022). $^{111}$In is widely used for SPECT diagnostics, emitting the cascade gamma rays with energies of 171 keV and 245 keV, and featuring and intermediate-state lifetime of 85 ns. Additionally, it is well known as one of the cascade radionuclides used in PAC spectroscopy in the field of materials sciences. The changes in angular correlation due to differences in the pH of the $^{111}$In solution have been reported in several papers (Akselrod *et al* 2000, Shukri *et al* 1988). We also measured the changes in angular correlation caused by differences in the pH of $^{111}$In solution using our originally developed GAGG pixel detectors. The eight GAGG detectors were arranged in a ring, allowing for the measurement of angles distribution between the two gamma rays in all directions from 0 to 180 degrees. Consistent with previous reports, we observed variations in the emission ratio of the second gamma ray at 90 degrees in the pH range of 3 to 5. Moreover, four GAGG detectors, each equipped with a parallel-hole collimator on the front, were arranged at 90-degree intervals, and we successfully achieved simultaneous imaging and pH sensing of $^{111}$In solutions in microtubes (Shimazoe *et al* 2022).

With the aim of improving detection efficiency than that using mechanical collimators and detecting angular correlation detection in all directions, we also demonstrated Compton imaging as pH sensing (Kim *et al* 2025). Since Compton imaging does not require any mechanical collimators, it can be used for both gamma ray imaging and angular correlation measurement. The 245 keV gamma rays were used for Compton imaging based on Compton scattering and subsequent photoelectric absorption of the scattered photons (hereafter referred to as Compton coincidence events), while the 171 keV gamma rays were utilized for angular correlation measurements via photoelectric absorption events, in coincidence with the Compton coincidence events of the 245 keV gamma rays. Since point sources were used, only coincidence events of 245 keV and 171 keV gamma rays that correspond to Compton cones reconstructed from the 245 keV events passing through the known source positions were used for angular correlation analysis. A challenge of this method is that it does not localize the position of each coincidence event, making three-dimensional imaging difficult and limiting its applicability to complex geometries. For medical applications, it is essential to enable accurate extraction of angular correlation at each position, even in complex distributions of radiopharmaceuticals.

**Advances in Science and Technology to Meet Challenges**

To accommodate complex distribution, it is necessary to accurately localize the position of each coincidence event between 245 keV and 171 keV gamma rays. In this regard, double-photon Compton imaging (Orita *et al* 2021, Uenomachi *et al* 2020) provides an effective solution. In the double-photon Compton imaging method, by simultaneously detecting Compton coincidence events from each of the two cascade gamma rays, the position of the radionuclide can be localized at the intersection point of the Compton cones reconstructed from each Compton coincidence event (Figure.2). To combine this double-photon coincidence imaging method with pH sensing using $^{111}$In, a measurement system capable of high timing resolution (below a few nanoseconds), high energy resolution, and detection of low-energy Compton scattering events below several tens of keV is required. One of ideal combinations is a Compton camera consisting of a silicon pixel semiconductor detector as the scatterer and a scintillator pixel detector such as GAGG or $CeBr_3$ as the absorber. By employing a silicon detector as the scatterer, high energy resolution and the ability to detect low-energy Compton scattering events can be achieved, while high timing resolution can be realized with a scintillator detector as the absorber. On the other hand, the use of silicon detectors as the scatterer is expected to result in significantly lower detection efficiency compared to Compton cameras composed of scintillator detectors. Furthermore, the double-photon Compton imaging itself also exhibits a detection efficiency approximately $10^{-3}$ times that of conventional single Compton imaging (Uenomachi *et al* 2020). Thus, detection efficiency is one of the major challenges, and one of most straightforward approaches is to increase the number of detectors. This can be achieved by deploying multiple Compton camera units or by stacking silicon detectors used as scatterers. Alternatively, other materials such as CdTe may have the potential to serve as more sensitive scatterers, provided they can be fabricated with sufficiently thin thicknesses. For future medical applications, it is essential to develop a system that achieves not only appropriate detection efficiency, but also high imaging capability and time resolution.

As advanced techniques of a method to induce local interactions by applying an external magnetic field or electric field gradient, we used an ultrasonic transducer and a magnetic field generator, and observed changes in the angular correlation (Sensui *et al* 2023, Ueki *et al* 2023). Since these methods induce local electromagnetic interactions, such as Larmor precession externally, they are expected to enable high-sensitive imaging using non-directional detectors in the future.

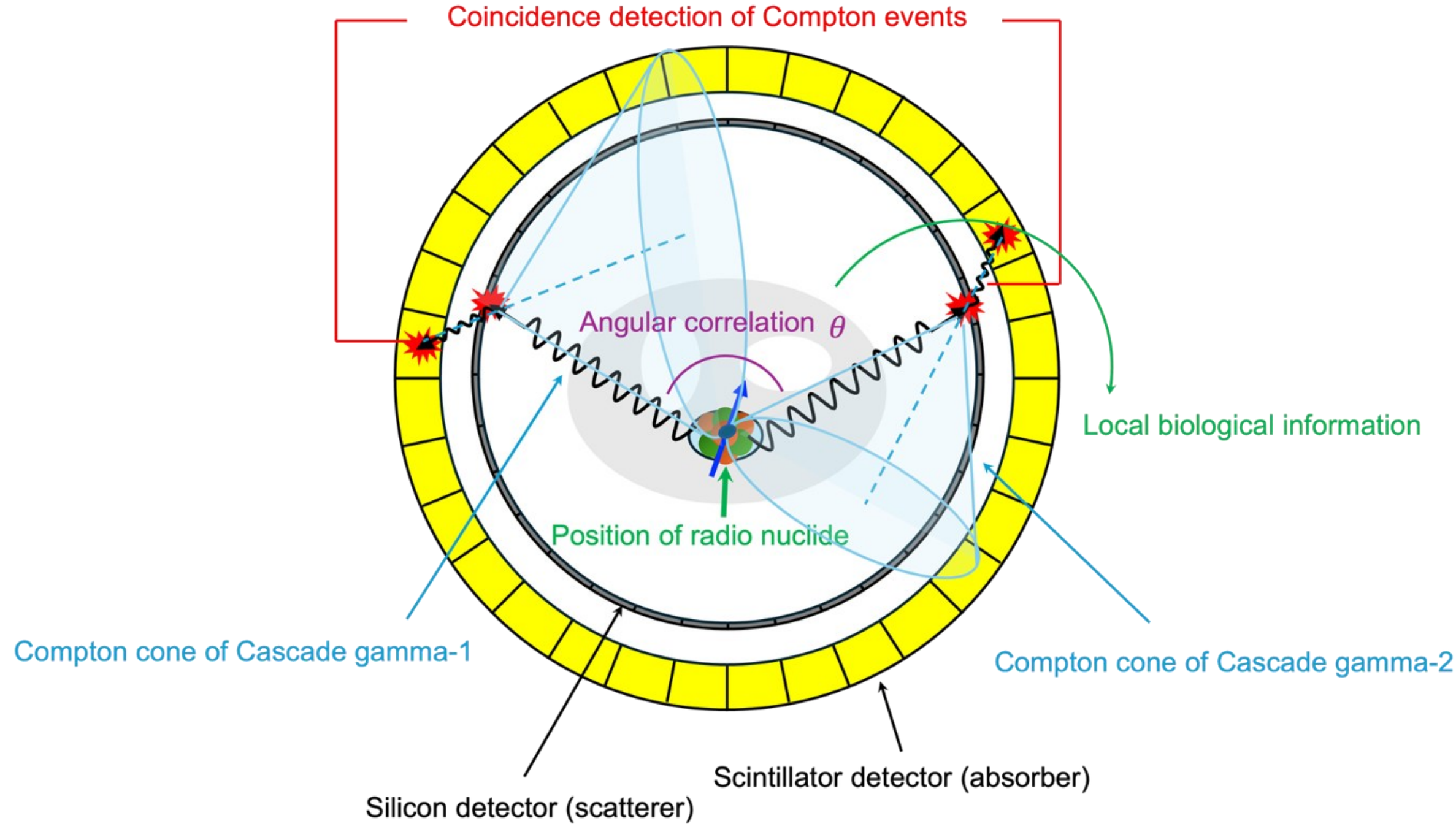

**Figure 2.** Concept of double-photon Compton imaging applied to pH sensing using a Compton camera composed of a silicon pixel detector as the scatterer and a scintillator pixel detector (e.g., GAGG or $CeBr_3$) as the absorber. This configuration enables efficient and omnidirectional detection of angular correlations, along with accurate localization of the radionuclide position.

## Concluding Remarks

We propose a novel nuclear medicine imaging approach that integrates double-photon coincidence imaging with PAC spectroscopy, enabling simultaneous acquisition of spatial and biological information using cascade-emitting radionuclides such as $^{111}$In. Simultaneous imaging and pH sensing were demonstrated using a parallel-hole collimator setup. To improve detection efficiency and achieve omnidirectional angular correlation measurement, Compton imaging was explored as an alternative to mechanical collimators, utilizing 245 keV Compton events for imaging and 171 keV photoelectric absorption events for angular correlation. Although precise event localization is a challenge, double-photon Compton imaging addresses this by localizing radionuclides at the intersection of Compton cones from each gamma ray. A practical system requires high timing and energy resolution, and the ability to detect low-energy Compton scattering. A promising configuration involves a Compton camera with a silicon pixel detector as the scatterer and a scintillator, such as GAGG, as the absorber. Additionally, we investigated pH-dependent angular correlation in DOTA-chelated $^{111}$In (Shimazoe *et al* 2024), relevant for medical and biological applications, as $^{111}$In is commonly labeled with chelators for drug delivery. Although further validation of angular correlation changes in different molecular forms is needed, this technique holds strong potential for advancing next-generation nuclear diagnostic imaging.

## Acknowledgements

We gratefully acknowledge financial support from JST, PRESTO Grant Number JPMJPR23F1, JPMJPR17G5 and JSPS KAKENHI Grant Number 22H05022.

# 16. Conclusions

*Paweł Moskal*
Marian Smoluchowski Institute of Physics, Jagiellonian University, Kraków, Poland
Center for Theranostics, Jagiellonian University, Kraków, Poland
p.moskal@uj.edu.pl

This roadmap describes the current status and perspectives of research aiming to transition Quantum-Entangled PET (QE-PET) from a theoretical curiosity of fundamental physics to a practical medical imaging modality. By moving beyond the simple detection of annihilation photon coincidences to the exploitation of their quantum entanglement and polarization correlations, QE-PET opens new perspectives for higher-quality PET diagnostics through the suppression of noise from random coincidences and by enabling the application of entirely new diagnostic indicators beyond the Standardized Uptake Value.

The transition to clinical QE-PET hinges on overcoming the low sensitivity inherent in multi-photon (fourfold) coincidence detection. This roadmap presents the status and perspectives of QE-PET systems based on crystal scintillators (e.g., LYSO, GAGG), semiconductors (e.g., CZT), and plastic scintillators. The experimental possibilities and theoretical descriptions of multiple Compton scattering events are explored as the core of a functional QE-PET system. Various system solutions are described, including Whole-Gamma Imaging, Compton Imaging, Double Photon Compton Imaging, combined PET and Compton Camera systems, and the design of a total-body QE-J-PET scanner utilizing plastic scintillators.

This roadmap discusses the implications for QE-PET development arising from recent observations that the quantum correlation between photons is not lost when one of them undergoes small-angle Compton scattering. Results from independent experimental groups and theoretical explanations for these observations are presented.

Finally, this roadmap explores the application of quantum entanglement between annihilation photons - not only to enhance standard PET imaging by refining the density distribution of annihilation sites but also to establish the degree of quantum entanglement as a diagnostic biomarker. This emerging application promises new insights into intramolecular architecture and tissue oxygenation. Furthermore, we discuss the potential for pH sensing within this modality. Whether entanglement-based imaging and pH mapping can be successfully translated into clinical practice remains an open question and a subject of ongoing research. This roadmap serves as an invitation to the scientific community to join this burgeoning field.